\documentstyle[12pt]{article}
\newcommand{\be}{\begin{equation}}
\newcommand{\ee}{\end{equation}}
\newcommand{\bea}{\begin{eqnarray}}
\newcommand{\eea}{\end{eqnarray}}
\newcommand{\sirc}[1]{\stackrel{\circ}{#1}}

\font\mybb=msbm10 at 12pt

\def\bb#1{\hbox{\mybb#1}}
\def\id{\protect{{1 \kern-.28em {\rm l}}}}

\def\Z {\bb{Z}}

\makeatletter
\@addtoreset{equation}{section}
\makeatother

\def\appendix#1{
  \setcounter{section}{1}
  \setcounter{equation}{0}
  \renewcommand{\thesection}{\Alph{section}}
  \section*{Appendix \thesection\protect\indent \parbox[t]{11.715cm} {} }
  \addcontentsline{toc}{section}{Appendix \thesection\ \ \ }}

\begin{document}
\begin{flushright}
YITP-SB-99-56\\
hep-th/9911238\\
\end{flushright}

\begin{center}
{\LARGE Consistency of the $AdS_7\times S_4$ reduction and\\
the origin of self-duality in odd dimensions}
\vskip1truecm

{\large\bf Horatiu Nastase,
 Diana Vaman and\\
Peter van Nieuwenhuizen}\footnote{Research supported by NSF Grant 9722101;
E-mail addresses:
hnastase@insti.physics.sunysb.edu,\protect\\
dvaman@insti.physics.sunysb.edu,
vannieu@insti.physics.sunsyb.edu
} \\ {\large   C.N.Yang
Institute for Theoretical Physics,\\ S.U.N.Y. Stony Brook, NY 11794-3840,USA\\}
\vskip1truecm
\end{center}
\abstract{
\parbox {4.75 in}{

~~~~~We discuss the full nonlinear Kaluza-Klein (KK)
reduction of the original formulation of d=11 supergravity on
$AdS_7\times S_4$ to gauged maximal ({\cal N}=4) supergravity in 7 dimensions.
We derive the full nonlinear embedding of the $d=7$ fields in the
$d=11$ fields
(``the ansatz'') and check the consistency of the ansatz by
deriving the d=7 supersymmetry laws from the d=11 transformation laws in
the various sectors.
The ansatz itself is nonpolynomial but the final $d=7$ results are polynomial.
The correct $d=7$ scalar potential is obtained.
For most of our results the explicit form of the matrix $U$ connecting
the $d=7$ gravitino to the Killing spinor is not needed, but we derive the
equation which $U$ has to satisfy and present the general solution.
Requiring that the expression $\delta F=d\delta A$ in $d=11$ can be written as
$\delta d(fields\;\; in \;\;d=7)$, we find the ansatz for the 4-form $F$.
It satisfies the Bianchi identities. The corresponding ansatz for the
3-form $A$ modifies
the geometrical proposal by Freed et al. by including
$d=7$ scalar fields.
A first order formulation for  $A$
in $d=11$ is needed to obtain the d=7 supersymmetry laws and the action for
the nonabelian selfdual antisymmetric tensor field
$S_{\alpha\beta\gamma,A}$. Therefore selfduality in odd
dimensions originates from a first order formalism in higher dimensions.
}}
\newpage
\section{Introduction}

${}$

The consistency of Kaluza Klein (KK) truncation at the
nonlinear level of a massless gravitational higher-dimensional field
theory to a massless gravitational lower-dimensional field theory has
been a fundamental problem for a long time.
In this article we present a complete solution for a particular model.
Namely, we consider the original formulation of d=11 supergravity \cite{cjs},
without any string extensions, and write all components of all d=11
fields, denoted by $\Phi(y,x)$, as nonlinear expressions in terms of the
massless d=7 fields $\phi(y)$ and spherical harmonics $Y(x)$ on $S_4$. All
spherical harmonics will be expressed in terms of (products of) Killing spinors
 $\eta^I (x)$ with $I=1,...,4$.
Hence we present below explicit expressions for $\Phi(y,x)=
\Phi(\phi(y),\eta(x))$. The consistency of this 'ansatz' amounts to
proving that the d=11 supersymmetry variations of $\Phi$ produce the
correct d=7 supersymmetry variations of $\phi$. A particular nonlinear
combination of $\phi$'s and $\eta$'s enters as a $4\times 4$ matrix
$U^{I'}_{\;\;I}(y,x)$.
It appears in the ansatz for all fermionic $\Phi$'s and in the relation
$\varepsilon(y,x)\sim
\epsilon_{I'} (y)U^{I'}_{\;\;I}(y,x)\eta^I(x)$ between the d=11
susy parameter $\varepsilon(y,x)$ and the d=7 parameter $\epsilon(y)$.
This matrix $U$ must satisfy a linear matrix equation, and we present
explicit solutions.
This is the first time a complete nonlinear KK
reduction of an original supergravity theory to all massless modes in
a lower dimension has been given.
 There exists in the literature also a modified version of
$d=11$ supergravity with a local $SU(8)$ symmetry and for this theory
de Wit and Nicolai \cite{dwn84,dwnw,dwn86,dwn87} have
shown that the KK reduction is consistent. Because in their version of the
theory the matrix $U$ only described gauge degrees of freedom, they did
not try to determine it. (In an earlier study of the KK reduction of the
original $d=11$ supergravity to $d=4$ they gave partial results on the matrix
$U$ \cite{dwn84,dwnw}, but then they abandoned this project. In fact, the
matrix U was first introduced in an implicit form in \cite{anp}, while the
contributions to U quadratic in scalars were computed in \cite{nilsson}.
Other KK reductions to a subset of massless fields have also in some cases
been shown to be consistent \cite{dnpw,dp,cetal,cllp}.

This article contains a detailed and self-contained derivation of our results,
and is an expanded version of a previous letter \cite{nvv}. The basic problem
we must analyze is that in relations like $\delta \phi (y)Y(x)
=\varepsilon (y,x)
\Phi (y,x)$ one finds products of $Y(x)$'s on the r.h.s. Hence one must prove
relations of the form $Y(x)=\Pi Y(x)$. (An example of such a relation is
(\ref{u1}).)  This requires a large number of properties of spherical
harmonics. For the reader who is not an expert in these matters we have added
a special section (section 3) in which  we give a derivation of all identities
we needed, starting from scratch. The basic object, in terms
of which we express
all spherical harmonics, is the Killing spinor. It is a square root of a
Killing vector and is the spherical harmonic of the local supersymmetry
parameters. The spherical harmonics of scalars, spinors and other
fields are expressed into Killing spinors. We shall analyze all bosonic
variations $\delta\Phi (\phi(y), \eta (x))$ up to terms with
three d=7 fermion fields, and all fermionic variations up to terms with two
d=7 fermion fields, but all results are to all orders in bosonic fields.
Previous work in this area has also made the same restriction
\cite{dwnw,dwn86,dwn87}. In principle one could also study the variations
containing higher orders in fermi fields with our methods, but this is
algebraically very complicated and given all small miracles we have
encountered in our work, we expect that consistency holds there
as well.

Initially, the aim of KK reductions of higher dimensional supergravities
(sugras) was to find realistic field theories in $d=4$ dimensions, and
the compactification on $AdS_4\times S_7$ seemed promising (\cite{dewit}, see
also \cite{trieste}).
However, it was later realized that it was impossible to obtain in this
way chiral fermions and/or a vanishing cosmological constant \cite{wittenul}.
Recent developments in the AdS-CFT correspondence have renewed interest in KK
compactifications on anti-deSitter (AdS) spaces
\cite{mald,gkp,witten,3point,4point,cfm,lmrs}
(for a recent review see \cite{asmo}).

While KK reduction (more precisely, truncation to the set
of massless fields) on tori is
always guaranteed to be consistent, this is not in general so.
A simple  example of an inconsistent truncation is
Einstein gravity, which cannot be truncated to
gravity plus gauge fields of the symmetry group K of the compact manifold
$M_K$. This was already noticed by Kaluza and others who studied the reduction
from 5 to 4 dimensions. They realized that the field equation for the scalar
contains source terms depending on the gauge fields; hence setting the
scalar to zero was inconsistent. But even keeping the scalars in the
theory is in general
not enough to obtain consistency. In the Einstein equation, the terms
involving the gauge fields will have as spherical harmonics $V_{\mu}^a
V^{\mu}_b$, whereas the rest of the terms have spherical harmonic 1.
However, if one restricts
the gauge group $K$ to a subgroup $K'$ such that
the subset of
Killing vectors satisfies  $V_{\mu}^{a}V^{\mu}_{b}=\delta^{a}_{b}$, then
the KK reduction is consistent
($a,b\in K'$) \cite{dnpw} (In sugra, more fields are present - for instance in
11 dimensions we have the antisymmetric tensor $A_{\Lambda\Pi\Sigma}$ - and
so the condition $V_{\mu}^aV^{\mu}_b=\delta^a_b$ gets modified in a
complicated way).
In the case of compactifications on a group
manifold and on spheres $S^{4n+3}$, one can find such  subsets of
Killing vectors, whereas on $S^{2n}$ one can not (as shown in \cite{dnpw}).
For the Maxwell-Einstein system, on the other hand,
the truncation to massless fields is  consistent as far as the gauge
symmetries are concerned \cite{ck}.

The study of KK truncations took a new direction with the advent of
supergravity.
The most intensively studied model was the N=8 sugra, which proved to be very
challenging.  In the course of its study,
other criteria for consistent truncations were found in
\cite{dp,ps}.
In \cite{dp} it was realized that
for a field theory  invariant under a group G, a consistent truncation can
be obtained by retaining only all fields which are singlets under a
subgroup $G'\in G$.
In general, this gives an infinite set of fields (the untruncated
set is of course always
 infinite). If G is the isometry group of the KK compactification
on a space $M_k$, and $G'$ acts transitively on $M_k$
(i.e. any point in $M_k$ can be reached from any other point by a $G'$ gauge
transformation), the subset of fields
is finite. This requirement implies in particular
that $M_k$ is a coset manifold, $G/H=G'/H'$. In the
case of the $AdS_4\times S_7$ compactification, this  mechanism gives a group
theoretical explanation  for the consistency of a particular   truncation
consisting of massless  gauged N=3
 supergravity coupled to some massive matter multiplets (this
system is not contained in N=8 sugra).
 In  \cite{ps} it was claimed that
 the truncation to 4d $N\ge 1$ sugra multiplets
with second order formulation for the fields is always consistent, and it was
speculated that the truncation to a sugra
multiplet is always consistent.

The study of the consistency of the KK reduction of d=11 sugra at the
nonlinear level was tackled in a series of papers by de Wit and Nicolai.
In their pioneering work, they followed
two approaches. One was to construct directly an ansatz
for the 11d metric as opposed to the vielbein; this yielded an ansatz for
the fermions only up to a matrix U, depending on the 4d scalar
fields. The dependence of this matrix on one of the scalars was even
determined, but they did not succeed in finding the dependence on all
other scalars. They then switched to
another approach in which they reformulated 11d sugra as a theory  with
a local $SO(1,3)\times SU(8)$
tangent space symmetry group, and dimensionally reduced
this theory. This latter approach
was used to prove consistency of the reduction by reproducing  the 4d susy
transformations laws and equations of motion, but left the ansatz of the
original 11 dimensional fields in terms of the 4d ones implicit.
The ansatz for $F_{\Lambda\Pi\Sigma\Omega}$ was not explicitly given in the
first approach. In the second approach no $F_{\Lambda\Pi\Sigma\Omega}$ appears,
but the equations needed to construct $F_{\Lambda\Pi\Sigma\Omega}$
from the second approach were highly involved.

Lately, suggested by the need to embed $AdS_4$ and $AdS_7$
black holes as solutions of 11 d sugra (and $AdS_5$ black holes as  solutions
of 10d IIB sugra), the authors of \cite{cetal, cllp}
 proposed nonlinear ans\"{a}tze
for the embedding in 11d sugra and 10d IIB sugra of a further truncation
of the gauged sugras in 4, 7 and 5 dimensions to a set of U(1) gauge
fields and scalars (in the $AdS_7$ case, two U(1) gauge fields and
two scalars). The consistency of the truncations was proven by reproducing
the lower dimensional equations of motion from the ones of d=11 and d=10
sugras (see next section). After that, a series of other papers looked at other
truncations of 11d sugra on $AdS_7\times S_4$ (a model with one scalar, one
SU(2) gauge field and one antisymmetric tensor in \cite{lupo}, a model with
4 scalars in \cite{cglp}). Consistent truncations of bosonic subsets of
fields to other dimensions were also considered in \cite{lpt,cjlp,volk,chavo}.
The authors of \cite{chsab} also worked out an ansatz or an $S_3$ truncation of
${\cal N}=1 \;\;d=10$ sugra which reproduced the bosonic action of ${\cal N}=2
\;\d=7$ gauged sugra; however, they did not explicitely checked consistency.

In a previous letter \cite{nvv},
we have given the full nonlinear ans\"{a}tze for the KK
reduction of 11d sugra on $AdS_7\times S_4$.
Here we present the details of the construction.

Crucial for the proof of consistency is a proper understanding of the
origin of the concept of self-duality
in odd dimensions -- seven in this case.
In sugra, this mechanism of selfduality in odd dimensions leads to a
consistent coupling of a multiplet of massive antisymmetric tensor fields
to nonabelian gauge fields. It was first discovered in a study
of the KK reduction of d=11 sugra on $S_4$ \cite{tpn}, and found to give
a formulation dual to Chern-Simons theory for the abelian case \cite{dj},
but not for the nonabelian case \cite{klr}.
It was found that the second order field equation for the field $A_{\alpha
\beta\gamma , A}$, which one obtains from KK reduction of the field $
A_{\Lambda\Pi\Sigma}$, was of the form $\Box A+m^2 A+m\epsilon\partial A=0$.
Adding in d=7 an auxiliary field ${\cal B}_{\alpha\beta\gamma , A}$ with
action ${\cal B}^2$,
a rotation from A,{\cal B} to S,b yielded the final massless field
$S_{\alpha\beta\gamma, A}$ and a massive mode $b_{\alpha\beta\gamma ,A}$,
\be
A=S+b, {\cal B}=b- O(S+b)\label{rotation}
\ee
where $ O$ is an operator involving derivatives of the form
$\epsilon D +m$, see (\ref{auxili}). We have found that
$\epsilon {\cal B}$
is obtained by KK reduction of the auxiliary field ${\cal B}_{\Lambda\Pi
\Sigma\Omega}$ in d=11, which is obtained by using a first-order formulation
for the 3-index photon in d=11.

This paper is organized as follows. In section 2.1 we discuss the issue of the
 consistency of KK truncations in general. We argue that it is sufficient
to obtain the susy transformation laws of the dimensionally reduced theory
from the susy transformation laws of the untruncated theory. In section 2.2
 we present d=11 sugra with a first-order formulation for the 3-index
photon $A_{\alpha\beta\gamma ,A}$, and prove its local supersymmetry.
 The linearized KK reduction is given  in section 2.3. Although it was
already given in \cite{pnt}, we summarize it here since it forms the starting
point for the nonlinear extension.
In section 3 we discuss Killing spinors and spherical harmonics. We begin
in section 3.1
with a description of the spherical harmonics used for  the massless multiplet
 and give various useful identities in section 3.2.
Section 4 is the heart of this paper: it contains
the proof of the consistency of KK reduction. In 4.1 we present the
 full ansatz for
the d=11 vielbein and gravitino and we discuss the susy transformation laws
of the vielbein. In 4.2 we derive an ansatz for the d=11
3-index antisymmetric tensor field $A_{\Lambda\Pi\Sigma}$ by a partial
analysis of the susy laws for $A_{\Lambda\Pi\Sigma}$ and by
a preliminary study of the susy
transformation laws of the gravitino. It leads to the d=7 'self-duality in odd
dimensions'. In 4.3 we give the reduction of the rest of the d=11 gravitino
transformation laws and derive the  transformation law of the 3-index d=7
tensor $S_{\alpha\beta\gamma ,A}$. In section 5 we derive for completeness
the seven dimensional bosonic equations of motion from
the 11 dimenisonal bosonic equations of motion, with the gauge fields set to
zero.  We also derive the seven
dimensional bosonic action and compare our ansatz with other ans\"{a}tze for
consistent truncation to bosonic subsets of fields.
We finish in section 6 with conclusions and discussions. In Appendix 1
we give our
conventions and some useful relations for Dirac matrices, including certain
completeness relations. In Appendix 2 we give
some details of the charge conjugation matrices in various dimensions and
 properties of modified Majorana spinors.

\section{First-order d=11 supergravity and linearized KK reduction}
\subsection{KK consistency}

Let us briefly discuss the issue of the
consistency of the KK reduction. In general,
if we truncate the fields in a Lagrangian by putting the 'massive' ones
$\{ \phi_{(n)}\}$ to zero and keeping only the 'massless' ones $\{ \phi_{(0)}
\}$, we have to check that the full equations of motion and
transformation laws (for all the symmetries of the system) are consistent
under this truncation.

More precisely, the equations of motion of the massive fields, $\delta S/
\delta \phi_{(n)}$ must not contain any term depending only on the massless
fields (at the linear level this is true because $\phi_{(n)}$  are
eigenfunctions of the kinetic operator) because otherwise setting
$\phi_{(n)}=0$ would be inconistent.
Similarly, in the transformation rules for the symmetries of the
system, the massive fields should not transform into a term with only
massless ones, again because
then we cannot put the massive fields to zero.

In the case of the KK reduction of a D dimensional sugra action, $S^{(D)}(
\{ \Phi \} )$, we usually know the Lagrangian  in lower (d) dimensions
$S^{(d)}( \{ \phi_{(0)} \} )$
which we should obtain after truncation, but we need  to find the
nonlinear ansatz which specifies how the final massless fields $\phi_{(0)}$
are embedded in the higher-dimensional fields $\Phi$.
For a  consistent truncation, there should be no term linear in massive fields
 in the untruncated action $\int d^dx \phi_{(n)}
f_n(\{ \phi_{(0)}\})$, because such a term would
 give an inconsistency of the equations of
motion. Let's assume that both the equations of motion and the susy laws are
inconsistent, i.e. there exists a massive field which
varies into a term involving
only massless fields, $\delta\phi_{(n)}=g_n(\{\phi_{(0)}\}) +$ more.
If we can reproduce the correct d-dimensional susy laws from the D dimensional
susy laws,
the variation of the term in the action which is linear in
massive fields gives an extra piece depending only on massless fields
\be
\delta S^{(D)}(\{ \phi_{(n)}\},
\{ \phi_{(0)} \} )=\delta S^{(d)}( \{ \phi_{(0)} \} )+\int g_n(\{ \phi_{(0)}\})
f_n(\{\phi_{(0)}\}) +{\cal O}(\phi_{(n)})=0
\ee
Since the d-dimensional theory is invariant, $\delta S^{(d)} (\{\phi_0\})$
vanishes. Hence, for consistency
$g_n f_n=0$, i.e. the equations of motion and susy laws
can't both contain purely massless terms for a given massive field.
Since the commutator of the susy transformations
gives the equations of motion, the two inconsistencies are presumably
always equivalent.
Therefore one criterion for a consistent truncation is to find a
nonlinear ansatz which gives the susy transformation laws of $S^{(d)}$
from the susy transformation laws of $S^{(D)}$.

In \cite{cetal,cllp}
the consistency of the KK reduction from the original $d=11$
supergravity to $AdS(7)\times S_4$ to a small subset (two scalars and two
$U(1)$ gauge fields) of our fields was studied using the other criterion,
namely consistency of the field equations.
Only bosonic fields were considered. More recently,
this work has been extended to other subsets of bosonic
fields (a model with one scalar, one SU(2) gauge field and  one antisymmetric
$S_{\alpha\beta\gamma}$, \cite{lupo} and a model with only 4 scalars \cite
{cglp}).
We do not
restrict our atttention to subsets, and consider both fermionic and bosonic
fields. For completeness we shall also consider consistency of the bosonic
field equations.
\subsection{Supergravity and first order formulations}

We start with the sugra Lagrangian in (10,1) dimensions with a  first order
formalism for the antisymmetric tensor field.
(For our conventions and notation, see the appendix). This formulation was
presented in our previous letter \cite{nvv}. After that, a paper appeared
which gave first order formulations of supergravities, in particular a
formulation of 11d sugra which is first order in
 both the spin connection and the antisymmetric tensor field \cite{julia}.

\bea
{\cal L}&=&-\frac{1}{2k^2}ER(E,\Omega )-\frac{E}{2}\bar{\Psi}_{\Lambda}\Gamma
^{\Lambda\Pi\Sigma}D_{\Pi}(\frac{\Omega +\hat{\Omega}}{2})\Psi_{\Sigma}
\nonumber\\&&
+\frac{E}{48}({\cal F}_{\Lambda\Pi\Sigma\Omega}{\cal F}
^{\Lambda\Pi\Sigma\Omega}-48
{\cal F}^{\Lambda\Pi\Sigma\Omega}\partial_{\Lambda}A_{\Pi\Sigma\Omega})
\nonumber\\&&
-k\frac{\sqrt{2}}{6}\epsilon^{\Lambda_0...\Lambda_{10}}\partial_{\Lambda_0}
A_{\Lambda_1\Lambda_2 \Lambda_3}\partial_{\Lambda_4}A_{\Lambda_5\Lambda_6
\Lambda_7}A_{\Lambda_8\Lambda_9\Lambda_{10}}\nonumber\\&&
-\frac{\sqrt{2}k}{8}E[\bar{\Psi}_{\Pi}\Gamma^{\Pi\Lambda_1...\Lambda_4\Sigma}
\Psi_{\Sigma}+12\bar{\Psi}^{\Lambda_1}\Gamma^{\Lambda_2\Lambda_3}\Psi^{
\Lambda_4}][\frac{1}{24}\tilde{F}_{\Lambda_1...\Lambda_4}]
\label{11action}
\eea
where ${\cal F}_{\Lambda\Pi\Sigma\Omega}$ is an independent field with field
equation ${\cal F}_{\Lambda\Pi\Sigma\Omega}=F_{\Lambda\Pi\Sigma\Omega}$,
and
\be
F_{\Lambda\Pi\Sigma\Omega}=\partial_{\Lambda}A_{\Pi\Sigma\Omega}+23
\; terms
\ee
Furthermore $E=\det{{E_\Lambda}^M}$ and
\bea
R(E,\Omega)&=&R_{\Lambda\Pi}\;^{MN}(\Omega)E_M^{\Pi}E_N^{\Lambda}\\
R_{\Lambda\Pi}\;^{MN}(\Omega)
&=&(\partial_{\Lambda}\Omega _{\Pi}\;^{MN}+
\Omega_{\Lambda}\;^M\;_{P}
\Omega_{\Pi}\;^{PN}-(\Lambda\leftrightarrow \Pi)\\
D_{\Pi}\left(\frac{\Omega+\hat{\Omega}}{2}\right)\Psi_{\Sigma}&=&\partial_{\Pi}
\Psi_{\Sigma}+\frac{1}{4}
\left\{\frac{\Omega_{\Pi}\;^{MN}+\hat{\Omega}_{\Pi}\;^{MN}}{2}\right\}
\Gamma_{MN}\Psi_{\Sigma}
\\
\tilde{F}_{\Lambda_1...\Lambda_4}=
(\frac{F+\hat{F}}{2})_{\Lambda_1...\Lambda_4}
&=&24\left[\partial_{[\Lambda_1}A_{\Lambda_2
\Lambda_3\Lambda_4]}+\frac{1}{16\sqrt{2}}\bar{\Psi}_{[\Lambda_1}\Gamma_
{\Lambda_2\Lambda_3}\Psi_{\Lambda_4]}\right]
\eea
Here $\hat{F}$ is the supercovariant curl of A
\footnote{By replacing  $24{\cal F}\partial A$ in
(\ref{11action}) by ${\cal F}\hat{F}$,
the terms $(\bar{\Psi}\Gamma\Gamma \Psi)\tilde{F}$ get absorbed.
Then the ${\cal
F}$ field equation reads ${\cal F}=\hat{F}$ and becomes supercovariant. We have
not been able to absorb the remaining four-fermi terms by using our new
first order formulation.} and
 we use 1.5 order formalism for the spin connection, i.e., $\Omega$ is
not independent, but rather it is the solution of its field equation (
to which only the $\Omega$ term, but not the $\hat\Omega$ term,
 in the gravitino action contributes). The
supercovariant spin connection $\hat{
\Omega}$ is obtained from $\Omega$ by adding terms bilinear in the fermions
such that there are no $\partial \epsilon$ terms in its susy transformations
\be
{\hat{\Omega}_{\Pi}}^{MN}={\Omega_{\Pi}}^{MN}(E)+\frac{k^2}{4}(\bar{\Psi}_{\Pi}
\Gamma^M\Psi^N-\bar{\Psi}_{\Pi}\Gamma^M\Psi^n+\bar{\Psi}^M\Gamma_{\Pi}\psi^N)
\ee
The relation between $\Omega$ and $\hat{\Omega}$ is then given by
\be
\Omega_{\Pi}\;^{MN}=\hat{\Omega}_{\Pi}\;^{MN}-\frac{1}{8}\bar{\Psi}
^{\Lambda}\Gamma_{\Lambda\Pi}\;^{MN}\;_{\Sigma}\Psi^{\Sigma}
\ee

It is useful to  redefine
\be
{\cal F}_{\Lambda\Pi\Sigma\Omega}=\partial_{\Lambda}A_{\Pi\Sigma\Omega}+23
\; terms+\frac{{\cal B}_{MNPQ}E_{\Lambda}^M...E_{\Omega}^Q}{\sqrt{E}}
\ee
Substituting this definition into
the terms in the action involving ${\cal F}$, we obtain
\be
-\frac{1}{48}E(\partial_{[\Lambda}A_{\Pi\Sigma\Omega]}+23\;\;\;terms)^2+
\frac{{\cal B}_{MNPQ}^2}{48}
\ee

The supersymmetry transformation rules which leave the action with the
${\cal B}_{MNPQ}^2$ term invariant read
\bea
\delta E_{\Lambda}^M&=&\frac{k}{2}\bar{\varepsilon}\Gamma^M\Psi_{\Lambda}
\label{dele}\\
\delta \Psi_{\Lambda}&=&\frac{1}{k} D_{\Lambda}(\hat{\Omega})\varepsilon
+\frac{\sqrt{2}}{12}(\Gamma^{\Lambda_1...\Lambda_4}\;_{\Lambda}-8\delta
_{\Lambda}^{\Lambda_1}\Gamma^{\Lambda_2\Lambda_3\Lambda_4})
\varepsilon(\frac{1}{24}\hat{F}_{\Lambda_1...\Lambda_4})\nonumber\\&&
+ \frac{1}{24}(b\Gamma_{\Lambda}\;^{
\Lambda_1...\Lambda_4}\frac{1}{\sqrt{E}}
{\cal B}_{\Lambda_1...\Lambda_4}-a\Gamma
^{\Lambda_1\Lambda_2\Lambda_3}\frac{1}{\sqrt{E}}
{\cal B}_{\Lambda\Lambda_1\Lambda_2
\Lambda_3})\varepsilon\\
\delta A_{\Lambda_1\Lambda_2\Lambda_3}&=&-\frac{\sqrt{2}}{8}\bar{\varepsilon}
\Gamma_{[\Lambda_1\Lambda_2}\Psi_{\Lambda_3]}\label{deltaa}\\
\delta
{\cal B}_{MNPQ}&=& \sqrt{E}\bar{\varepsilon}
(a\Gamma_{MNP}E_{Q}^{\Lambda}
R_{\Lambda}(\Psi )+b \Gamma_{MNPQ\Lambda}R^{\Lambda}
(\Psi ))
\eea
where $R_{\Lambda}(\Psi )$ is the gravitino field equation,
\bea
R^{\Lambda}(\Psi )=\frac{1}{E}\frac{\delta{\cal L}}{\delta\bar{\Psi}_
{\Lambda}}&=&-\Gamma^{\Lambda\Pi\Sigma}D_{\Pi}\Psi_{\Sigma}\nonumber\\
-\frac{\sqrt{2}}{4}k
(\frac{1}{24}\hat{F}_{\Lambda_1...\Lambda_4})
\Gamma^{\Lambda\Lambda_1...\Lambda_5}\Psi_{\Lambda_5}&-& 3\sqrt{2}k(\frac{1}{24}
\hat{F}^{\Lambda\Pi\Sigma\Omega})\Gamma_{\Pi\Sigma}\Psi_{\Omega}
\eea
The expression $\hat{F}_{\Lambda\Pi\Sigma\Omega}$ denotes the usual
supercovariantization of $\partial_{\Lambda}A_{\Pi\Sigma\Omega}+23\; terms$
and $a$ and $b$ are free
parameters. They will be fixed by the requirement of consistency of the
KK truncation on $S_4$. The gravitational constant $k$ has dimensions 11/2,
hence $kA$ is dimensionless and $k\Psi$ has dimension 1/2. Below we set $k$
equal to unity.

New in the transformation laws
are the ${\cal B}$ terms in $\delta \Psi _{\Lambda}$ and
the expression for $\delta {\cal B}$. Of course $\delta {\cal B}$
is proportional to the gravitino field
equation, because ${\cal B}=0$ is a field equation and
field equations (usually) transform into field equations.

This theory admits a background solution which satisfies the equations of
motion and which describes a geometry  of the type $AdS_7\times S_4$. The
source is given by
\be
F_{\mu\nu\rho\sigma}=\frac{3}{\sqrt{2}}m(det \sirc{e}_{\mu}^m (x))
\epsilon_{\mu\nu\rho\sigma}
\ee
where  $\sirc{e}_{\mu}^m$ is the vielbein on $S_4$ and  $\sirc{e}_{\alpha}^a$
the vielbein on $AdS_7$. The parameter $m$ is the inverse radius of the
sphere as will now be show.
The field equations for the background geometry read then
\be
R_{\mu\nu}-\frac{1}{2}\sirc{g}_{\mu\nu}R=-\frac{1}{6}\left(F_{\mu\cdot\cdot\cdot}
{F_\nu}^{\cdot\cdot\cdot}-\frac{1}{8}\sirc g_{\mu\nu}F^2\right)
=-\frac{9}{4}\sirc g_{\mu\nu}m^2
\ee
and
\be
R_{\alpha\beta}-\frac{1}{2}\sirc g_{\alpha\beta}R=\frac{1}{48}\sirc g_{\alpha
\beta}F^2=\frac{9}{4}\sirc g_{\alpha\beta} m^2
\ee
The maximally symmetric solution is given by
\bea
R_{\mu\nu}\;^{mn}(\sirc{e}^{(4)})&=&m^2\{\sirc{e}_{\mu}^m(x)\sirc{e}_{\nu}^n(x)
-\sirc{e}_{\nu}^m(x)\sirc{e}_{\mu}^n(x)\}\label{curv}\\
R_{\alpha\beta}\;^{ab}(\sirc{e}^{(7)})&=&-\frac{1}{4}m^2(\sirc{e}_{\alpha}
^a(y)\sirc{e}_{\beta}^b(y)-\sirc{e}_{\alpha}^b(y)\sirc{e}_{\beta}^a(y))
\label{adsul}\\
R&=& R^{(4)} +R^{(7)}=-\frac{3}{2}m^2
\eea
All other fields vanish in the background
\be
\Psi_{\Lambda}^M=0\;;\;
F_{\alpha\Lambda\Pi\Sigma}=0\;;\;
{\cal B}_{\Lambda\Pi\Sigma\Omega}= 0
\ee

First order formulations in sugra have been given before in d=4. Soon
after the N=1 sugra was formulated with a second-order formulation
for the spin connection \cite{fvf},
a first-order formulation for the spin connection (with an independent
field $\omega_{\mu}^{mn}$ and an expression for $\delta\omega_{\mu}^{mn}$ )
was given \cite{dz}. For d=11 sugra, a first-order formulation  for the spin
connection was given in \cite{cfgpp}.

The reason we start with a first order formulation for the antisymmetric
tensor field in d=11 is that  at the linearized level one needs
in 7 dimensions a 3-index auxiliary
field to combine with the 3 index tensor to give the 7 dimensional
 supergravity field $S_{\alpha\beta\gamma ,A}$ found in \cite{
ppn}. Since the 3-index auxiliary field can also be written as a 4-index
auxiliary field by a duality transformation, this suggested to us
that the auxiliary field in d=11 gives by reduction the
 auxiliary field in d=7. Our results confirm this suggestion.

Could we have chosen another first order formulation which would
give us at the linearized level this auxiliary field? The other
obvious choice is the first order formulation for the spin connection
 (Palatini formalism). The auxiliary field would then be
 obtained as the totally
antisymmetric part of the difference between the independent spin connection
and the solution of its equation of motion, i.e.
\be
{\cal B}_{\Lambda\Pi\Sigma}=(\Omega_{[\Lambda}\;^{MN}-
\hat{\Omega}_{[\Lambda}\;^{MN}(E,\Psi)){E_{M\Pi}} {E_{N\Sigma]}}
\ee
When we dimensionally reduce, we obtain an auxiliary antisymmetric tensor
field ${\cal B}_{\alpha\beta\gamma}$.
We will see later (in section 3) that this approach does not produce the
7 dimensional auxiliary field we need.

\subsection{Linearized Kaluza-Klein reduction on $S_4$}

Any 11 dimensional field is written as a sum of the corresponding
background field and the fluctuations. In the fluctuations we keep
only 'massless' modes, i.e., fields which in 7 dimensions belong to
the maximal
(N=4) sugra multiplet. In full generality, the massless 7 dimensional
fields may occur nonlinearly in the expansion of 11 dimensional fields.

The linearized terms in the expansion of the bosons have the generic
form
\be
\Phi_{\alpha\mu}(y,x)=\sum_{I}\phi_{\alpha,I}(y)Y_{\mu}^I(x)
\ee
where $\phi$ are 7 dimensional fields and Y are spherical harmonics. The
 7 dimensional indices are attached to 7 dimensional fields, 4
dimensional indices to the spherical harmonics Y of the internal space.
For fermions one finds a similar decomposition,
\be
\Psi^A(y,x)=\sum_{I}\Psi_I^{\tilde{a}}(y)Y^{I,\tilde{\tilde{a}}}
\ee
where the spinor index A=1,32 of an 11 dimensional spinor decomposes as
$A=\tilde{a}\otimes\tilde{\tilde{a}}$ with $\tilde{a}$ spinor indices
on $AdS_7$ and $\tilde{\tilde{a}}$ spinor indices on $S_4$.
We shall suppress these
spinor indices , and thus $\tilde{a}$ and $\tilde{\tilde{a}}$
will not be used below.

The linearized ansatz has been given in \cite{pnt}.
We reproduce it here for completeness.
The vielbein $E_{\Lambda}^M(y,x)$ produces the metric
 $g_{\Lambda\Pi}=E_{\Lambda}^M E_{M\Pi}
=\sirc{g}_{\Lambda\Pi}+kh_{\Lambda\Pi}$, where $h_{\Lambda\Pi}$ are the
fluctuations given by the
following expressions in the various sectors

(i) In $AdS_7$ spacetime
\be
h_{\alpha\beta}(y,x)=h_{\alpha\beta}(y)-\frac{1}{5}\sirc{g}_{\alpha\beta}
(y)(h_{\mu\nu}(y,x)\sirc{g}^{\mu\nu}(x))\label{delta}
\ee
where the redefinition of the graviton is
needed in order to diagonalize its kinetic term.
(The linearized kinetic term for the graviton is the Fierz-Pauli action for a
massless spin 2 field and reads in any dimension ${\cal L}=1/2 h_{\mu\nu,\rho}^2
+h_\mu^2-h^\mu h_{,\mu}+1/2h_{,\mu}^2$ where $h_\mu=\partial^\nu h_{\nu\mu}$
and $h=h^\mu_\mu$.)

(ii) In the mixed sector,
\be
h_{\mu\alpha}(y,x)=B_{\alpha,IJ}(y)V_{\mu}^{IJ}(x);\;I,J=1,4
\ee
where $V_{\mu}^{IJ}(x)=V_{\mu}^{JI}(x)$ are the ten Killing vectors on $S_4$.

(iii) In the $S_4$ sector
\be
h_{\mu\nu}(y,x)=S_{IJKL}(y)\eta_{\mu\nu}^{IJKL}(x)
\ee
 where the scalars
$S_{IJKL}$ are in the {\bf 14} representation of USp(4) and $\eta_{\mu\nu}^{
IJKL}$ is the corresponding spherical harmonic. It has the symmetry of the Riemann
tensor, but its symplectic trace and its totally antisymmetric part vanish
(see below (\ref{s111}) and (\ref{s112})).

The gravitino $\Psi_{\Lambda}$ splits into $\Psi_{\mu}$ and $\Psi_{\alpha}$,
where the spin 1/2 fields are written as

\be
\Psi_{\mu}(y,x)=\lambda_{J,KL}(y)\gamma_5^{1/2}\eta_{\mu}^{JKL}(x)
\ee
The $\lambda_{J,KL}$ with J,K,L=1,...4
 are the fermions in the {\bf 16} representation  of
USp(4), with $\eta_{\mu}^{J,KL}(x)$ the corresponding spherical harmonic.
$\lambda_{J,KL}$ is antisymmetric and symplectic traceless in KL, and its
totally antisymmetric part vanishes (and so all its symplectic traces are
zero). By definition, $\sqrt{\gamma_5}=\frac{
i-1}{2}(1+i\gamma_5)$, see appendix A1.

We took the scalars to be in the {\bf 14} representation and the spinors
in the {\bf 16} representation because of the following reason.
First, the total
number of scalars has to be 14, and the total number of fermions 16, a fact
which we know for instance from the case of toroidal compactification.
Second, the complete mass  spectrum on $S_4$ was found in \cite{pvn2}, and
the lowest mass modes are the singlet and the {\bf 14}. The singlet has
bigger mass, and is found to belong to another multiplet. The spinors have
as lowest modes a {\bf 4} and the {\bf 16}, but again the {\bf 4} has
higher mass and belongs to the same multiplet as the scalar singlet.

The gravitini $\Psi_{\alpha}$ also have to be redefined in order to diagonalize
their kinetic term
\be
\Psi_{\alpha}(y,x)=\psi_{\alpha I}(y)\gamma_5^{\pm 1/2}\eta^I(x)-\frac{1}{5}
\tau_{\alpha}\gamma_5\gamma^{\mu}\Psi_{\mu}(y,x)\label{psi}
\ee
where $\eta^I(x)$ is the Killing spinor. Clearly, $\gamma_5^{-1/2}=
\frac{1+i}{2}(i\gamma_5-1)$.

The antisymmetric tensor $F_{\Lambda\Pi\Sigma\Omega}=\sirc{F}_
{\Lambda\Pi\Sigma\Omega}
+f_{\Lambda\Pi\Sigma\Omega}$ (where $f_{\Lambda\Pi\Sigma\Omega}$
is the fluctuation and $\sirc{
F}_{\Lambda\Pi\Sigma\Omega}$ the background solution) decomposes
at the linearized level as follows

\bea
\frac{\sqrt{2}}{3}f_{\mu\nu\rho\sigma}&=&\sqrt{det\sirc{g}_{\mu\nu}}\epsilon
_{\mu\nu\rho\sigma}h_{\lambda}^{\lambda},\;\;h_{\lambda}^{\lambda}\equiv
h_{\mu\nu}(y,x)\sirc{g}^{\mu\nu}(x)\label{epsil}\\
\frac{\sqrt{2}}{3}f_{\alpha\nu\rho\sigma}&=&\sqrt{det\sirc{g}_{\mu\nu}}
\epsilon_{\nu\rho\sigma\tau}[\frac{1}{10}D^{\tau}D_{\alpha}h_{\lambda}^
{\lambda}-h_{\alpha}^{\tau}] \label{alnurosi}\\
\frac{\sqrt{2}}{3}f_{\mu\nu\alpha\beta}&=&\frac{i}{3}\partial_{[\alpha}
B_{\beta ]}^{IJ}\bar{\eta}^I\gamma_{\mu\nu}\gamma_5\eta^J  \\
\frac{\sqrt{2}}{3}f_{\nu\alpha\beta\gamma}&=&\frac{\sqrt{2}}{3}
A_{\alpha\beta\gamma ,IJ}\partial_{\nu}\phi_5^{IJ}(x)\label{fmual}\\
\frac{\sqrt{2}}{3}f_{\alpha\beta\gamma\delta}&=&\frac{\sqrt{2}}{3}4
\partial_{[\alpha}A_{\beta\gamma\delta ],IJ}\phi_5^{IJ}(x)\label{fmu}
\eea
so that
\bea
A_{\mu\nu\rho}(y,x)&=&\frac{\sqrt{2}}{40}\sqrt{\sirc{g}}\epsilon_{\mu\nu
\rho\sigma}D^{\sigma}h_{\lambda}^{\lambda}\label{a1}\\
A_{\alpha\mu\nu}(y,x)&=&\frac{i}{12\sqrt{2}}B_{\alpha ,IJ}(y)\bar{\eta}^I(x)
\gamma_{\mu\nu}\gamma_5\eta^J(x)\\
A_{\alpha\beta \mu}&=&0\\
A_{\alpha\beta\gamma}(y,x)&=&\frac{1}{6}A_{\alpha\beta\gamma ,IJ}(y)
\phi_5^{IJ}(x)
\eea
where $A_{\alpha\beta\gamma , IJ}$ is a set of antisymmetric
symplectic-traceless tensors in
the {\bf 5} representation of USp(4) and $\phi_5^{IJ}=\bar{\eta}^I\gamma_5
\eta^J=-\bar{\eta}^J\gamma_5\eta^I$ the corresponding
scalar spherical harmonic. The factor 1/10 in (\ref{alnurosi})
was determined in \cite{pnt} by solving the
Bianchi identities, but we have here given the linearized KK
formulas for the fields themselves.To show that (\ref{a1}) reproduces
(\ref{epsil}) one needs $\Box \sirc{g}^{\mu\nu}\eta_{\mu\nu}^{IJKL}=
-10\sirc{g}^{\mu\nu}\eta_{\mu\nu}^{IJKL}$, see (\ref{s112}). The rest of the
result in (\ref{alnurosi}) then follows if one uses $\sirc{D}_{\mu} \eta
=\frac{i}{2}\gamma_{\mu}\eta$ and $V_{\mu}=\bar\eta \gamma_{\mu}\eta$, see
(\ref{killing}).

When massive modes are put to zero, one would expect that $A_{\alpha\beta
\gamma ,IJ}$ becomes equal to
the 7 dimensional supergravity field $S_{\alpha\beta\gamma, IJ}$.
However,  the field $A_{\alpha\beta\gamma ,IJ}$ has an action with
2 derivatives, whereas the action of $S_{\alpha\beta\gamma, IJ}$ is
linear in derivatives. That means that
 we need to introduce by hand an auxiliary field ${\cal B}_{\alpha
\beta\gamma ,IJ}$, and rotate the $A$ and ${\cal B}$ fields such
that the sum of the action of $A$ with 2 derivatives and the action
of ${\cal B}$ with none gives
two decoupled actions each linear in derivatives, one for $S_{\alpha\beta\gamma
,IJ}$ and one for a massive field b. After setting the
massive field b to zero, the dependence of the
  auxiliary field ${\cal B}_{\alpha\beta\gamma , IJ}$ on
$S_{\alpha\beta\gamma ,IJ}$ reads
\be
{\cal B}_{\alpha\beta\gamma ,IJ}=\frac{1}{5}(S_{\alpha\beta\gamma ,IJ}
+\frac{1}{6}\epsilon_{\alpha\beta\gamma}\;^{\delta\epsilon\eta\zeta}
D_{\delta}S_{\epsilon\eta\zeta ,IJ})\label{auxili}
\ee
Note that if we want to understand ${\cal B}$ as coming from a KK reduction,
its spherical harmonic should be the same as for the field $A$.
In general two fields in lower dimensions with the same index structure
cannot have the same spherical harmonic, but in our case ${\cal B}$
and $A$ have
different mass dimensions, and therefore can have the same spherical harmonic.

\section{Spherical harmonics and Killing spinors}

\subsection{Spherical harmonics for the massless multiplet}

The coset representatives of a reductive coset manifold
$L(x)=exp(-x^{\alpha}K_{\alpha})$
satisfy the equation $L^{-1}dL=e^aK_a+\omega^iH_i$. From $dL^{-1}=
-(L^{-1}dL)L^{-1}$ with coset generators $K_{\alpha}$ and subgroup generators
$H_i$, one then obtains $(d+\omega^i H_i)L^{-1}=-(e^mK_m)
L^{-1}$. Using the spinor representation of SO(5)
with $H_i=\frac{1}{4}[\gamma_m,\gamma_n]$ and $K_m=-ic\gamma_m$
we define $\eta^I(x)\equiv
L^{-1}\eta^I(0)$ where $\eta^{I}(0)$ is a constant spinor with
$\eta^{I}(0)^{{\bf \alpha}}=\Omega
^{{\bf\alpha} I}$. (The index ${\mathbf\alpha}$
is a spinor index and runs from
1 to 4, while $\Omega$ is a constant antisymmetric $4\times 4$ matrix. We
suppress the spinor index ${\bf \alpha}$ in most of the text.)
In general, the difference between $\omega^i H_i$ and the
spin connection term $1/2\;\; \omega(spin)_a\;^{bc}J_{bc}$ is $1/4\bar{c}_{ab}
\;^c J_c\;^b$, where ${\bar{c}_{ab}}^c $ is defined by $[K_a,K_b]=c_{ab}\;^i
H_i+c_{ab}\;^d K_d, \bar{c}_{db}\;^a= c_{db}\;^a+\delta^{aa'}(c_{a'd}\;
^{b'}\delta_{bb'}+b\leftrightarrow d)$ \cite{japan}.
Since $S_4$=SO(5)/SO(4) is a symmetric manifold, ${c_{ab}}^d=0$ and
thus $\omega^i H_i$ is the spin connection. For a treatment of
spherical harmonics on $S_4$ and on general coset manifolds, where
these issues are siscussed, see \cite{gwn,pvn2}.

On $S_4$ we use $\mu =1,4$ instead of $\alpha$ as curved coset indices,
and $m=1,4$ as flat coset indices.
Then the spinors $\eta^I(x)$ are Killing spinors
as they satisfy
\be
(\partial_{\mu}+\frac{1}{4}\omega_{\mu}^{mn}(x)
\gamma_{mn})\eta^I(x)=c e_{\mu}^m(x)(i\gamma_m)\eta^I(x)\label{kspeq},
\ee
where the scale parameter $c$ is fixed in terms of the curvatures. For a
sphere $S_4$ of radius $m$ as chosen in (\ref{curv}),
$c=\pm\frac{1}{2}m$. This can be derived from the
integrability condition on the Killing spinor
\footnote{It follows from this integrability condition that
in general Killing spinors exist only on Einstein manifolds. On
spheres there are two sets of Killing spinors $\eta^\pm$ satisfying
$\sirc D_\mu \eta^\pm = \pm c \gamma_\mu \eta^\pm$
which are related by $\eta^+=\gamma_5\eta^-$ in even dimensions.}:
$[\sirc{D}_{\mu},
\sirc{D}_{\nu}]\eta=\sirc{R}_{\mu\nu}\;^{mn}\frac{1}{4}\gamma_{mn}
\eta$, when we substitute the background curvature.
In the following we choose to work with Killing spinors for which
$c=\frac{1}{2}m$; we shall also drop the factors of $m$, understanding
that we should replace everywhere $\partial_{\mu}$ by $\partial_{\mu}/m$
to get the correct dimensions.

The Killing spinors can be given an explicit form
\be
\eta^{{\bf \alpha}I} = \{exp(\frac{-i}{2}x^\mu \delta{_\mu ^m} \gamma^m
)\}^{{\bf\alpha}}\;_{{\bf\beta}}\Omega^{{\bf\beta} I}\label{killingspinor}
\ee
where $x^{\mu}$ are general coordinates in patches covering $S_4$.
They satisfy
\be
\sirc{D}_{\mu}\eta^I= \frac{i}{2}
\gamma_{\mu}\eta^I; \;\;\;\bar{\eta}^I\eta^J=\Omega^{IJ};\;\;\;
\sirc{D}_{\mu}\bar{\eta}^I=-\frac{i}{2}\bar{\eta}^I\gamma_{\mu}\label{killing}
\ee
(In appendix A2 we discuss the definition
$\bar{\eta}^I=\eta^{I,T}C$, where the charge conjugation
matrix $(C_4^{-})_{\alpha\beta}$ on $S_4$ is equal to the symplectic metric
$\Omega_{\alpha\beta}$. It is shown there that $\Omega^{I\alpha}=-(C^{-1})^{I
\alpha}$).
The first relation
follows from our earlier discussion, while the second one is easy to check.
A direct consequence of the second equation is a completeness type of relation
satisfied by the Killing spinors
\be
\eta_J^\alpha \bar\eta^J_\beta=-\delta^\alpha_\beta\;\;{\rm with}\;\;\eta^
{\alpha}_J\equiv \eta^{\alpha I}\Omega_{IJ}
\label{compltn}
\ee
One can directly verify the Killing equation by constructing the vielbein
$e^m$ and spin connection $\omega^i$ from $L^{-1}dL$, expressing L in terms
of a cosine and sine as in (\ref{coord}),
and substituting into (\ref{kspeq}).

All spherical harmonics can now be built from Killing spinors:
\begin{itemize}

\item Scalars {\bf 5}, $\phi_5^{IJ}=-\phi_5^{JI}=\bar{\eta}^I\gamma_5\eta^J,
\phi_5^{IJ}\Omega_{IJ}=0$, or $Y^A\equiv
\frac{1}{4} \phi_5^{IJ}(\gamma^A)_{IJ}$ with I,J=1,4  and $\gamma ^A $ defined
in (\ref{5gammas}). (The matrices $(\gamma^A)
_{IJ}$ are antisymmetric and obtained from $\{i\gamma_m\gamma_5, \gamma_5\}^I
\;_J$ by lowering the index according to the rule (\ref{lower})).

\item Conformal Killing vectors {\bf 5}, $C_{\mu}^{IJ}=-C_{\mu}^{JI}=
\bar{\eta}^I\gamma_{\mu}
\gamma_5\eta^J, C_{\mu}^{IJ}\Omega_{IJ}=0$, or $C_{\mu}^{A}=\frac{i}{4}
C_{\mu}^{IJ}(\gamma^A)_{IJ}=\sirc{D}_{\mu}Y^A$, where $\sirc{D}_{\mu}Y^A
=\partial_{\mu}Y^A$. They satisfy $\sirc{D}_{(\mu}C_{\nu)}^A=\frac{1}{4}
g_{\mu\nu}(\sirc{D}^{\rho}C_{\rho}^A)$. (In fact $\sirc{D}_{\mu}C_{\nu}^A
=\sirc{D}_{(\mu}C_{\nu)}^A$.)

\item Killing vectors {\bf 10}, $V_{\mu}^{IJ}=V_{\mu}^{JI}=
\bar{\eta}^I\gamma_{\mu}\eta^J$ or $V_{\mu}^{AB}=-V_{\mu}^{BA}=-\frac{i}{8}
V_{\mu}^{IJ}(\gamma^{AB})_{IJ}$, satisfying
$\sirc{D}_{(\mu}V_{\nu)}^{AB}=0$ and $V_{\mu}^{IJ}=iV_{\mu}^{AB}(\gamma_{AB})
^{IJ}$. We show in (\ref{vmu}) that $V_{\mu}^{AB}=Y^{[A}\partial_{\mu} Y^{B]}$

\item Symmetric tensors $\eta_{\mu\nu}^{IJKL}=\eta_{\nu\mu}^{IJKL}$,
with Young tableau in the form of a $2\times 2$ box,
\be
\eta_{\mu\nu}^{IJKL}=\eta_{\mu\nu}^{IJKL}(-2)- \frac{1}{3}
\eta_{\mu\nu}^{IJKL}(-10)\label{unupetrei}
\ee
where $\Box\eta_{\mu\nu}^{IJKL}(-2)=-2\eta_{\mu\nu}^{IJKL}$ and
 $\Box\eta_{\mu\nu}^{IJKL}(-10)=-10\eta_{\mu\nu}^{IJKL}$. The first harmonic
is traceless, while the second one is a pure trace. In terms of Killing
spinors one has
\bea
\eta_{\mu\nu}^{IJKL}(-2)&=&C_{(\mu}^{IJ}C_{\nu)}^{KL}-\frac{1}{4}
\sirc{g}_{\mu\nu} C_{\lambda}^{IJ}C^{\lambda KL}\label{s111}\\
\eta_{\mu\nu}^{IJKL}(-10)&=&\sirc{g}_{\mu\nu} (\phi_5^{IJ}\phi_5^{KL}
+\frac{1}{4}C_{\lambda}^{IJ}C^{\lambda KL})\label{s112}
\eea
(Use $\Box \phi_5 ^{IJ}=-4\phi_5^{IJ}$ and $\Box C_{\mu}^{IJ}=-C_{\mu}^{IJ}$.)
Both (\ref{s111}) and (\ref{s112}) are symplectic-traceless (use (\ref{compltn}))
and without a totally antisymmetric part (use Fierz rearrangements).

\item Vector-spinor $\eta_{\mu}^{JKL}$ with Young tableau in the form of a gun,
\be
\eta_{\mu}^{JKL}=\eta_{\mu}^{JKL}(-2)+\eta_{\mu}^{JKL}(-6)\label{vecsp}
\ee
The first harmonic satisfies $\gamma^{\nu}\sirc{D}_{\nu}\eta_{\mu}^{JKL}(-2)
=-2\eta_{\mu}^{JKL}(-2)$ and is gamma traceless, while the second one is  a
pure gamma trace and satisfies  $\gamma^{\nu}\sirc{D}_{\nu}\eta_{\mu}^{JKL}(-6)
=-6\eta_{\mu}^{JKL}(-6)$. Again these spherical harmonics can be expressed in
terms of Killing spinors
\bea
\eta_{\mu}^{JKL}(-2)&=&3(\eta^J C_{\mu}^{KL}-\frac{1}{4}\gamma_{\mu}\gamma^
{\nu}\eta^J C_{\nu}^{KL})\\
\eta_{\mu}^{JKL}(-6)&=& \gamma_{\mu}(\eta^J \phi_5^{KL}-\frac{1}{4}\gamma^{\nu}
\eta^J C_{\nu}^{KL})
\eea

\end{itemize}
The relative factors $1/3$ in (\ref{unupetrei}) and 1 in (\ref{vecsp}) are
not fixed by properties
of the spherical harmonics themselves, but rather by consistency of the
susy transformation rules and equations of motion at the linearized level
\cite{pnt}.

\subsection{Identities involving spherical harmonics}
Substituting (\ref{killingspinor}) into the definition of $Y^A$ in terms of
 the Killing spinors,
one obtains the  vectors $Y^A$ in terms of the coset parameters as:
\be
Y^A =-\frac{1}{4} Tr\left[\gamma^A \gamma_5 exp(-i\not\! x)\right]
\label{coord}
\ee
or, in components:
\be
Y^m =-{\delta_\mu}^m\frac{x^\mu}{x} \sin{x}
\ee
\be
Y^5 = -\cos{x}
\ee
where $x^2=x^\mu x^\nu \delta_{\mu \nu}$. Clearly, $\sum (Y^A)^2=1$, and $x$ has
a geometrical meaning: $-x$ is the azimuthal angle of a point on the unit sphere
with coordinates $Y$.

On a sphere one can also define stereographic coordinates
$\xi$ related to $Y^m$ by a conformal mapping. If the unit sphere lies on top
of the stereographic plane one obtains
\be
\xi^\mu ={\delta_m}^\mu \frac{2 Y^m}{1-Y^5}.
\ee
In these $\xi$ coordinates the vielbein has a very simple form
\be
{e_\mu}^m (\xi)= {\delta_\mu}^m \frac{4}{4+ \xi^2}
\ee
Also the spin connection and Killing spinors are simple rational functions
of the $\xi$'s
\be
{\omega_\mu}^{mn}=\frac{1}{1+\xi^2}(-\delta_\mu^m\xi^n +\delta^n_\mu \xi^m)
\ee
\be
\eta^{\pm}(\xi)=\frac{4}{4+\xi^2}(1\pm \frac{i}{2}\gamma_m
\xi^\mu \delta_\mu^m )\eta^\pm (0)
\ee

We have now 3 different coordinate systems on $S_4$: the coset coordinates x,
the Euclidean coordinates $Y^A$ and the stereographic coordinates $\xi$. We
shall mostly use the Euclidean coordinates $Y^A$, but note that the explicit
expressions for the Killing spinors (which we do not use) are simplest in
stereographic coordinates.

The scalars $Y^A$ parameterizing the sphere
in a 5 dimensional space satisfy the identity
\be
Y^A Y_A=1\label{y1}
\label{yaya}
\ee
It follows that
\be
Y^A\sirc{D}_{\mu}Y_A=0, \;\;{\em or\;}\; Y^A C_{A \mu}=0\label{y2}
\label{ydy}
\ee
It is easy to check that
\be
\sirc{D}_{\mu}\sirc{D}_{\nu} Y^A=-\sirc{g}_{\mu\nu}Y^A
\label{ddy}
\ee
Hence we obtain the completeness relation
\be
\sirc{D}_{\mu}Y^A\sirc{D}_{\nu}Y_A=\sirc{g}_{\mu\nu},\;\; {\em or\;}\;
C_{\mu}^A C_{A\nu}=\sirc{g}_{\mu\nu}
\label{dydy}
\ee
Another useful identity is
\be
\sirc{D}_{\mu}Y^A\sirc{D}^{\mu}Y^B+ Y^A Y^B=\delta^{AB}\label{compl}
\ee

It follows by using that $\partial_mY^A=C_m^A $ and $Y^A\equiv C_5^A$ form
an orthogonal $5\times 5$
matrix according to (\ref{y1}), (\ref{y2}) and (\ref{dydy}). We also give another
proof because it illustrates the techniques we will use. Acting with
$\sirc{D}_{\lambda}$ on both sides of (\ref{compl}); the
r.h.s. is then zero trivially, and the l.h.s. is zero
 upon using (\ref{ddy}). So the
l.h.s. is constant, and at x=0, where $\eta^{I\alpha}=\Omega^{\alpha I}
$, we obtain
\bea
\sirc{D}_{\mu}Y^A|_{x=0}&=&\frac{1}{4}Tr(Ci\gamma_5\gamma_{\mu}\gamma^A
\tilde{\Omega})=-
\delta_{\mu}^A\nonumber\\
Y^A|_{x=0}
&=&\frac{1}{4}Tr(C\gamma_5\gamma^A\tilde{\Omega})=-\delta_5^A\label{x=0}
\eea
This proves (\ref{compl}).

In the same way one may prove that
\be
V_{\mu}^{AB}=Y^{[A}\sirc{D}_{\mu}Y^{B]}\label{vmu}
\ee
namely one may show that the r.h.s. satisfies  the Killing vector equation,
$\sirc{D}
_{(\mu}V_{\nu)}^{AB}=0$, by using (\ref{ddy}) and the antisymmetry
in AB. The proportionality constant is fixed by taking $x=0$ for a particular
case, e.g. A=5, B=$\nu$. Using the definition of $V_{\mu}^{AB}$, the l.h.s.
gives at x=0 $-\frac{1}{2}g_{\mu\nu}$,
and the r.h.s. gives with (\ref{x=0}) also $-\frac{1}{2}g_{\mu\nu}$.

Similarly, we may derive
\be
4\sqrt{\sirc{g}}\epsilon_{\mu\nu\rho\sigma}\partial^{\rho}Y^{[A_1}\partial^
{\sigma}Y^{A_2]}=+\bar{\eta}^I\gamma_{\mu\nu}\gamma_5\eta^J (\gamma^{A_1A_2})
_{IJ}\label{yg2}
\ee
To prove this, we act with $\sirc{D}_{\lambda}$ on the last term
and get
$+i\bar{\eta}^I\gamma_5\gamma_{\mu\nu\lambda}\eta^J (\gamma^{A_1A_2})_{IJ}$.
We obtain the same result if we act with $\sirc{D}_{\lambda}$ on the
first term.
At x=0, using (\ref{x=0}), the l.h.s. gives
$4\sqrt{\sirc{g}}\epsilon_{\mu\nu A_1A_2}$
and the r.h.s. gives
$Tr(C\gamma_{\rho\sigma}\gamma_5\gamma^{A_1A_2}\tilde{\Omega})$. Evaluating
the trace one finds that both results are equal. This proves (\ref{yg2}).

>From (\ref{compl}) and (\ref{vmu}) we find also
\bea
V_{\mu}^{AB}V^{\nu}_{AB}&=&\frac{1}{2}\delta_{\mu}^{\nu}
; V_m^{IJ}V^n_{IJ}=-4\delta_m^n\label{vortog}\\
V_{\mu}^{AB}Y_A&=&\frac{1}{2}\sirc{D}_{\mu}Y^B\\
\frac{1}{2}[V_{\mu}^{AB}Y_BY^C-A\leftrightarrow C]&=&\frac{1}{2}V_{
\mu}^{AC}\label{vyy}\\
V_{\mu}^{AB}V^{\mu}_{CD}&=&Y^{[A}Y_{[C}\delta^{B]}_{D]}\label{vv}
\eea
>From SO(5) symmetry \footnote{The l.h.s. has to be an invariant tensor
of SO(5) with 2 indices because the Haar measure $d^4x\sqrt{\sirc{g}}$ on
$S_4$ is SO(5) invariant. This allows only $\delta^{AB}$. The constant
of proportionality is found by taking a trace with $\delta^{AB}$.}
we deduce
\be
\int_{S_4}d^4x\sqrt{\sirc{g}}Y^A Y^B=\frac{V_4}{5}\delta^{AB}.
\ee
where $V_4$ is the volume of $S_4$. In the same way we find
\be
\int_{S_4}d^4x\sqrt{\sirc{g}}Y^A Y^B Y^C Y^D=\frac{V_4}{5\cdot 7}(\delta^{AB}
\delta^{CD}+\delta^{AC}\delta^{BD}+\delta^{AD}\delta^{BC})
\ee
and integrals of any even number of $Y^A$'s will be proportional to
symmetrized products of delta functions in a similar way. Integrals of an
odd number of $Y^A$'s will give zero. See also the Appendix in \cite{lmrs}.
A basic identity which we will use repeatedly is given by
\be
\epsilon_{A_1...A_5}=5\sqrt{\sirc{g}}\epsilon_{\mu\nu\rho\sigma}
\partial^{\mu}Y^{[A_1}...\partial^{\sigma}Y^{A_4} Y^{A_5]}
\label{basic}
\ee
It also follows from the fact that $C_m^A$ and $C_5^A$ form an orthogonal matrix.
Another proof is obtained by acting with $\sirc{D}_{\lambda}$
on both sides. The l.h.s. gives zero, and because of (\ref{ddy}), on the
r.h.s. when
 $\sirc{D}_{\lambda}$ hits the $\partial Y$'s or $Y^{A_5}$ gives zero
 by antisymmetry. Being constant, the r.h.s. should be proportional
 to the l.h.s., and the proportionality constant is fixed by looking at x=0.
At x=0, $\eta^{\alpha I}=\Omega^{\alpha I}$, so from (\ref{x=0}) we get on the
r.h.s. $5\sqrt{\sirc{g}}\epsilon_{\mu\nu\rho\sigma}\delta^{\mu [A_1}...\delta
^{\sigma A_4}\delta^{5A_5]}$= $\epsilon_{A_1...A_4}\delta_{5A_5}+
\epsilon_{A_2...A_5}\delta_{5A_1}+...+\epsilon_{A_5...A_1}\delta_{5A_4}=
\epsilon_{A_1...A_5}$.

By repeatedly using (\ref{yaya}),(\ref{ydy}),(\ref{dydy}), one derives from
(\ref{basic}) the following  further identities
\bea
\epsilon_{A_1...A_5}dY^{A_1}&=&4
\sqrt{\sirc{g}}\epsilon_{\mu\nu\rho\sigma}
dx^{\mu}\partial^{\nu}Y^{[A_2}\partial^{\rho}Y^{A_3}\partial^{\sigma}Y^{A_4}
Y^{A_5]}\label{edy}\\
\epsilon_{A_1...A_5}dY^{A_1}\wedge dY^{A_2}&=&3\sqrt{
\sirc{g}}\epsilon_{\mu\nu\rho\sigma}dx^{\mu}dx^{\nu}\partial^{\rho}Y^{[A_3}
\partial^{\sigma}Y^{A_4}Y^{A_5]}\label{edydy}\\
\epsilon_{A_1...A_5}dY^{A_1}\wedge dY^{A_2}\wedge dY^{A_3}&=&2
\sqrt{\sirc{g}}\epsilon_{\mu\nu\rho\sigma}dx^{\mu}\wedge dx^{\nu}\wedge
dx^{\rho}\partial^{\sigma}Y^{[A_4}Y^{A_5]}\nonumber\\
&&=-2
\sqrt{\sirc{g}}\epsilon_{\mu\nu\rho\sigma}dx^{\mu}\wedge dx^{\nu}\wedge
dx^{\rho} V^{\sigma[A_4 A_5]}\label{edydydy}\\
\epsilon_{A_1...A_5}dY^{A_1}\wedge...\wedge dY^{A_4}&=&
\sqrt{\sirc{g}}\epsilon_{\mu\nu\rho\sigma}dx^{\mu}\wedge...\wedge
dx^{\sigma} Y^{A_5}\label{edydydydy}
\eea
Contracting these identities with $Y^{A_5}$, leads to a further chain
of identities.
\bea
\epsilon_{A_1...A_5}dY^{A_1}\wedge...\wedge dY^{A_4}Y^{A_5}&=&
\sqrt{\sirc{g}}\epsilon_{\mu\nu\rho\sigma}dx^{\mu}\wedge...\wedge
dx^{\sigma} \label{edydydydyy}\\
\epsilon_{A_1...A_5}dY^{A_1}\wedge dY^{A_2}\wedge dY^{A_3}Y^{A_5}&=&
\sqrt{\sirc{g}}\epsilon_{\mu\nu\rho\sigma}dx^{\mu}\wedge dx^{\nu}\wedge
dx^{\rho}\partial^{\sigma}Y^{A_4}\label{edydydyy}\\
\epsilon_{A_1...A_5}dY^{A_1}\wedge dY^{A_2}Y^{A_5}&=&\sqrt{
\sirc{g}}\epsilon_{\mu\nu\rho\sigma}dx^{\mu}dx^{\nu}\partial^{\rho}Y^{[A_3}
\partial^{\sigma}Y^{A_4]}\label{edydyy}\\
\epsilon_{A_1...A_5}dY^{A_1}Y^{A_5}&=&\sqrt{\sirc{g}}\epsilon_
{\mu\nu\rho\sigma}
dx^{\mu}\partial^{\nu}Y^{[A_2}\partial^{\rho}Y^{A_3}\partial^{\sigma}Y^{A_4]}
\label{edyy}\\
\epsilon_{A_1...A_5}Y^{A_5}&=&\sqrt{\sirc{g}}\epsilon_{\mu\nu
\rho\sigma}\partial^{\mu}Y^{[A_1}...\partial^{\sigma}Y^{A_4]}\label{ey}
\eea
The last identity follows from (\ref{basic}).
 As a check on the normalization of the identities
(\ref{edy}) to (\ref{ey}), we consider the point
 x=0 and use (\ref{x=0}). For example, the l.h.s. of (\ref{edydydy}) yields
 $\epsilon_{A_1... A_5}\delta_{\mu}^{A_1}\delta_
{\nu}^{A_2}\delta_{\rho}^{A_3}$ which agrees with $2\epsilon_{\mu\nu\rho\sigma}
\delta_{[A_4}^{\sigma}\delta_{A_5]}^5$ on the r.h.s.

By Fierz rearrangements, we get other identities.
 Fierzing $\eta^{[K}(\bar{\eta}^{J]}\eta^I)$, we get
\be
\gamma_5\eta^I\phi_5^{JK}-\gamma_{\mu}\gamma_5\eta^I C_{\mu}^{JK}=
4\eta^{[K}\Omega^{J]I}-\eta^{I}\Omega^{JK}\label{fierz}
\ee
from which we retrieve (\ref{compl}) after multiplying with $\bar{\eta}^L$.
 Fierzing  $\eta^{(K}(\bar{\eta}^{J)}\eta^I)$, we obtain
\be
\gamma_{\mu}\eta^I V_{\mu}^{JK}-\frac{1}{2}\gamma_{\mu\nu}\eta^I(\bar{\eta}^J
\gamma_{\mu\nu}\eta^K)=4\eta^{(K}\Omega^{J)I}
\ee
In particular, (\ref{fierz}) will be heavily used.
\section{Derivation of the complete nonlinear Kaluza Klein reduction.}
\subsection{The dimensional reduction of the 11 dimensional vielbein
transformation law}

With the tools developed in the previous section, we  turn to the main
problem of this article, finding the complete nonlinear Kaluza-Klein
ansatz.

For completeness, we give below the action and transformation laws of
the dimensionally reduced theory, namely maximal gauged sugra in $d=7$.
The model has a local $SO(5)_g$ gauge group for which A,B,... =1,5 are vector
indices, while I,J,... =1,4 are spinor indices.
The scalars ${\Pi_A}^i$ parameterize the coset $SL(5,{\bf R})/SO(5)_c$
but in the gauged model the  $SL(5,{\bf R})$ rigid symmetry of the
action is lost and replaced by the $SO(5)_g$ gauge invariance.
The subscripts $g$ stand for gauge, and $c$ for composite. The indices
i,j,...=1,5 are $SO(5)_c$ vector indices and $I',J'$,...=1,4  are spinor
indices.
The model has the following fields: the vielbein ${e_\alpha}^a$, the
4 gravitinos $\psi^{I'}_\alpha$, the $SO(5)_g$ vector $B_\alpha^{AB}=-
B_\alpha^{BA}$, the scalars ${\Pi_A}^i$, the antisymmetric tensor
$S_{\alpha\beta\gamma,A}$ and the spin 1/2 fields $\lambda_i^{I'}$
(vector-spinors under $SO(5)_c$, satisfying $\gamma^i\lambda_i^{I'}=0$).
They have the correct mass-terms which
ensure 'masslessness' in d=7 AdS space \cite{mtn,pvn2}. The action reads
\bea
&&e^{-1}{\cal L}=-\frac{1}{2}R+\frac{1}{4}m^2(T^2-2T_{ij}T^{ij})-\frac{1}{2}
P_{\alpha ij}P^{\alpha ij}-\frac{1}{4}({\Pi_A}^i{\Pi_B}^j
F_{\alpha\beta}^{AB})^2
\nonumber\\
&+&\frac{1}{2}({{\Pi^{-1}}_i}^AS_{\alpha\beta\gamma ,A})^2
+\frac{1}{48}me^{-1}\epsilon^{\alpha\beta\gamma\delta\epsilon\eta\zeta}
\delta^{AB}S_{\alpha\beta\gamma ,A}F_{\delta\epsilon\eta\zeta ,B}-
\frac{1}{2}\bar{\psi}_{\alpha}\tau^{\alpha\beta\gamma}\bigtriangledown_{\beta}
\psi_{\gamma}
\nonumber\\
&-&\frac{1}{2}\bar{\lambda}^i\tau^{\alpha}\bigtriangledown_{\alpha}
\lambda_i
-\frac{1}{8} m(8T^{ij}-T\delta^{ij})\bar{\lambda}_i\lambda_j+\frac{1}{2}
mT^{ij}\bar{\lambda}_i\gamma_j\tau^{\alpha}\psi_{\alpha}+\frac{1}{2}\bar{\psi}
_{\alpha}\tau^{\beta}\tau^{\alpha}\gamma^i\lambda^jP_{\beta ij}
\nonumber\\
&+&\frac{1}{8}mT\bar{\psi}_{\alpha}\tau^{\alpha\beta}\psi_{\beta}
+\frac{1}{16}\bar{\psi}_{\alpha}(\tau^{\alpha\beta\gamma\delta}-2g^{\alpha
\beta}g^{\gamma\delta})\gamma_{ij}\psi_{\delta}{\Pi_A}^i{\Pi_B}^j
F_{\beta\gamma}^{AB}
\nonumber\\
&+&\frac{1}{4}\bar{\psi}_{\alpha}\tau^{\beta\gamma}
\tau^{\alpha}\gamma_i\lambda_j{\Pi_A}^i{\Pi_B}^jF_{\beta\gamma}^{AB}
+\frac{1}{32}\bar{\lambda}_i\gamma^j\gamma_{kl}\gamma^i\tau^{\alpha\beta}
\lambda_j{\Pi_A}^k{\Pi_B}^lF_{\alpha\beta}^{AB}+\frac{im}{8\sqrt{3}}
\bar{\psi}_{\alpha}(\tau^{\alpha\beta\gamma\delta\epsilon}
\nonumber\\
&+&6g^
{\alpha\beta}\tau^{\gamma}g^{\delta\epsilon})\gamma^i\psi_{\epsilon}
{{\Pi^{-1}}_i}^A S_{\beta\gamma\delta ,A}
-\frac{im}{4\sqrt{3}}\bar{\psi}_{\alpha}(\tau^{\alpha\beta\gamma\delta}-3
g^{\alpha\beta}\tau^{\gamma\delta})\lambda^i{{\Pi^{-1}}_i}^A S_{\beta
\gamma\delta ,A}
\nonumber\\
&-&\frac{im}{8\sqrt{3}}\bar{\lambda}^i\tau^{\alpha\beta\gamma}
\gamma^j\lambda_i{{\Pi^{-1}}_j}^A S_{\beta\gamma\delta ,A}
+\frac{ie^{-1}}{16\sqrt{3}}\epsilon^{\alpha\beta\gamma\delta\epsilon\eta\zeta}
\epsilon_{ABCDE}\delta^{AG}S_{\alpha\beta\gamma, G}F_{\delta\epsilon}^{BC}
F_{\eta\zeta}^{DE}
\nonumber\\
&+&\frac{m^{-1}}{8}e^{-1}\Omega_5[B]-
\frac{m^{-1}}{16}e^{-1}\Omega_3[B]\label{7action}
\eea
The local supersymmetry transformation rules are given by
\bea
&&\delta e_{\alpha}^a =\frac{1}{2}\bar{\epsilon}\tau^a\psi_{\alpha}\label{grav}
\\
&&{\Pi_A}^i{\Pi_B}^j\delta B_{\alpha}^{AB}=\frac{1}{4}\bar{\epsilon}
\gamma^{ij}\psi_{\alpha}+\frac{1}{8}\bar{\epsilon}\tau_{\alpha}\gamma^k
\gamma^{ij}\lambda_k\label{gauge}
\\
&&\delta S_{\alpha\beta\gamma ,A}=-\frac{i\sqrt{3}}{8m}{\Pi_A}^i(2
\bar{\epsilon}\gamma_{ijk}\psi_{[\alpha}+\bar{\epsilon}\tau_{[\alpha}\gamma^l
\gamma_{ijk}\lambda_l){\Pi_B}^j{\Pi_C}^kF_{\beta\gamma]}^{BC}
\nonumber\\
&&-\frac{i\sqrt{3}}{4m}\delta_{ij}{\Pi_A}^j D_{[\alpha}
(2\bar{\epsilon}\tau_{\beta}\gamma^i\psi_{\gamma]}+\bar{\epsilon}\tau_{\beta
\gamma ]}\lambda^i ) \nonumber\\&&
+\frac{i\sqrt{3}}{12}\delta_{AB}{{\Pi^{-1}
}_i}^B(3\bar{\epsilon}\tau_{[\alpha\beta}\gamma^i\psi_{\gamma]}-\bar{\epsilon}
\tau_{\alpha\beta\gamma}\lambda^i )\label{S} \\
&&{{\Pi^{-1}}_i}^A\delta {\Pi_A}^j=\frac{1}{4}(\bar{\epsilon}\gamma_i
\lambda^j+\bar{\epsilon}\gamma^j\lambda_i)\label{scalar}\\
&&\delta\psi_{\alpha}=\bigtriangledown_{\alpha}\epsilon-\frac{1}{20}mT
\tau_{\alpha}\epsilon -\frac{1}{40}(\tau_{\alpha}\;^{\beta\gamma}-8
\delta_{\alpha}^{\beta}\tau^{\gamma})\gamma_{ij}\epsilon {\Pi_A}^i{\Pi_B}^j
F_{\beta\gamma}^{AB}
\nonumber\\
&&+\frac{im}{10\sqrt{3}}(\tau_{\alpha}\;^{\beta\gamma\delta}
-\frac{9}{2}\delta_{\alpha}^{\beta}\tau^{\gamma\delta})\gamma^i\epsilon
{{\Pi^{-1}}_i}^AS_{\beta\gamma\delta ,A} \label{gravi}\\
&&\delta\lambda_i=\frac{1}{16}\tau^{\alpha\beta}(\gamma_{kl}\gamma_i-\frac{1}{5
}\gamma_i\gamma_{kl})\epsilon {\Pi_A}^k{\Pi_B}^l F_{\alpha\beta}^{AB}
+\frac{im}{20\sqrt{3}}\tau^{\alpha\beta\gamma}(\gamma_i\; ^j-4\delta_i^j)
\epsilon{{\Pi^{-1}}_j}^A S_{\alpha\beta\gamma ,A}
\nonumber\\
&&+\frac{1}{2}m(T_{ij}-\frac{1}{5}T\delta_{ij})
\gamma^j\epsilon+\frac{1}{2}\tau^{\alpha}
\gamma^j \epsilon P_{\alpha ij}
\eea

The contractions over indices $I',J'$ are always as in $\bar{\epsilon}_{I'}
\tau^a\psi_{\alpha }^{I'}$ and $\bar{\epsilon_{I'}} {(\gamma^i )^{I'}}_{J'}
\psi_{\alpha}^{J'}$.
Here $T_{ij}={{\Pi^{-1}}_i}^A{{\Pi^{-1}}_j}^B\delta_{AB}$ and
$\Omega_3[B]$ and  $\Omega_5[B]$ are the Chern-Simons forms for $B_{\alpha
}^{AB}$ (normalized to $d\Omega_3[B]=(TrF^2)^2$ and $d\Omega_5[B]=(TrF^4)$).
The tensor $P_{\alpha\,ij}$ and the connection $Q_{\alpha\,ij}$ (appearing in
the covariant derivatives as $\nabla_\alpha= \partial_\alpha+
Q_{\alpha ..}$ and $D_\alpha =\partial_\alpha +Q_{\alpha ..}+
P_{\alpha ..}$ in (\ref{S})) are the symmetric and antisymmetric parts of $
{(\Pi^{-1})_i}^A\left({\delta_A}^B \partial_\alpha + g B_{\alpha\,A}\;^B\right)
{\Pi_B}^k \delta_{kj}$ respectively. The curl of $S_{\alpha\beta\gamma, A}$
has strength 4, so $F_{\alpha\beta\gamma\delta ,A}=4\nabla_{[\alpha}S_{
\beta\gamma\delta ],A}$. Further, $g=2m$ (or rather $g=2mk$ but $k=1$) and
(\ref{adsul}) is the $AdS_7$ solution of the field equations of
(\ref{7action}) with $T_{ij}=\delta_{ij}$.  We put $m=1$ most of the time,
so the
the field strength of the gauge field is defined as $F_{\alpha\beta}^{AB}=
\partial_\alpha B_\beta^{AB} + 2 B_\alpha^{AC} B_\beta^{CB} - (\beta
\leftrightarrow \alpha)$. However, note that the limit $m\rightarrow 0$ is
singular due to the factors of $m^{-1}$ in front of the Chern-Simons terms.

At the nonlinear level, we need an ansatz for the vielbein rather
than the metric as we did in section 2.3. The  ansatz for the vielbein
$E_{\Lambda}^M$ is constructed as follows.
In order to fix the off-diagonal part of the Lorentz group and remain with
local $SO(6,1)\times SO(4)$ invariance,
we impose the gauge choice $E_{\mu}^a =0$.
The natural extension of the rescaling in (\ref{delta}), needed
to obtain the usual Einstein actionin d=7 without extra powers of $\Delta$,
is\footnote{More generally, when dimensionally reducing a D dimensional
supergravity theory  to d dimensions, one has $E_{\alpha}^a(y,x)=
e_{\alpha}^a(y)\Delta^{-1/(d-2)}(y,x)$} :
\be
E_{\alpha}^a(y,x)=e_{\alpha}^a(y)\Delta^{-1/5} (y,x), \;\Delta(y,x)\equiv
\frac{det E_{\mu}^m}{\det\sirc{e}_{\mu}^m}\label{ealfaa}
\ee
This is a standard result in KK reduction of theories with gravity.
The ansatz for $E_{\alpha}^m(y,x)$ is also standard. It contains the gauge
bosons
\bea
E_{\alpha}^m(y,x)&=&B_{\alpha}^{\mu}(y,x)E_{\mu}^m(y,x)\label{bansatz}\\
B_{\alpha}^{\mu}(y,x)&=&-2B_{\alpha}^{AB}(y)V^{\mu AB}(x)
\eea
where $B_{\alpha}^{AB}(y)$ is the 7 dimensional SO(5) gauge field and
$V^{\mu AB}$ the corresponding Killing vector.
The factor of (-2) in the ansatz of $B_\alpha^\mu$ comes from $g=2m$, and
the minus sign is related to the sign convention in the definition of the
covariant derivative $D_\alpha =\partial_\alpha +g B_\alpha$.
The only nontrivial step is finding an ansatz for $E_{\mu}^m$.

Before we do that, let's analyze the fermions. The gravitinos
$\Psi_M\equiv E_M^\Lambda\Psi_{\Lambda}$ split into
$\Psi_a\equiv E_a^\Lambda\Psi_{\Lambda}$ and $\Psi_m\equiv E_m^{\Lambda}\Psi_{
\Lambda}$ (We suppress the  32- dimensional spinor
index.) In order to obtain diagonal kinetic terms for the fermions in $d=7$,
we proceed analogously to the vielbein and begin by introducing fields
$\Psi_a(y,x)$ and $\Psi_m(y,x)$ as follows:
\bea
\Psi_a&=&\Delta^{1/10}(\gamma_5)^{-p}\psi_a -A\frac{1}{5}\tau_a\gamma_5
\gamma^m
\Delta^{1/10}(\gamma_5)^q\psi_m\label{useful}\\
\Psi_m&=&\Delta^{1/10}(\gamma_5)^{q}\psi_m\label{useful2}\\
\varepsilon&=&\Delta^{-1/10}(\gamma_5)^{-p}\epsilon,
\bar{\varepsilon}=\Delta^{-1/10}\bar{\epsilon}(\gamma_5)^{-p}
\eea
where A, p and q will be fixed.
The inverse relations, together with our ansatz are given by
\bea
\psi_{\alpha}(y,x)&=&\Delta^{-1/10}(y,x)(\gamma_5)^p e_{\alpha}^a(y)
(\Psi_a(y,x)+A\frac{1}{5}\tau_a\gamma_5\gamma^m\Psi_m(y,x))\nonumber\\&=&
\psi_{\alpha I'}(y)U^{I'}\;_I(y,x)\eta^I(x)
\label{psia}\;\; where\;\; \psi_a=e_a^{\alpha}(y)\psi_{\alpha}(y,x)\\
\psi_m(y,x)&=&\Delta^{-1/10}(y,x)(\gamma_5)^{-q}\Psi_m(y,x)\nonumber\\
&=&\lambda_{J'K'L'}(y)U^{J'}\;_J(y,x)U^{K'}\;_K(y,x)U^{L'}\;_L(y,x)
\eta_m^{JKL}(x)
\label{psim}
\\
\epsilon (y,x)&=&\Delta^{1/10}(y,x)(\gamma_5)^p\varepsilon(y,x)
=\epsilon_{I'}(y)U^{I'}\;_I\eta^I(x)\label{epsi}\\
\bar{\epsilon}(y,x)&=&\bar{\epsilon}_{I'}(y)U^{I'}\;_I\bar{\eta}^I(x)
\;{\rm where}\;\; \bar{\epsilon}_{I'}=\epsilon_{I'}^TC^{(7)}\;\;{\rm and}
\;\;\bar{\eta}^I=\eta^{I,T}C^{(4)}
\eea
Substituting this
ansatz for $\Psi_a$ and $\Psi_m$ into the 11d action, and requiring
that the action for $\psi_\alpha$ and $\psi_m$ becomes diagonal
 we find that A=+1 and $p =\pm \frac{1}{
2}$, $q=\pm \frac{1}{2}$ (four combinations). However, by trying to
match the 7 dimensional transformation laws, we will
find that only $p=q$ works. The freedom in the signs of
$p,q$ corresponds to a rescaling  by $\gamma_5$.
This freedom will be fixed in the following by the requirement of consistent
truncation. We will find that only $p=q=-1/2$ works. (See for instance
(\ref{fixes}) below.) Incidentally we note that in the linearized ansatz
used in \cite{pnt} the other sign was chosen. Since in \cite{pnt} only
the leading fermionic transformation laws were studied, it did not
matter at that point which rescaling by $\gamma_5$ was used.

We also note that the rotation in (\ref{useful}, \ref{useful2}) is the only
one which diagonalizes the action for the seven dimensional gravitini and
spin 1/2 fermions. If we try the more general rotation
\bea
\Psi_a&=&A(\gamma_5)^p\psi_a +B\tau^a\gamma^m(\gamma_5)^q\psi_m
\nonumber\\
\Psi_m&=&C(\gamma_5)^r\psi_m +D\gamma_m(\gamma_5)^s\tau^a\psi_a
\eea
by requiring a diagonal kinetic action for the gravitino and spin 1/2
fermions we recover (\ref{useful}) and (\ref{useful2}).

In the field redefinitions, $\gamma_5^{-1/2}$in $\psi_{\alpha}$ is
needed to cancel the $\gamma_5$ coming from $\Gamma^{\alpha\beta\gamma}
=\tau^{\alpha\beta\gamma}\otimes
\gamma_5$ in the gravitino action; $\Delta^{-1/10}$ is needed to bring
the gravitino action to the usual Rarita-Schwinger form with no extra
powers of $\Delta$ (which would come from
$det E=det e^{(7)}\Delta^{1-7/5}det\sirc{e}^{(4)}$); finally, the
rotation of $\Psi_a, \Psi_m$ into $\psi_a, \psi_m$ is needed to
cancel the mixed terms $\Psi_{\alpha}\Gamma
^{\alpha\beta m} D_{\beta}\Psi_m$. Since $\delta \Psi_{\Lambda}=
D_{\Lambda}\varepsilon$+ more, we need factors $\gamma_5^{1/2}$ and
$\Delta^{-1/10}$ in $\varepsilon$. For the
linearized case we reobtain (\ref{psi}).

The $U$ matrix in the expansions of $\psi_{\alpha}, \psi_m$ and $\epsilon$
is a local (x and y dependent) SO(5) matrix in the spinor representation
 depending on the scalar
fields in 7 dimensions (${\Pi_A}^i$).
Various authors \cite{dwn84,anp,nilsson} found it necessary to add a U matrix
 for consistency
of the truncation of the susy transformation laws, and we shall find the
same need (see below (\ref{u1})). One can either
 rotate the label index $I$ or the spinor
index of the Killing spinors. We choose the former, which is perhaps
more natural, since
the fermions have an SO(5) composite index, whereas the Killing
spinors have naturally an SO(5) gauge index.  (de Wit and Nicolai
took the alternative rotation of spinor indices
by U because they reformulated d=11 sugra in a form with local SU(8) invariance
parameterized by an arbitrary $U(x,y)$, and in d=11 there are only
 spinorial indices are available.
When one compactifies, the d=11 SU(8) invariance
is fixed, only the 4d part remains unfixed. There is a difference between
the gauge $U=1$ which gives the usual d=11 sugra (in the triangular gauge
$E_\mu^a=0$) and the gauge needed for
$S_7$ compactification. Therefore, the gauge fixing produces also for us a
given field-dependent matrix $U(y,x)$ which acts on the Killing spinors.)

In d=7 one can always
make $SO(5)_c$ gauge transformations on the fermions, and this would modify
the y dependence of $U(y,x)$ by a gauge factor.
As we shall see in (\ref{susygrav}),  consistency of our results for the
transformation rules  requires the matrix $U$ to satisfy the
condition
\bea
&&U^{I'}\;_I\tilde{\Omega}^{IJ}{U^{J'}}_J=\tilde{\Omega}^{I'J'}
\rightarrow (\tilde{\Omega}\cdot U^T\cdot\Omega)^I_{\;I'}=-{(U^{-1})^I}_{I'}
\nonumber\\&&
({\rm so\;\; that}\;\;U\tilde{\Omega}U^T=\tilde{\Omega}\;\; ;\;\; U^T\tilde{
\Omega}U=\tilde{\Omega})
\label{uinv}
\eea
The need for this relation will occur in various other places as well.
Since $\Omega$ is
the charge conjugation matrix (see Appendix A.1) and
$\tilde\Omega=-\Omega^{-1}$,
this proves that $U$ is a SO(5) matrix in the spinor representation.
(Namely, an USp(4) matrix as it is explained in the Appendix A.2).

Let us now return  to $E_{\mu}^m$. We will determine the ansatz for
$E_\mu^m$ by matching
$\delta B_{\alpha}^{[AB]}$ with the expression in 7 dimensional gauged
supergravity. The result is given in (\ref{viel2}).

Since we are imposing the gauge $E_{\mu}^a=0$,
we need a compensating SO(10,1) Lorentz transformation characterized
by an antisymmetric matrix $\Omega_m\;^a$ in order to stay in this
gauge
\bea
\delta E_{\mu}^a&=&0=\delta_{SUSY}E_{\mu}^a+E_{\mu}^m\Omega_m\;^a\Rightarrow
\nonumber\\
\Omega_m\;^a&=&-\frac{1}{2}\ \bar{\varepsilon}\Gamma^a\Psi_m=-\Omega^a\;_m
\label{Lorentz}
\eea
This Lorentz transformation acts on all vielbein components, for instance
\be
\delta E^{\mu}_a=\delta_{SUSY}E^{\mu}_a+E^{\mu}_m\Omega^m\;_a, etc.
\ee
We can now require that we get the correct graviton transformation law
in d=7. Using (\ref{dele}), we get from (\ref{ealfaa})
\bea
\delta ' e_{\alpha}^a&=&\Delta^{1/5}[\delta(d=11) E_{\alpha}^a+\frac{1}{5}
E^{\mu}_m(\delta(d=11)
 E_{\mu}^m)E_{\alpha}^a +E_{\alpha}^m\Omega_m\;^a]\nonumber\\
&=&\frac{1}{2}\Delta^{-1/10}\bar{\epsilon}\gamma_5^{-p}[\tau^a
\gamma_5\Psi_b e^b_{\alpha}+\frac{1}{5}\gamma^m\Psi_m e_{\alpha}^a]
\nonumber\\
&=&\frac{1}{2}  \bar\epsilon_{I'}\tau_a \psi_{\alpha J'}
U^{I'}_{\;\;I} U^{J'}_{\;\;J} \bar\eta^I \eta^J+...
\label{susygrav}
\eea
Since according to (\ref{killing}) $\bar\eta^I \eta^J$ equals $\Omega^{IJ}$,
we see the condition (\ref{uinv}) appearing.
Because this is not yet the final form of the vielbein law, we have introduced
the notation $\delta(d=11)$ for the d=11 susy transformations and $\delta '$
for the intermediate transformation.

In order to obtain agreement with (\ref{grav}), we add
 a field dependent SO(6,1) rotation to the d=11 transformation laws.
 Adding a term $\frac{1}{5}\tau^{ab}\gamma^m
\Psi_m e_{\alpha}^b$ inside the brackets,  we obtain
\be
\delta e_{\alpha}^a=\frac{1}{2}\bar{\epsilon}\tau^a\psi_{\alpha}=
\frac{1}{2}\bar{\epsilon}_I\tau^a\psi_{\alpha J}\Omega^{IJ}=\frac{1}{2}
\bar{\epsilon}_I\tau^a\psi_{\alpha}^I
\ee
This is the correct 7 dimensional transformation law of the d=7 vielbein.
\footnote{The $d=7$ susy results are thus a combination of $d=11$ susy and
$d=7$ Lorentz symmetry. Excluding 3-fermion terms, the $d=7$ Lorentz
transformations do not show up anywhere else.}

The transformation law  for the gauge fields follows from (\ref{bansatz}).
We need the compensating Lorentz transformation in (\ref{Lorentz}) on both
vielbeins in (\ref{bansatz}),
but there is no SO(6,1) Lorentz transformation since $B_{\alpha}^{\mu}$ has
no Lorentz index. One finds
\bea
\delta B_{\alpha}^{\mu}(y,x)
&=&\delta(d=11)(E_{\alpha}^m E_m^{\mu})+ E_{\alpha}^m(E^{\mu}_a
\Omega^a\;_m)+(E_{\alpha}^a\Omega_a\;^m)E_m^{\mu}\nonumber\\
&=&i\frac{\Delta^{-1/5}}{2}E_m^{\mu}\{-\bar{\epsilon}\gamma^m\psi_{\alpha}
-i\bar{\epsilon}\tau_{\alpha}(\delta^{mn}+\frac{\gamma^m\gamma^n}{5})
\gamma_5\psi_n\}\label{dbmua}
\eea
where we used (\ref{gamma5}) in the last step.
But in 7 dimensions, we have from (\ref{gauge})
\be
\delta B_{\alpha}^{AB}(y)={(\Pi^{-1})_i}^A{(\Pi^{-1})_j}^B
[\frac{1}{4}\bar{\epsilon
}\gamma^{ij}\psi_{\alpha}+\frac{1}{8}\bar{\epsilon}\tau_{\alpha}
\gamma^k\gamma^{ij}\lambda_k]\label{gf7}
\ee
By multiplying (\ref{gf7}) with $-2V^{\mu}_{AB}$ and equating the first
term in (\ref{dbmua}) and (\ref{gf7}), one obtains
\be
-E^{\mu}_m\bar{\eta}^I\gamma^m\eta^J {U^{I'}}_I {U^{J'}}_J
\bar{\epsilon}_{I'}\psi_{\alpha J'}\frac{i\Delta^
{-1/5}}{2}=\frac{1}{2}{(\Pi^{-1})_i}^A{(\Pi^{-1})_j}^B V^{\mu}_{AB}
(\gamma^{ij})^{I'J'}\bar{\epsilon}_{I'}\psi_{\alpha J'}\label{emum}
\ee
Dropping a common factor $1/2\bar{\epsilon}_{I'}\psi_{\alpha J'}$,
we find the equation
\be
iE^\mu _m (U V^m U^T)^{I'J'}=-\Delta^{1/5}{(\Pi^{-1})
_i}^A{(\Pi^{-1})_j}^B V^{\mu}_{AB} (\gamma^{ij})^{I'J'}\label{vie}
\ee
where $V^{m,IJ}=\bar{\eta}^I\gamma^m\eta^J$.

This equation can be solved for $E^\mu _m$ by multiplication with $(UV^nU^T)_{
I'J'}$ and using (\ref{uinv}) and (\ref{vortog})
\be
E_m^{\mu}=\frac{i}{4}\Delta^{1/5}{(\Pi^{-1})_i}^A{(\Pi^{-1})
_j}^B V^{\mu}_{AB} Tr(\gamma^{ij} U V_m U^T\Omega)\label{viel}
\ee
(We used that $U\tilde{\Omega}U^T=\tilde{\Omega}$ implies that
$U^T\Omega U=\Omega$ as follows from eliminating $U^T$).
To obtain an explicit expression for $E_{\mu}^m$, we move the factors
$\Pi^{-1}$ and $\gamma^{ij}$ in the r.h.s. of (\ref{vie}) to the left.
Contracting with $V_{\nu}^{AB}$ and $E_{\mu}^m$
and using (\ref{vortog}), we find
\be
E^m_\nu=\frac{i}{4}\Delta^{-1/5}
Tr(\gamma_{ij} UV^m U^T\Omega){\Pi_A}^i {\Pi_B}^j V^{AB}_\nu\label{viel2}
\ee

As a check on our results developed so far, we note that
substituting (\ref{viel}) into (\ref{vie}) one should get an identity. We find
\be
\frac{1}{4}{(\Pi^{-1})_i}^A{(\Pi^{-1})
_j}^B V^{\mu}_{AB} [Tr(\gamma^{ij} U V_m U^T\Omega)](U V^m U^T)^{I'J'}
={(\Pi^{-1})
_i}^A{(\Pi^{-1})_j}^B V^{\mu}_{AB} (\gamma^{ij})^{I'J'}\label{u1}
\ee
The matrices $U^{I'}\;_{I}$ are really needed for the consistency of
(\ref{vie}). If one sets $U=1$  in (\ref{u1}) and removes the factors
$\Pi^{-1}(y)$ one is left with an x-dependent equation for the Killing vectors
which is incorrect. When $U$ is y-dependent, one cannot factor off
the $\Pi^{-1}$
and the relation becomes a condition on $U$. It will follow from the
fundamental condition on U which we obtain in (\ref{urel}).

We should now check if with our ansatz for the vielbein in (\ref{viel})
we also reproduce the second term in the transformation law (\ref{gf7})
of the gauge field. We already know at this point $E^{\mu}_m$, hence first we
should make sure that the r.h.s. of (\ref{dbmua}) is proportional to
$V_{\mu}^{AB}$. For this purpose we work on the combination of spherical
harmonics on the r.h.s. of (\ref{dbmua}). They can be rewritten as follows
\bea
\bar{\eta}^I(\delta^{mn}+\frac{1}{5}\gamma^m\gamma^n)\gamma_5\eta_n^{JKL}
&=&3(\phi_5^{IJ}C_m^{KL}-C_m^{IJ}\phi_5^{KL})\nonumber\\
&&+\frac{24}{5}V_m^{I[L}\Omega^{K]J}-\frac{6}{5}V_m^{IJ}\Omega^{KL}
\label{etajkl}
\eea
where we have substituted the expression of $\eta_m^{JKL}$ in (\ref{vecsp})
and used the Fierz relation (\ref{fierz}). However, when projected on
$\lambda_{J'K'L'}$ only the first two terms on the r.h.s of (\ref{etajkl})
will contribute because $\lambda$ is symplectic traceless.
Thus, the $\psi_m$ term in the $d=11$ transformation law of the gauge field is:
\bea
{\delta B_{\alpha}^{\mu}|}_{\psi_m \;term}&=&\frac{3}{2} E_m^\mu \Delta^{-1/5}
\bar{\epsilon}_{I'} \tau_\alpha\lambda_{J'K'L'}{U^{I'}}_I {U^{J'}}_J
{U^{K'}}_K {U^{L'}}_L \nonumber\\&&
(\phi_5^{IJ}(C^m)^{KL}-\phi_5^{KL}(C^m)^{IJ})\label{cevaul}
\eea
where we have used the ansatz in (\ref{psim}) for $\psi_m$.
Using $Y^A=\frac{1}{4} \phi_5^{IJ}(\gamma^A)_{IJ}$ and $C_{\mu}^B
=\partial_{\mu} Y^B$, it follows from (\ref{vmu}) that this expression
is proportional to the Killing vector $V^{\mu}_{AB}$, but there are still
the extra $U$ matrices. By starting with (\ref{fierz}) contracted with
$\bar{\eta}^I \gamma_m$, we obtain the second line in (\ref{cevaul}) and a
complicated term of the form $(\bar{\eta}\gamma_{mn}\gamma_5\eta)^{KL} C_n^{IJ}
$, which however
vanishes since $\lambda_{J'K'L'}$ is antisymmetric in $K'L'$. We get
\be
\delta B_{\alpha}^{\mu}|_{\psi_m term}=
-3 E_m^\mu \Delta^{-1/5} \bar{\epsilon}_{I'}
\tau_\alpha\lambda_{J'K'L'} \Omega^{I'L'}(U V^m U^T)^{J'K'}
\ee
 Using (\ref{vie}) we arrive at
\be
\delta B_{\alpha}^{\mu}|_{\psi_m term}=
-3i {(\Pi^{-1})_i}^A {(\Pi^{-1})_j}^B (\gamma^{ij})^{J'K'}\Omega^{I'L'}
\bar{\epsilon}_{I'}\tau_\alpha\lambda_{J'K'L'} V^\mu_{AB}\label{2term}
\ee

Since we have arrived at an expression proportional to $V^{\mu}_{AB}$
we can compare with (\ref{gf7}).
The 7 dimensional transformation law is equal to
\be
-\frac{1}{4} {(\Pi^{-1})_i}^A {(\Pi^{-1})_j}^B \bar\epsilon\tau_\alpha\gamma^k
\gamma^{ij}\lambda_k V^\mu_{AB}
\ee
and this should agree with (\ref{2term}).
The two terms are equal if we assume the following normalization of the seven
dimensional spin 1/2 fields:
\be
\lambda^k_{I'}=3 i(\gamma^k)^{J'K'}\lambda_{I'J'K'}\label{normaliz}
\ee
At this point we have learned that if (\ref{vie}) is correct then
the transformation of the gauge field $B_\alpha^{AB}$ follows
from Kaluza-Klein reduction.

Next, we compute
$\Delta=det(E_\mu^m)/det({\sirc e}^m_\mu)$.
To remove the dependence on $U$, we square (\ref{vie}). Using (\ref{uinv}),
we obtain
\be
\Delta^{-2/5} g^{\mu\nu}=2 V^{\mu}_{AB} V^{\nu}_{CD}
{(\Pi^{-1})_i}^A {(\Pi^{-1})_j}^B  {(\Pi^{-1})_i}^C
{(\Pi^{-1})_j}^D\label{metric} \ee The matrix $g^{\mu\nu}$ is the
inverse of the metric $g_{\mu\nu}=G_{\mu\nu} =E_{\mu}^ME_{\nu
M}=E_{\mu}^mE_{\nu m}$. It is given by $g^{\mu\nu}=E_m^{\mu}
E^{m\nu}$ and differs from $G^{\mu\nu}=E_M^{\mu}E^{M\nu}$ by a
term $E_a^{\mu} E^{a\nu}$. Using (\ref{edydydydy}) and the fact
that $\det T =1$ we can evaluate directly the determinant of
$g_{\mu\nu}=\Delta^{4/5} \partial_\mu Y^A (T^{-1})^{AB}
\partial_\nu Y^B$
\bea &&\det g_{\mu\nu}\equiv \det T^{-1}_{AB}\partial^{\mu}Y^A
\partial^{\nu}Y^B\equiv \epsilon^{\mu_1...\mu_4}\epsilon^{\mu '_1...\mu '_4}
g_{\mu_1\mu '_1}...g_{\mu_4\mu '_4}\nonumber\\
&&\Delta^{16/5}\epsilon^{\mu_1...\mu_4}\partial_{\mu_1}Y^{A_1}...\partial_{\mu_4}Y^{A_4}
\epsilon^{\nu_1...\nu_4}\partial_{\nu_1}Y^{B_1}...\partial_{\nu_4}Y^{B_4}
T^{-1}_{A_1B_1}...T^{-1}_{A_4B_4}\nonumber\\&&
=\sirc{g}\Delta^{16/5}
\epsilon^{A_1...A_5}\epsilon^{B_1...B_5}Y_{A_5}Y_{B_5}
T^{-1}_{A_1B_1}...T^{-1}_{A_5B_5}=\sirc{g} T^{AB}Y_{A}Y_{B} \eea
conclude that
\be
\Delta^{-6/5}= {(\Pi^{-1})_i}^A  {(\Pi^{-1})_j}^B \delta^{ij} Y_A Y_B
\equiv T^{AB}Y_AY_B\label{del}
\ee

The equation (\ref{del}) is the starting point for obtaining the
nonlinear metric ansatz for the compact dimensions.

Substituting the 7-$d$ transformation law of the scalar fields in (\ref{del})
we get:
\bea
&&\delta(\Delta^{-6/5})=\delta T^{AB} Y_A Y_B= 2\delta  {(\Pi^{-1})_i}^A \cdot
 {(\Pi^{-1})_j}^B \delta^{ij} Y_A Y_B \nonumber\\
&&=\frac{-1}{2}{(\Pi^{-1})_i}^A {(\Pi^{-1})_j}^B
(\bar\epsilon \gamma^i\lambda^j + \bar\epsilon \gamma^j\lambda^i) Y_A Y_B
\label{del1}
\eea
where we recall the definition
$T_{AB}={(\Pi^{-1})_i}^A  {(\Pi^{-1})_j}^B
\delta^{ij}$ and used (\ref{scalar}).
On the other hand, from 11-$d$ sugra we obtain
\be
\delta(\Delta^{-6/5})=\frac{6}{5}\Delta^{-6/5}
\delta E_m^\mu\cdot E_\mu^m=-\frac{3}{5}\Delta^{-6/5}\bar{\varepsilon}\Gamma^m
\Psi_m \label{del65}
\ee
Again we find two expressions which should be equal, but whose spherical
harmonics are not yet manifestly the same. We rewrite the spherical
harmonic on the r.h.s. of (\ref{del65})  by
using the Fierz relation (\ref{fierz})
\be
\bar{\eta}^I\gamma_m\gamma_5\eta_m^{JKL}=-(5\phi_5^{IJ}\phi_5^{KL}
+\Omega^{IJ}\Omega^{KL}-4\Omega^{I[L}\Omega^{K]J})
\ee
Again, only the terms with $\phi_5\phi_5$ contribute because $\lambda_{I'J'K'}
$ is symplectic-traceless and one finds ($\phi_5\sim Y_A \gamma^A$)
\be
\delta(\Delta^{-6/5})=-3i\Delta^{-6/5} Y_A Y_C \bar\epsilon_{I'}
\lambda_{J'K'L'} (U\gamma^A \tilde\Omega U^T)^{I'J'} (U\gamma^C
\tilde\Omega U^T)^{K'L'}\label{del2}
\ee
There are four $U$ matrices, three from $\psi_m$ and one from $\epsilon$.
Substituting (\ref{del}) we arrive
at
\bea
&&\delta  {(\Pi^{-1})_i}^C \cdot
 {(\Pi^{-1})_j}^D \delta^{ij} Y_C Y_D\nonumber\\&&
 =-i\frac{3}{2}\Delta^{-6/5} Y_A Y_C
\bar\epsilon_{I'} \lambda_{J'K'L'} (U\gamma^A \tilde\Omega U^T)^{I'J'}
(U\gamma^C \tilde\Omega U^T)^{K'L'}
\eea
In order that this result for $\delta \Pi^{-1}$ agrees with (\ref{del1}), the
 $U$-matrices must satisfy:

\be
Y_A (U \gamma^A\tilde\Omega U^T)^{I'J'}= \pm \Delta^{3/5}  {(\Pi^{-1})_i}^A
(\gamma^{i})^{I'J'} Y_A \label{urel}
\ee
where we used the normalization relation (\ref{normaliz}) to remove
the spinors. We wil choose the minus sign, because we want that our matrix
U is equal to 1 on the background (and not to $-1$).
Equation (\ref{urel}) will turn out to be the most important tool in
our endeavor to find a consistent truncation of the 11 dimensional
supergravity.

At this point we have come in 7 dimensions as far as de Wit and
Nicolai in their first approach without a local SU(8). (Actually, a
relation corresponding to our (\ref{urel}) is not present in their
work, but they did find the one corresponding to (\ref{u1}).)
We have been able to find solutions
 for $U$ without having to extend the d=11 theory. First of all, we
have been able to show that (\ref{u1}) follows from (\ref{urel}), so that
(\ref{urel}) is the crucial equation. Secondly, we have found solutions to
(\ref{urel}).

The proof that (\ref{urel}) implies (\ref{u1}) goes as follows.
One begins with the l.h.s. of (\ref{u1})
\be
\frac{1}{4} {(\Pi^{-1})_i}^A {(\Pi^{-1})_j}^B  V_{\mu ,AB} Tr(\gamma^{ij}
U\gamma^{CD} U^{-1}) (U\gamma^{EF} U^{-1})^{I'J'} V_{m\;CD} V^m\;_{EF}
\ee
where one substitutes (\ref{vv}) to sum over the Killing vectors.
Using $V_{\mu , AB}=Y^{[A}C_{\mu}^{B]}$ and writing $2\gamma^{ij}=\gamma^i
\gamma^j-\gamma^j\gamma^i$, the r.h.s. of (\ref{urel}) appears twice, and using
(\ref{urel}) once we obtain
\be
-\frac{1}{4}\Delta^{-3/5} {(\Pi^{-1})_i}^B C_{\mu\;B} Y_A Y_C Y_E
Tr(\gamma^i U [\gamma^A, \gamma^{CD}] U^{-1}) (U\gamma^{ED} U^{-1})^{I'J'}
\ee
In the product of gamma matrices
only the term $\delta^{AC}\gamma^D$ survives due to
symmetry arguments. Splitting $( U\gamma^{ED} U^{-1})^{I'J'}$ into
$\left(U (\gamma^E \gamma^D -\delta^{ED}) U^{-1}\right)^{I'J'}$, eliminating
the first $U^{-1}$ by (\ref{uinv}) and
applying (\ref{urel}), but now in the opposite direction, we get:
\bea
&&\frac{1}{4} {(\Pi^{-1})_i}^A {(\Pi^{-1})_j}^E C_{\mu\;A} Y_E 4 \delta^{ij}
\tilde\Omega^{I'J'}-\nonumber\\
&&\frac{1}{4} {(\Pi^{-1})_i}^A {(\Pi^{-1})_j}^E C_{\mu\;A} Y_E
Tr(\gamma^i U \gamma^D U^{-1}) (\gamma^j U \gamma_D U^{-1})^{I'J'}
\eea
What is left to do is the sum over $\gamma^D$ (use (\ref{gg})) in order to get
directly the r.h.s. of (\ref{u1}) (with $\gamma^{ij}$ written again as
$\gamma^i \gamma^j- \delta^{ij}$).

Finally, we look at the solutions of equation (\ref{urel}) for the matrix $U$.
The $4\times 4$
matrix $U$ can be expanded into the complete basis $1,\gamma_A ,\gamma_{AB}$
as follows
\be
U=N+N_A \gamma_A +N_{AB}\gamma_{AB}
\ee
Then, the condition that it is an SO(5) matrix in the spinor representation,
equation (\ref{uinv}), says that
\be
U^{-1}=-\tilde{\Omega}U^T\Omega =N+N_A\gamma_A- N_{AB}\gamma_{AB}
\ee
Unitarity of $U$ implies that $N, N_A$ and $N_{AB}$ are real.
The fact that $UU^{-1}=U^{-1}U= \id$ gives then the conditions
\bea
&&N^2+ N_A^2+2N_{AB}^2=1\label{beta}\\
&&N_A  N_{AB}=0\label{alfa}\\
&&2NN_A=\epsilon_{ABCDE}N_{BC}N_{DE}\label{gama}
\eea
On the other hand (\ref{urel}), which we can rewrite as
\bea
UY\llap/ &=&v\llap/ U,\\
v_i&=&{{\Pi^{-1}}_i}^AY_A\Delta^{3/5}\;\; ;\;\; v^2=Y^2=1
\eea
implies the equations
\bea
&&N_A(Y_A-v_A)=0\label{unu}\\
&&N(Y_A-v_A)=-2N_{AB}(Y_B+v_B)\label{doi}\\
&&N_{[A}(Y+v)_{B]}=\frac{1}{2}\epsilon_{ABCDE}(Y_C-v_C)N_{DE}\label{trei}
\eea

The first observation is that if U is an SO(5) solution of $UY\llap/
=v\llap/U$ then also $UY\llap/$ is a solution. But we want solutions which are
equal to 1 on the background, so we will disregard the solutions which are
equal to $Y\llap/$ on the background. From (\ref{doi}) we get that
\be
N=\frac{2N_{AB}v_AY_B}{1-v\cdot Y}
\ee
whereas by multiplying (\ref{trei}) with $Y_B+v_B$ we get
\be
N_A=\frac{\epsilon_{ABCDE}v_BY_CN_{DE}}{1+v\cdot Y}+N'\frac{Y_A+v_A}{2(1+
v\cdot Y)}
\ee
where at this point $N'$ is arbitrary.
This satisfies (\ref{unu}) automatically. If we take $N\neq 0$, then from
(\ref{alfa}) we obtain after contraction with $Y_B$
\bea
N'&=&-\frac{2}{N_{AB}v_AY_B}\epsilon_{ABCDE}v_BY_CN_{DE}N_{AF}Y_F
\label{notice}\\
N_A&=&\frac{\epsilon_{FBCDE}v_BY_CN_{DE}}{1+v\cdot Y}(\delta_{AF}
-\frac{(Y_A+v_A) N_{FG}Y_G}{N_{A'B'}v_{A'}Y_{B'}})
\eea
(Notice that if $N=0$ one divides in (\ref{notice}) by zero, so the
case $N=0$ should be treated separately)).
The solutions which are equal to 1 on the background (i.e. have
$N\neq 0$) are parameterized
by an antisymmetric matrix $N_{AB}$. The SO(5) conditions
become now constraints on $N_{AB}$
\bea
&&\epsilon_{ABCDE}v_BY_CN_{DE}N_{AG}\left(\delta_{FG}-\frac{Y_G(Y_{A'}+v_{A'})
N_{A'F}}{N_{A''B''}v_{A''}Y_{B''}}\right)=0\label{so51}\\
&&\epsilon_{ABCDE}N_{DE}\left(\frac{1-(v\cdot Y)^2}{2(N_{A''B''}v_{A''}Y_{B''}
)}N_{BC}-2v_BY_C\right)\nonumber\\
&&=-2(Y_A+v_A)\frac{\epsilon_{FBCDE}v_BY_CN_{DE}N_{FG}Y_G}{N_{A''B''}v_{A''}
Y_{B''}}\label{so52}\\
&&N^2+N_A^2+2N_{AB}^2=1\label{so53}
\eea
where the last condition fixes the normalization. The $UY\llap/ =v\llap/ U$
conditions yield further constraints on $N_{AB}$.
\bea
&&\frac{N_{BC}v_BY_C}{1-v\cdot Y}(Y_A-v_A)+ N_{AB}(Y_B+v_B)=0\label{doi'}\\
&&(Y+v)_{[F}\frac{\epsilon_{A]BCDE}v_BY_CN_{DE}}{1+v\cdot Y}
=\frac{1}{2}\epsilon_{AFCDE}(Y_C-v_C)N_{DE}\label{trei'}
\eea

The most general solution of these remaining five equations can
be obtained by noting
that in 5d space we have, besides $Y_A$ and $v_A$, another 3 independent
vectors $Z_1, Z_2, Z_3$. We can then make out of them
an orthonormal set, also orthogonal
to Y and v. Then the most general solution for $N_{AB}$ will be written as
\be
N_{AB}=\sum_{m\neq n=1}^5 a_{mn}X_m^{[A}X_n^{B]}
\ee
where $\{X_i^A\}=\{Y^A,v^A, Z_1^A, Z_2^A, Z_3^A\}$.  We define
\bea
S_A&=&\frac{Y_A+v_A}{\sqrt{2(1+v\cdot Y)}}\\
D_A&=&\frac{Y_A-v_A}{\sqrt{2(1-v\cdot Y)}}
\eea
satisfying $S^2=D^2=1, S\cdot D=0$. Then (\ref{doi'}) and (\ref{trei'})
yield (after multiplying (\ref{trei'}) with $\epsilon_{AFC'D'E'}$)
\bea
&&(S\cdot N\cdot D)D_A+N_{AB} S_B=0\label{snd}\\
&&D_{[C'}N_{D'E']}+4S_DN_{D[E'}S_{C'}D_{D']}=0\label{snsd}
\eea
Substituting (\ref{snd}) into (\ref{snsd}) implies $D_{[C'}N_{D'E']}=0$,
of which  the most general solution is
\be
N_{AB}=D_{[A}X_{B]}\label{gensol}
\ee
with $X_A$ an arbitrary vector. Then (\ref{so51}) and (\ref{so52}) are
satisfied trivially, because $\epsilon_{ABCDE}v^AY^BD^CX^D=0$, whereas
(\ref{so53}) fixes the normalization. We notice that $D_A, S_A, Z_{1A},
Z_{2A}, Z_{3A}$ make an orthonormal set, and any component of $X_A$ along
$D_A$ will drop out of $N_{AB}$, so that we have
\be
X_A=aS_A+b_iZ^i_A
\ee
The normalization condition (\ref{so53}) gives
\be
\sum b_i^2+a^2 v\cdot Y =1
\ee
Thus the most general solution contains 3 parameters.
We will now look
at particular cases. We note that the covariant vectors at our disposal are
of the form
\be
X_i^{(z)}=\frac{{[(\Pi^{-1})^z]_i}^AY_A}{Y\cdot T^z\cdot Y}
\ee
where $z\in \Z$ and if ${\Pi_A}^i\neq \delta_A^i$, then the $X^{(z)}$'s
will generate the whole sphere (they will not lie in a lower dimensional
hyperplane), so we can build $Z_1, Z_2, Z_3$ out of 3 suitably chosen
$X_i^{(z)}$'s ($z\neq 0,1$).

The simplest possibility is to
build U only out of $Y^A$ and $v^A$. Then $N_{AB}=av_{[A}Y_{B]}$, so
that $N=a(1+v\cdot Y)$ and $N_A=0$. The normalization condition (\ref{beta})
fixes then $a=\pm(2(1+v\cdot Y)^{-1/2}$. The unique covariant solution
built out of only $Y_A$ and $v_A$ then reads
\be
U_{(1)}=\frac{1+v\cdot Y+v_AY_B\gamma_{AB}}{\sqrt{2(1+v\cdot Y)}}=\frac{1+
v\llap/Y\llap/}{\sqrt{2(1+v\cdot Y)}}
\ee
It clearly satisfies $UY\llap/=v\llap/U$ and is and $USp(4)$ matrix.

We note that the general solution (\ref{gensol}) can be rewritten as
\be
U_{(X)}=\pm \frac{X\cdot (Y+v)+X^A(Y^B-v^B)\gamma_{AB}}{\sqrt{2}\sqrt{1-v\cdot
Y+2 X\cdot Y X\cdot v}}=\frac{X\llap/ Y\llap/+ v\llap/ X\llap/}{\sqrt{2}
\sqrt{1-v\cdot Y+2 X\cdot Y X\cdot v}}
\ee
where X now  is still arbitrary, but has unit norm.
We can easily check that if $\vec{X}$ is in the (Y,v) plane we reproduce
$U_{(1)}$. However, now we can take X to be a noncovariant vector, which we can
take to be $X=(0, 0, 0, 0, 1)$. Then
\be
U_{noncov}=\frac{\gamma_5 Y\llap/ +v\llap/ \gamma_5}{\sqrt{2}\sqrt{1-v\cdot Y
+2 Y_5v_5}}=\frac{Y_5+v_5-i\gamma_m(Y_m-v_m)}{\sqrt{(Y_5+v_5)^2+(Y_m-v_m)^2}}
,\;\;m=1,4
\ee
which can be written as an exponent
\bea
U&=&exp[-i\gamma_m(Y^m-v^m)u]; \nonumber\\
u&=&[(Y^m-v^m)^2]^{-1/2}\arctan-\frac{\sqrt{
(Y^m-v^m)^2}}{Y^5+v^5}
\eea

We can also generate the most general noncovariant solution made out of only
$Y^A$ and $v^A$,
by choosing
\be
X=\frac{Y+\beta v+\alpha (\vec{0},1)}{\sqrt{1+\beta^2+\alpha^2+2\alpha\beta
v\cdot Y+ 2\alpha Y_5+2\alpha\beta v_5}}
\ee
to obtain
\be
U_{(X)}^{noncov}=\frac{1+ v\llap/ Y\llap/ +\tilde{\alpha}(\gamma_5 Y\llap/
+v\llap/ \gamma_5)}{\sqrt{2}\sqrt{1+v\cdot Y +2\tilde{\alpha}(Y_5+v_5)
+\tilde{\alpha}^2(1-v\cdot Y +2Y_5 v_5)}}
\ee
where $\tilde{\alpha}=\alpha /(\beta +1)$
is an arbitrary parameter.  We note that taking $\tilde{\alpha}$ to infinity,
we get the maximally noncovariant solution $U_{noncov}$.

Maybe it's interesting to analyze what happens at the linearized level.
Since ${(\Pi^{-1})_i}^A={(e^{-\delta\pi})_i}^A$, the vectors
$X_i^{(z)}= {[(\Pi^{-1})^z]_i}^AY_A/Y\cdot T^z \cdot Y$ become $X_i^{(z)}
\simeq Y^A -z\delta v_A$, where $\delta v_A=Y_A Y\cdot \delta\pi\cdot Y-
\delta \pi_{AB}Y_B$. And so the only vectors at our disposal are $Y^A$ and
$\delta v^A$, so the most general solution for U which is equal to 1 on the
background is obtained from $U_{(X)}^{noncov}$
\bea
U&\simeq& 1+\frac{1}{2(1+\tilde{\alpha}Y_5)} (\delta v_AY_B\gamma_{AB}+\alpha '
\delta v_A \gamma_{A5})\nonumber\\
&=&1+\frac{1}{2}\delta v_A Y_B\gamma_{AB} +\frac{\alpha_1}{2}Y^mC_m^A\delta v_B
\gamma _{AB}
\eea
with $\alpha_1$ parameterizing the deviation from covariance. Actually the
most
general linearized solution is found by adding also a term $\beta Y^m C_m^A
\delta v^B\gamma_{AB} Y\llap/$; because of the linearity of the equations
we can add the arbitrary noncovariant piece coming from the $U Y\llap/$
solution.

We end this section with a discussion of the metric. It is independent of $U$
and has simple geometrical properties as we shall see.
>From (\ref{viel}) and (\ref{urel}) one can in principle obtain an expression
 for the vielbein.
 Squaring this result would then lead
to an expression for the metric $g_{\mu\nu}$.
It is easier to obtain the result for $g_{\mu\nu}$ by directly verifying
that it is the inverse of $g^{\mu\nu}$ in (\ref{metric}).
\bea
g_{\mu\nu}&=&\Delta^{4/5} {C_\mu}^A {C_\nu}^B T^{-1}_{AB}\label{gmunu}\\
g^{\mu\nu}&=&\Delta^{2/5} \left( C^\mu_A C^\nu_B T^{AB} Y_C Y_D T^{CD}
-C^\mu_A Y_B T^{AB} C^\nu_C Y_D T^{CD}\right)
\eea
Since we are in the gauge where $E_\mu^a=0$, one easily checks that
$G_{\mu\nu}g^{\nu\rho}=g_{\mu\nu}g^{\nu\rho}=\delta_{\mu}^{\rho}$.
Recalling that $C_{\mu}^A=D_{\mu}Y^A=\partial_{\mu}Y^A$, we see that the
metric $g_{\mu\nu}=G_{\mu\nu}$ describes
an ellipsoid with a conformal factor $\Delta^{4/5}$,
whose axes at a specific point $y$ in the d=7 space time are
determined by the eigenvalues of $T^{-1}_{AB}$.

To summarize, the metric of the internal space reads
\bea
ds_4^2&\equiv& G_{\mu\nu}dx^{\mu}dx^{\nu}=\Delta^{4/5}T_{AB}^{-1}
\partial_{\mu}Y^A\partial_{\nu}Y^B dx^{\mu}dx^{\nu}\nonumber\\
&=&\Delta^{4/5}T^{-1}_{AB}dY^AdY^B\nonumber\\
G_{\alpha\mu}dx^{\alpha}dx^{\mu}&=&2\Delta^{4/5}T^{-1}_{AB}Y^CB^{BC}dY^A
\eea
while the metric of the 7 dimensional space-time is
\bea
ds_7^2&\equiv& G_{\alpha\beta}dy^{\alpha}dy^{\beta}\nonumber\\
&=&\Delta^{-2/5}g_{\alpha\beta}dy^{\alpha}dy^{\beta}+4\Delta^{4/5}
Y_AT^{-1}_{BD}B^{BA}Y_AB^{DC}Y_C
\eea
where $G_{\Lambda\Pi}$ denotes the 11 d metric. Then we can write the
11 dimensional metric in a concise form
\bea
ds_{11}^2&\equiv& G_{\Lambda \Pi}dx^{\Lambda}dx^{\Pi}
=\Delta^{-2/5}g_{\alpha\beta}dy^{\alpha}dy^{\beta}\nonumber\\
&+&\Delta^{4/5}T^{-1}_{AB}(dY^A+2B^{AC}Y_C)(dY^B+2B^{BD}Y_D)
\label{nicemetric}
\eea
Note that, since after compactification the gauge coupling is $g=2m$, and
we set $m=1$, the gauge covariant
derivative is $dY+2B\cdot Y\equiv DY$.
 Thus the metric is manifestly gauge invariant
\be
ds_{11}^2=\Delta^{-2/5}g_{\alpha\beta}dy^\alpha dy^\beta+
\Delta^{4/5}(DY)^A T_{AB}^{-1}(DY)^B
\ee

For later use for the gravitino transformation law,
we rewrite the vielbein and its inverse in a simpler form,
involving only the conformal Killing vectors
\be
E_\mu^m=\frac{1}{4}\Delta^{2/5}\Pi_A\;^i C_\mu^A C^{mB}
Tr(U^{-1} \gamma^i U \gamma_B)\label{vielsim}
\ee
\be
E^{\mu m}=\frac{1}{4}{(\Pi^{-1})_i}^AC^{\mu}_AC^{m}_B\Delta^{-2/5}
Tr(U^{-1} \gamma^i U \gamma^B)
\label{vielinvsim}
\ee

For the derivation of this result one may use (\ref{EC}).
\subsection{The ansatz for the 11 dimensional antisymmetric tensor and
 auxiliary field and self-duality in odd dimensions}

At this point we have obtained the complete nonlinear ansatz for the
metric and the gravitino and we checked the transformation rules of the
d=7 bosonic fields which are embedded in the d=11 vielbein,
viz. the graviton, the gauge fields and the scalars. We
now turn to the antisymmetric tensor $A_{\Lambda\Pi\Sigma}$.

We begin by presenting the final form for our ansatz for the KK
reduction on $S_4$ of the field strength $F_{\Lambda\Omega\Pi\Sigma}$;
a discussion of how we arrived at this result will be given afterwards.        \bea
\frac{\sqrt{2}}{3m}
F_{\mu\nu\rho\sigma}
&=&\epsilon_{\mu\nu\rho\sigma}\sqrt{\det \sirc{g}}\left[1+
\frac{1}{3}\left(\frac{T}{Y_A Y_B T^{AB}}-5\right)\right.\nonumber\\&&\left.
-\frac{2}{3} \left(\frac{Y_A (T^2)^{AB} Y_B}{(Y_A T^{AB} Y_B)^2}-1\right)
\right]
\label{mnpr}
\\
\frac{\sqrt{2}}{3m}
F_{\mu\nu\rho\alpha}&=&
\partial_{[\mu} \left(\epsilon_{ABCDE}B_\alpha^{AB}
C_\nu^C C_{\rho ]}^D \frac{T^{EF} Y_F}{Y\cdot T\cdot Y}\right)\nonumber\\
&&+\sqrt{\sirc{g}}\epsilon_{\mu\nu\rho\sigma}C^\sigma_A
\frac{1}{3}\left(\frac{\partial_\alpha T^{AB} Y_B}{Y_A T^{AB} Y_B}-
\frac{T^{AB} Y_B}{(Y_A T^{AB} Y_B)^2} (Y_C \partial_\alpha T^{CD} Y_D)
\right)
\label{mnra}\\
\frac{\sqrt{2}}{3m}
F_{\mu\nu\alpha\beta}&=&\frac{2}{3}\left[\partial_{[\alpha}
\left(\epsilon_{ABCDE}
B_{\beta ]}^{AB} C_\mu^C C_\nu^D \frac{T^{EF} Y_F}{Y\cdot T\cdot Y}\right)
\right.\nonumber\\
&+&2\left.\partial_{[\mu}\left(\epsilon_{ABCDE} B^{AF}_{[\alpha}
Y_F B^{BC}_{\beta ]}
C_{\nu ]}^D\frac{T^{EG} Y_G}{Y\cdot T\cdot Y}\right)\right]
\label{mnab}\\
\frac{\sqrt{2}}{3m}
F_{\mu\alpha\beta\gamma}&=&\partial_\mu {\cal A}_{\alpha\beta\gamma}\nonumber\\
&+&\frac{4}{3}\partial_{\mu}\left(\epsilon_{ABCDE} B_{[\alpha}^{AB}
B_\beta^{CF}Y_F
B_{\gamma ]}^{DG}Y_G\frac{T^{EH} Y_H}{Y\cdot T\cdot Y}\right)\nonumber\\
&-&2\partial_{[\alpha}\left(\epsilon_{ABCDE} B_{\beta}^{AB} B_{\gamma ]}^{CF}
Y_F C_\mu^D \frac{T^{EG} Y_G}{Y\cdot T\cdot Y}\right)\nonumber\\
&+&\partial_\mu\left(\epsilon_{ABCDE} (\partial_{[\alpha}B_\beta^{AB}
+\frac{4}{3} B_{[\alpha}^{AF} B_{\beta}^{FB})
B_{\gamma ]}^{CD} Y_E
\right)
\label{mabc}\\
\frac{\sqrt{2}}{3m}
F_{\alpha\beta\gamma\delta}&=&4\partial_{[\alpha} {\cal A}_{\beta\gamma
\delta ]}\nonumber\\
&+&4\partial_{[\alpha}\epsilon_{ABCDE}\left(\frac{4}{3}B_\beta^{AB}
B_{\gamma}^{CF} Y_F
B_{\delta ]}^{DG} Y_G \frac{T^{EH} Y_H}{Y\cdot T\cdot Y}\right.
\nonumber\\
&+&\left.(\partial_\beta B_\gamma^{AB} +\frac{4}{3} B_\beta^{AF}
B_\gamma^{FB})
B_{\delta ]}^{CD} Y^E\right)
\label{abcd}
\eea

The independent fluctuation, ${\cal A}_{\alpha\beta\gamma}$, mixes with the
auxiliary field
\be
{\cal B}_{\alpha\beta\gamma\delta}=
k\epsilon_{\alpha\beta\gamma\delta}\;^{\epsilon\eta\zeta}
\tilde{\cal B}_{\epsilon\eta\zeta}
\ee
where $k$ is an arbitrary scale factor which will be fixed below (see below
(\ref{detk}). All the other components of ${\cal B}_{\Lambda\Pi\Sigma\Omega}$
are set to zero because in d=11 they vanish on-shell, so that, in order that
they also vanish on d=7 on-shell, they would have to be proportional to
d=7 field equations. Field equations of the form $\partial ^2\phi$ are
ruled out because they would lead to $(\partial^2\phi)^2$ terms in the action.
 In principle,
the other components of ${\cal B}_{\Lambda\Pi\Sigma}$ could
still depend on the field
equation for $S_{\alpha\beta\gamma}$ because that one is linear in
derivatives. However, this produces incorrect answers for the susy
transformation rules of the fermions, as we shall show.

Both ${\cal A}_{\alpha\beta\gamma}$ and $\tilde{\cal {B}}_
{\alpha\beta\gamma}$ will
be written as ${\cal A}_{\alpha\beta\gamma}\sim S_{\alpha\beta\gamma, A}(y)
Y^A(x)$ and $B_{\alpha\beta\gamma}\sim T^{AB}S_{\alpha\beta\gamma ,A}Y^B$.
Since they contain the same spherical harmonic $Y^A$ they will mix. This
mixing of ${\cal A}_{\alpha\beta\gamma}$ with $\tilde{\cal {B}}_
{\alpha\beta\gamma}$
is needed to convert the second order field equation obeyed by
${\cal A}_{\alpha\beta\gamma}$ into the first order one obeyed by the
antisymmetric tensor field of 7 dimensional gauged sugra,
$S_{\alpha\beta\gamma, A}$.

The ansatz for $F_{\Lambda\Omega\Sigma\Pi}$ is obtained by requiring
consistency of susy laws.
It is the same as the geometrical ansatz of \cite{hmm} if the scalars
are set to zero. One would expect that this should be the case, since the
ansatz proposed
by the authors of \cite{hmm} was constructed such that the Chern-Simons
terms in the 7-dimensional action are obtained after integrating the
Chern-Simons term of
the 11-dimensional action on $S_4$.
The two terms in $F_{\mu\nu\rho\sigma}$ and $F_{\mu\nu\rho\alpha}$ which only
depend on scalar fields  are components
of a separately closed form. We note that any closed form $f_{(4)}$
 which appears only in the $F_{\mu\nu\rho
\sigma},F_{\mu\nu\rho\alpha}$ sector, must be of the form
$f_{\mu\nu\rho\sigma}=\epsilon_{\mu\nu\rho\sigma} f,\;\;
f_{\mu\nu\rho\alpha}=\epsilon_{\mu\nu\rho\tau}D^{\tau}D_{\alpha}\frac{1}{
\Box}f$ and will not contribute in the 11 dimensional
Chern-Simons action $\epsilon FFA$. This
is so because, if we work for simplicity
in 12 dimensions (where the Chern-Simons form
becomes $\epsilon FFF$), then $f_{(4)}$ contributes only to a
term of the type
\be
\epsilon^{\mu\nu\rho\sigma}\epsilon^{\alpha_1 ...\alpha_8}(f_{\mu\nu\rho
\sigma}F_{\alpha_1...\alpha_4}F_{\alpha_5 ...\alpha_8}-32f_{\mu\nu\rho\alpha_1
}F_{\sigma\alpha_2\alpha_3\alpha_4}F_{\alpha_5...\alpha_8})
\ee
which vanishes if we partially integrate the $D_{\alpha_1}$ in
$f_{\mu\nu\rho\alpha_1}$, use the Bianchi identity, and then partially
integrate back the resulting $D_{\sigma}$.

The precise expression of the ansatz in the
$F_{\mu\nu\rho\sigma}$ sector is highly constrained. It must
reproduce the linearized term in (\ref{epsil}), and it must yield the correct
scalar potential in d=7 after integrating over the compact space.
In order to perform this integral to which both the Einstein action and the
kinetic action of the 3-index photon contribute we start with
the d=4 scalar curvature associated with the conformal metric
$\tilde g_{\mu\nu} = \Delta^{-4/5} g_{\mu\nu}$, namely
\footnote{We used a symbolic manipulation program to obtain (\ref{curvature}).}
\bea
\tilde R^{(4)} &=& \tilde g^{\mu\nu} R_{\lambda\mu\nu\;^\lambda}\nonumber\\
&=&\frac{-2 Y_A Y_B (T^3)^{AB} +2T Y_AY_B (T^2)^{AB} + Y_AY_BT^{AB}
(Tr(T^2)- T^2)}{(Y_AY_BT^{AB})^2}\label{curvature}
\eea
and its relation with the d=4 scalar curvature
\be
R^{(4)}= \Delta^{-4/5} \tilde R^{(4)} -6 g^{\mu\nu} D_\mu D_\nu
\ln(\Delta^{-2/5}) -6 g^{\mu\nu} D_\mu \ln(\Delta^{-2/5}) D_\nu ln(\Delta^{-2/5})
\ee
Using (\ref{del}) to convert $\Delta$ into $T^{AB}$, and adding the other
contributions from the Einstein action,
 the integral over the compact space of the Einstein action, when
setting the gauge fields to zero, and disregarding terms with d=7 space
time derivatives yields
\be
\frac{-1}{2} \int d^4x \sqrt{\det\sirc{g}(x)} \left(\Delta^{-6/5}
\tilde R^{(4)} +2 \frac{Y_AY_B (T^3)^{AB} Y_C Y_D T^{CD} - \left(
Y_AY_B (T^2)^{AB}\right)^2}{(Y_AY_B T^{AB})^2}\right)
\ee
We further use that
\bea
\int d^4x \sqrt{\det\sirc{g}(x)} \frac{Y_A Y_B}{Y_C Y_D T^{CD}} &=&
\frac{1}{5} T^{-1}_{AB}\; Vol(S_4)\label{YY/YTY}\\
\int d^4x \sqrt{\det\sirc{g}(x)} \frac{Y_AY_BY_CY_D}{(Y_E Y_F T^{EF})^2}
&=&
\frac{Vol(S_4)}{5\cdot 7}(T^{-1}_{AB} T^{-1}_{CD}\nonumber\\&&
+T^{-1}_{AC} T^{-1}_{BD}+
T^{-1}_{AD} T^{-1}_{BC})\label{4Y/YTY^2}
\eea
(This becomes clear after diagonalizing $T^{CD}$.) At this point,
 the integrated Einstein action contribution
is of the desired form, namely a linear combination of $T^2$ and $Tr(T^2)$:
$(-31/70) Tr(T^2)+(23/70) T^2$.

On the other hand, the integrated kinetic action of the 3 index photon has
the form $F_{\mu\nu\rho\sigma}^2\sim
 (Y_E Y_F T^{EF})^2(1+{\cal S})^2$,
where $\frac{3}{\sqrt{2}}\sqrt{det\sirc{g}(x)}
\epsilon_{\mu\nu\rho\sigma}(1+{\cal S}) =F_{\mu\nu\rho\sigma}$.
(Use that $(\det g^{\mu\nu})(\det \sirc{g}_{\mu\nu})=\Delta^{-2}$
and substitute (\ref{del}).)
The function ${\cal S}$ must be homogeneous of degree zero in
$T$, because in d=7 the potential is proportional to $T^2-2T_{ij}T^{ij}=
T^2-2T_{AB}T^{AB}$ and the leading term $(Y_EY_FT^{EF})^2$ has already
two T's. Furthermore, S depends only on the scalar fluctuations in
${(\Pi^{-1})_i}^A$. In fact, the most general expression ${\cal S}$ can
have is
\be
{\cal S}=a (\frac{Y T^2 Y}{(Y T Y)^2}-1) +b(\frac{T}{Y T Y}-5)\label{calS}
\ee
because higher powers of $T$ appearing in the ratios would yield terms
proportional to $Tr(T^{-1})$ in the 7-dimensional action. To be more explicit,
let's consider a term of the form ${Y T^3 Y}/(Y T Y)^3$. Then, when
integrating ${\cal S}^2$, such a term also generates a integral of
the form $\int d^4x (Y T^3 Y)^2/(Y T Y)^4$ which clearly will produce unwanted
$Tr(T^{-1})$'s (apply (\ref{4Y/YTY^2})).
One requires then that the linearized limit of (\ref{mnpr}) yields the
results of \cite{pnt}, and that when adding the Einstein action contribution
we recover upon integration the d=7 scalar potential.
The last term in (\ref{mnpr}) has no linearized contributions, while the
coefficient of the second term gets fixed to 1/3 by requiring agreement with
\cite{pnt}. Hence, $b=1/3$ while $a$ is still a free parameter. However, $a$
gets also fixed after evaluating the integral over $S_4$ of the square of
$(1+{\cal S})$ using (\ref{YY/YTY}), (\ref{4Y/YTY^2}) and (\ref{calS}) and
requiring that after adding the Einstein action contribution one obtains
$(T^2-2Tr(T^2))/4$. Thus, $a$ has to satisfy a quadratic equation, namely
$a(a+2/3)=0$.
It will turn out that the consistency of the gravitino transformation law
excludes the first solution ($a=0$), and requires the second ($a=-2/3$).

The result for $F_{\mu\nu\rho\alpha}$ is rather surprising. Initially, we made
the ansatz
\be
F_{\mu\nu\rho\alpha}=\sqrt{\sirc{g}}\epsilon_{\mu
nu\rho\sigma}\sirc{D}
^{\sigma}\frac{1}{\sirc{\Box}}\partial_{\alpha}(1+{\cal S})+gauge\;\; field
\;\;dependent\;\;terms
\ee
where $1+{\cal S}$ contains the ansatz for $F_{\mu\nu\rho\sigma}$ in
{\ref{mnpr}). This ansatz satisfies the Bianchi identity
\be
4\sirc D_{[\sigma} F_{\mu\nu\rho ]\alpha} -
\partial_\alpha F_{\mu\nu\rho\sigma}=0
\ee
However, our original expression for ${\cal S}$ contained no term with
$(Y\cdot T\cdot Y)^{-2}$ (it corresponded to the solution $a=0$), and for
this choice of ${\cal S}$, the ansatz for $F_{\mu\nu\rho\alpha}$ was found to
be truly nonlocal already in a perturbative expansion. With our present ansatz
($a=-2/3$), the apparently nonlocal expression is, in fact, local and given by
(\ref{mnra}). It seems very unlikely that the first solution, containing the
$d=11$ nonlocal term
$\sirc{D}^\sigma \frac{1}{\sirc\Box} D_\alpha (1+{\cal S})$ would lead to the
correct $d=7$ action, since its presence requires that we expand $\frac{1}
{\sirc\Box} D_\alpha (1+{\cal S})$ into an infinite series of spherical
harmonics, and though the factor $  \frac{1}{\sirc\Box}$ by itself may not be
fatal, being integrated over $S_4$, the fact that infinitely many
spherical harmonics would enter could lead to inconsistencies in the KK
reduction.

The gauge field dependence of the $F_{\Lambda\Sigma\Pi\Omega}$ is dictated
by the 11-dimensional susy laws.
We begin with the $F_{\mu\nu\rho\alpha}$ sector, and for simplicity,
after substituting the 11-dimensional susy variation in terms of our
fermionic ans\"atze, we keep only the gravitino terms.
Then, we get
\bea
&&\delta F_{\mu\nu\rho\alpha} |_{\psi} = -\frac{\sqrt{2}}{8} 3!
\left( \partial_{[\mu} \bar\varepsilon \Gamma_{\nu\rho ]} \Psi_\alpha
-\partial_{\alpha} \bar\varepsilon \Gamma_{[\nu\rho}\Psi_{\mu ]}
 -2\partial_{[\mu }\Gamma_{\alpha\rho}\Psi_{\nu ]}\right)|_{\psi}\\
&=&
-\frac{3\sqrt 2}{4} \partial_{[\mu}\left(\Delta^{-1/5}\; \bar\epsilon_{I'}
\psi_{\alpha , J'} \;\bar\eta^I \gamma_{np}\gamma_5 \eta^J \;U^{I'}_{\;I}
U^{J'}_{\;J}\; E_\nu^n E_{\rho ]}^p\right)\label{step1}\\
&=&\frac{3\sqrt 2}{4} \partial_{[\mu}\left(\bar\epsilon \gamma^{ijk}\psi_\alpha
C_\nu^A C_{\rho ]}^B \Pi_A^{\;i} \Pi_B^{\;j}(\Pi^{-1})_k^{\;E} \frac{Y_E}
{Y\cdot T\cdot Y}\right)
\label{step2}\\
&&=\frac{3\sqrt 2}{2}\partial_{[\mu}\left(\delta B_\alpha^{AB} C_\nu^C
C_{\rho ]}^D \frac{T^{EF}Y_F}{Y\cdot T\cdot Y}\right)
\epsilon_{ABCDE}\nonumber\\
&&=\delta\left[\frac{3\sqrt 2}{2}\partial_{[\mu}\left(B_\alpha^{AB} C_\nu^C
C_{\rho ]}^D \frac{T^{EF}Y_F}{Y\cdot T\cdot Y}\right)
\epsilon_{ABCDE}\right]\label{step3}
\eea
where to go from (\ref{step1}) to (\ref{step2}) we used (\ref{compltn}) and
the identity
\be
E_\mu^m \Delta^{-2/5} U^{I'}\;_I U^{J'}\;_J C_m^{IJ}
= -i{\Pi_A}^i (\gamma_i)^{I'J'} C_\mu^A\label{EC}
\ee
(Substitute into (\ref{viel2}) that $V^m\sim \gamma^{CD}Y_C\partial^m Y_D$,
replace $2\gamma^{CD}$ by $\gamma^C\gamma^D-\gamma^D\gamma^C$ and use
(\ref{urel}) to eliminate $\gamma^C$. One obtains then the commutator $[\gamma
_{ij},\gamma_k]\sim\delta_{jk}\gamma_i-\delta_{ik}\gamma_j$. The $\Pi^{-1}$
from (\ref{urel}) cancels then one of the $\Pi$'s in $E_{\mu}^m$). The total
susy variation $\delta$ could be pulled out in (\ref{step3}) because we
are keeping only gravitino dependent terms, and scalars vary into spin 1/2
fields.

Hence (\ref{step3}) allows us to read off the gauge field dependence of
$F_{\mu\nu\rho\alpha}$ (see eq. (\ref{mnra})). Another nontrivial check of
this ansatz is thatthe susy variation of the scalars in the l.h.s. of
(\ref{mnra}) reproduces the $B_\alpha^{AB}$-dependent terms which we
get from the 11-dimensional susy variation of $F_{\mu\nu\rho\alpha}$.

The ansatz in the next sector, namely, $F_{\mu\nu\alpha\beta}$ will be
fixed again by
requiring the consistency of susy laws. The complete answer is given in
(\ref{mnab}). The procedure to determine the ansatz remains the same:
vary $F_{\mu\nu\alpha\beta}$ under 11-dimensional susy laws, look for
simplicity only at gravitino dependent terms, and rewrite (after substituting
the various fermionic and vielbein ans\"atze) the 11-dimensional susy
variation in terms of a total 7-dimensional susy variation of 7-dimensional
bosonic (gauge and scalar) fields. So, varying the r.h.s. one gets
\bea
&&\delta F_{\mu\nu\alpha\beta}|_\psi= -\frac{\sqrt 2}{8}(2!)^2 \left(
2\partial_{[\mu} \bar\varepsilon \Gamma_{\nu ] [\alpha}\Psi_{\beta ]}-
\partial_{[\mu}\bar\varepsilon \Gamma_{\beta\alpha}\Psi_{\nu ]}\right.
\nonumber\\
&&\left.-\partial_{[\alpha} \bar\varepsilon \Gamma_{\nu\mu}\Psi_\beta
+2\partial_{[\beta} \bar\varepsilon \Gamma_{\alpha ][\nu}\Psi_{\mu ]}
\right)|_\psi\nonumber\\
&&=-\sqrt{2}\partial_{\mu}\left(-i\bar\epsilon_{I'}\tau_\alpha\psi_{\beta\; J'}
\;\Delta^{-2/5}\;\bar\eta^I\gamma_n\gamma_5\eta^J \; U^{I'}_{\;I} U^{J'}_{\;J}
E_\nu^n\right.\nonumber\\
&&\left.-2\bar\epsilon_{I'}\psi_{\beta\;J'} \; \Delta^{-1/5} \;
\bar\eta^I \gamma_{np}\gamma_5\eta^J \;U^{I'}_{\;I} U^{J'}_{\;J}\;
E_\nu^n E_\rho^p \;B_\alpha^{AB} V_{AB}^\rho\right)\nonumber\\
&&+\frac{\sqrt 2}{2} \partial_{[\alpha}\left(\bar\epsilon_{I'}\;
\psi_{\beta ]\;J'}\;U^{I'}_{\;I} U^{J'}_{\;J}
\;\Delta^{-1/5}\;\bar\eta^{I}\gamma_{nm}\gamma_5\eta^J \;E_\nu^n E_\mu^m\right)
\label{ss1}\\
&&=\sqrt{2} \partial_{[\mu}\left(C_{\nu ]}^A \Pi_A^{\; i}
\bar\epsilon\gamma^i\tau_{[\alpha}\psi_{\beta ]}-4\epsilon_{ABCDE}
\delta B_{[\beta}^{AB}B_{\alpha ]}^{CF}Y_F C_{\nu ]}^D \frac{T^{EG} Y_G}
{Y\cdot T\cdot Y}\right)\nonumber\\
&&-\sqrt{2}\partial_{[\alpha}\left(\epsilon_{ABCDE}\delta B_{\beta}^{AB}
C_\nu^C C_\mu^D \frac{T^{EF} Y_F}{Y\cdot T\cdot Y}\right)\label{ss2}
\eea
To go from (\ref{ss1}) to (\ref{ss2}) we again made use of (\ref{compltn})
and (\ref{EC}). Furthermore, in (\ref{ss2}) the first term vanishes
trivially, and the last one can be written as a total 7-dimensional susy
variation. With the use of the Schouten identity
\be
(\delta (B_{[\beta}^{AB}) B_{\alpha ]}^{CF} C_\nu^D \frac{T^{EG} Y_{G}}
{Y\cdot T\cdot Y}Y_{[F}\epsilon_{ABCDE ]}=0
\ee
which yields
\bea
\delta(B_{[\beta}^{AB}) B_{\alpha ]}^{CF} C_\nu^D \frac{T^{EG} Y_{G}}
{Y\cdot T\cdot Y}Y_F\epsilon_{ABCDE}&=&\left((\delta(B_{[\alpha})\cdot Y)^A
B_{\beta ]}^{BC}  C_\nu^D\frac{T^{EG} Y_{G}}
{Y\cdot T\cdot Y}\right.\nonumber\\
&+&\left.\frac{1}{2}\delta(B_{[\alpha}^{AB}) B_{\beta ]}^{CD}
C_\nu^E\right)
\epsilon_{ABCDE}
\eea
we get that
\bea
&&\delta (B_{[\beta}^{AB}) B_{\alpha ]}^{CF} C_\nu^D \frac{T^{EG} Y_{G}}
{Y\cdot T\cdot Y}Y_F\epsilon_{ABCDE}=
\frac{1}{2}\left(\delta(B_{[\beta}^{AB}) (B_{\alpha ]}\cdot Y)^C C_\nu^D
\frac{T^{EG} Y_{G}}
{Y\cdot T\cdot Y}\right.\nonumber\\
&&+\left.(\delta(B_{[\alpha})\cdot Y)^A B_{\beta ]}^{BC} C_\nu^D
\frac{T^{EG} Y_{G}}{Y\cdot T\cdot Y}+\frac{1}{2} \delta (B_{[\alpha}^{AB})
B_{\beta ]}^{CD} C_\nu^E\right)\epsilon_{ABCDE}\label{sch}
\eea
After acting with a $\partial_{[\mu}$ on (\ref{sch}), the remaining terms
can be cast into a total susy variation. Then we substitute back into
(\ref{ss2}) and we recognize that the 11-dimensional susy variation of
$F_{\alpha\beta\mu\nu}$ is the same as the 7-dimensional susy variation of
the ansatz given in (\ref{mnab}).

To derive the ansatz of the next sector $F_{\mu\alpha\beta\gamma}$ is a bit
more laborious. We follow the procedure outlined previously, and vary under
susy the r.h.s. of (\ref{mabc})
\bea
&&\delta F_{\mu\alpha\beta\gamma}|_\psi=\frac{-\sqrt{2}}{8}3!
\left(\partial_\mu \bar\varepsilon\Gamma_{[\alpha\beta} \Psi_{\gamma ]}
-2\partial_{[\alpha}\bar\varepsilon\Gamma_{\mu\beta}\Psi_{\gamma ]}-
\partial_{[\alpha} \bar\varepsilon\Gamma_{\beta\gamma ]}\Psi_\mu\right)
|_\psi\nonumber\\
&&=\frac{-3\sqrt2}{4}\partial_{\mu}\left[\bar\epsilon
\Delta^{-1/10}\sqrt{\gamma_5}\left(\tau_{[\alpha\beta}\Delta^{-2/5}
+2\gamma_m\tau_\beta\gamma_5 E_{[\alpha}^m \Delta^{-1/5}+
\gamma_{mn}E_{[\alpha}^m E_\beta^n\right)\Delta^{-1/10}
\right.\nonumber\\
&&\left.\sqrt{\gamma_5}
\psi_{\gamma ]}\right]
+\frac{3\sqrt 2}{2}\partial_{[\alpha}\left[\bar\epsilon\Delta^{-1/10}
\sqrt{\gamma_5}\left(2\gamma_m\tau_\beta\gamma_5 E_\mu^m\Delta^{-1/5}
+\gamma_{mn}E_\mu^m E_\beta^n\right)\Delta^{-1/10}\sqrt{\gamma_5}
\psi_{\gamma]}\right]\nonumber\\\label{ss3}
\eea
Then we substitute our ans\"atze in the l.h.s. of (\ref{ss3}) and with
the same tricks as we used before we arrive at
\bea
&&\delta F_{\mu\alpha\beta\gamma}|_\psi=\frac{3\sqrt 2}{4}(\Pi^{-1})_i^{\;\;A}
C_\mu^A \bar\epsilon\tau_{[\alpha\beta}\gamma^i\psi_{\gamma ]}\nonumber\\
&&+\frac{3\sqrt{2}}{2} D_{[\alpha} \left(\bar\epsilon\tau_\beta\gamma^i
\psi_{\gamma ]}\Pi_A^{\;\;i}\right)C_\mu^A\nonumber\\
&&+6\sqrt{2}\partial_\mu\left[\epsilon_{ABCDE}\delta(B_{[\gamma}^{AB})
(B_\alpha\cdot Y)^C (B_{\beta ]}\cdot Y)^D \frac{T^{EF} Y_F}{Y\cdot T\cdot Y}
\right]\nonumber\\
&&+6\sqrt{2}\partial_{[\alpha}\left[\epsilon_{ABCDE}\delta(B_{[\gamma}^{AB})
(B_{\beta ]}\cdot Y)^C C_\mu^D \frac{T^{EF} Y_F}{Y\cdot T\cdot Y}
\right]\label{ss4}
\eea

To get the correct gauge-field dependence in the $F_{\mu\alpha\beta\gamma}$
sector we take into account the susy variation of the independent fluctuation
${\cal A}_{\alpha\beta\gamma}$ whose ansatz in terms of 7-dimensional fields
is (see below) $-\frac{2i}{\sqrt{3}} S_{\alpha\beta\gamma ,A}(y) Y^A (x)$. Then
the 7-dimensional susy variation of the gauge field dependent part of
$F_{\mu\alpha\beta\gamma}$ will be the difference between (\ref{ss4}) and
the 7 dimensional susy variation of $\partial_\mu{\cal A}_{\alpha\beta\gamma}$.
Note that the first two terms in (\ref{ss4}) are already parts of the susy
variation of $-i\sqrt{6} S_{\alpha\beta\gamma}$. We are ready now to
extract the susy variation of the gauge field dependent piece of $F_{\mu
\alpha\beta\gamma}$:
\bea
&&\left((\delta (F_{\mu\alpha\beta\gamma})|_{B\;\;terms\;\;only}\right)|_\psi
\nonumber\\&
=&6\sqrt{2}\left[\partial_\mu\left(\epsilon_{ABCDE}
\delta(B_{[\gamma}^{AB})(B_\alpha\cdot Y)^C (B_{\beta ]}\cdot Y)^D
\frac{T^{EF} Y_F}{Y\cdot T\cdot Y}\right)
\right.\nonumber\\
&&+\left.\partial_{[\alpha}\left(\epsilon_{ABCDE}\delta(B_{\gamma}^{AB})
(B_{\beta ]}\cdot Y)^C C_\mu^D \frac{T^{EF} Y_F}{Y\cdot T\cdot Y}
\right)\right]\nonumber\\
&&+\frac{3\sqrt{2}}{2}\epsilon_{ABCDE} \delta (B_{[\alpha}^{AB})
F_{\beta\gamma ]}^{CD} C_\mu^E\label{ss5}
\eea
We are going to rewrite the first two terms as total 7-dimensional susy
variations. Then we shall ``peel off'' a $\delta$ from both sides of
(\ref{ss5}) and read off the ansatz for $(F_{\alpha\beta\gamma\mu})|_{B\;\;
terms\;\;only}$.

We begin by working out the first term. First, we shall pull outside a total
$\delta$ susy variation, then we use a Schouten identity to release
a susy gauge field variation from its contraction with the spherical harmonic.
\bea
&&\partial_\mu\left( \epsilon_{ABCDE}\delta(B_{[\gamma }^{AB}) (B_{\alpha}
\cdot Y)^C (B_{\beta ]}\cdot Y)^D \frac{T^{EF} Y_F}{Y\cdot T\cdot Y}
\right)\nonumber\\
&&=\delta\partial_\mu\left(\epsilon_{ABCDE}B_{[\gamma }^{AB} (B_{\alpha}
\cdot Y)^C (B_{\beta ]}\cdot Y)^D \frac{T^{EF} Y_F}{Y\cdot T\cdot Y}
\right)\nonumber\\
&&-2\partial_{\mu}\left(\epsilon_{ABCDE}B_{[\gamma }^{AB} \delta(B_{\alpha}
\cdot Y)^C (B_{\beta ]}\cdot Y)^D \frac{T^{EF} Y_F}{Y\cdot T\cdot Y}
\right)\nonumber\\
&&=\frac{1}{3}\delta\partial_\mu\left(\epsilon_{ABCDE}B_{[\gamma }^{AB}
(B_{\alpha}\cdot Y)^C (B_{\beta ]}\cdot Y)^D \frac{T^{EF} Y_F}{Y\cdot T\cdot Y}
\right)\nonumber\\
&&+\frac{1}{3}\partial_\mu\left( \epsilon_{ABCDE} B_{[\gamma }^{AB} \delta
(B_{\alpha}^{CD}) (B_{\beta ]}\cdot Y)^E\right)\label{ss6}
\eea

Schoutenizing the second term of (\ref{ss5}) we get:
\bea
&&\partial_{[\alpha}\left(\epsilon_{ABCDE}\delta(B_{\gamma}^{AB})
(B_{\beta ]}\cdot Y)^C C_\mu^D \frac{T^{EF} Y_F}{Y\cdot T\cdot Y}
\right)\nonumber\\
&&=\frac{1}{2}\epsilon_{ABCDE}\left[\delta\partial_{[\alpha}\left(
B_\gamma^{AB}(B_{\beta ]}\cdot Y)^C C_\mu^D \frac{T^{EF} Y_F}{Y\cdot T\cdot Y}
\right)\right.\nonumber\\
&&+\frac{1}{2}\left.\partial_{[\alpha} \left( B_\gamma^{AB}
\delta(B_{\beta ]}^{CD})C_\mu^E\right)\right]\label{ss7}
\eea
The last step is to substitute (\ref{ss6}) and (\ref{ss7}) into (\ref{ss5})
\bea
&&\left((\delta (F_{\mu\alpha\beta\gamma})|_{B\;\;terms\;\;only}\right)|_\psi
=\epsilon_{ABCDE}\left\{ 3\sqrt{2} \delta \left[ \partial_\mu \left(
\frac{2}{3} B_{[\gamma}^{AB}(B_\alpha \cdot Y)^C (B_{\beta ]}\cdot Y)^D
\right.\right.\right.\nonumber\\
&&\left.\left.\left.\frac{T^{EF} Y_F}
{Y\cdot T\cdot Y}\right)
+\partial_{[\alpha}\left( B_\gamma^{AB}(B_{\beta ]} \cdot Y)^A
C_\mu^D  \frac{T^{EF} Y_F}{Y\cdot T
\cdot Y}\right)\right]+3\sqrt 2 \partial_{[\alpha }\left(
B_\gamma^{AB}\delta(B_{\beta ]}^{CD} C_\mu^E\right)\nonumber\right.\\
&&\left.+2\sqrt{2}\partial_\mu (B_{[\gamma}^{AB}\delta(B_\alpha^{CD})
(B_{\beta ]}\cdot Y)^E+3\sqrt{2}
\delta(B_{[\alpha}^{AB}) \left(\partial_\beta
B_{\gamma ]}^{CD}+2(B_{\beta}\cdot B_{\gamma ]})^{CD}\right) C_\mu^E\right\}
\nonumber\\
&&=
\epsilon_{ABCDE}\delta\left[2\sqrt 2\partial_\mu\left(B_{[\gamma}^{AB}
(B_\alpha\cdot Y)^C (B_{\beta ]}\cdot Y)^D \frac{T^{EF} Y_F}{Y\cdot T
\cdot Y}\right)\right.\nonumber\\
&&\left.-3\sqrt{2}\partial_{[\alpha}\left(B_\beta^{AB}(B_{\gamma ]}\cdot Y)^C
B_{\gamma ]}^{BC}C_\mu^D \frac{T^{EF} Y_F}{Y\cdot T\cdot Y}\right)\right.
\nonumber\\
&&+\left.\frac{3\sqrt{2}}{2}
\partial_\mu\left(\partial_{[\alpha} B_\beta^{AB} +
\frac{4}{3} (B_{[\alpha }\cdot B_\beta)^{AB}\right) B_{\gamma ]}^{CD}
Y^E\right]\label{ss8}
\eea

To achieve the last expression on the l.h.s. of (\ref{ss8}) we need to
perform one more trick, namely to use the Schouten identity to rewrite
\bea
&&\epsilon_{ABCDE} B_{[\gamma}^{AB} \delta(B_\alpha^{CD}) B_{\beta ]}^{EF}
C_\mu^F=\epsilon_{ABCDE}\left[\delta\left(B_{[\gamma}^{AB} (B_\alpha\cdot
B_{\beta ]})^{CD}  C_\mu^E \right)\right.\nonumber\\
&&\left.-3 \delta (B_{[\gamma})^{AB} (B_\alpha \cdot
B_{\beta ]})^{CD} C_\mu^E\right]
\eea

We conclude that, indeed, from the requirement that the susy laws are
consistent we obtain the ansatz for $F_{\mu\alpha\beta\gamma}$ given
in (\ref{mabc}).

At this moment, we can (again for simplicity of the argument) solve
$F_{\alpha\beta\gamma\delta}$ from the Bianchi identities.
The Bianchi identities
\be
\partial_{[\Lambda}F_{\Pi\Sigma\Omega\Delta]}=0
\ee
are trivially satisfied in the $F$ sectors which we derived so far, because
the gauge field dependent terms $F$ can be easily cast into an exact form,
while the purely scalar sectors are components of a separately closed form.
Thus we use the last Bianchi identity
\be
\partial_\mu F_{\alpha\beta\gamma\delta}=
4\partial_{[\alpha}F_{\mu\beta\gamma\delta ]}
\ee
which yields
\bea
&&0=\partial_\mu\left[F_{\alpha\beta\gamma\delta}-4\epsilon_{ABCDE}
\partial_{[\alpha}
\left(2\sqrt{2}B_{\delta}^{AB} (B_\beta\cdot Y)^C
(B_{\gamma ]}\cdot Y)^D \frac{(T\cdot Y)^E}{Y\cdot T\cdot Y}\right.\right.
\nonumber\\&&+\left.\left.
\frac{3\sqrt{2}}{2}\left(\partial_\beta B_{\gamma}^{AB}+
\frac{4}{3} (B_\beta\cdot B_\gamma)^{AB}\right)B_{\delta ]}^{CD} Y^E\right)
\right]
\eea
to solve for $F_{\alpha\beta\gamma\delta}$. The result, which
is given in (\ref{abcd}) is compatibile with the susy laws, and satisfies
trivially $\partial_{[\epsilon}F_{\alpha\beta\gamma\delta ]}=0$.

A brief inspection of the ansatz given in (\ref{mnpr}-\ref{abcd}) suggests that
we can also give the ansatz for the 11-dimensional 3-index field,
$A_{\Lambda\Pi\Sigma}$.
\bea
A_{\mu\nu\rho}&=&-\frac{1}{6\sqrt{2}} \epsilon_{ABCDE} C_\mu^A C_\nu^B C_\rho^C
Y^D \frac{(T\cdot Y)^E}{Y\cdot T\cdot Y}
-\frac{1}{2\sqrt 2}\frac{\sirc{D}_\sigma}
{\sirc\Box}(\epsilon_{\mu\nu\rho\sigma}\sqrt{\sirc g})\nonumber\\
A_{\nu\rho\alpha}&=&\frac{1}{6\sqrt{2}} \epsilon_{ABCDE} B_\alpha^{AB} C_\nu^C
C_\rho^D \frac{(T\cdot Y)^E}{Y\cdot T\cdot Y}
\nonumber\\
A_{\nu\alpha\beta}&=&\frac{1}{3\sqrt{2}} \epsilon_{ABCDE}
(B_{[\alpha} \cdot Y)^A B_{\beta ]}^{CD} C_\nu^D \frac{(T\cdot Y)^E}
{Y\cdot T\cdot Y}\nonumber\\
A_{\alpha\beta\gamma}&=&-\frac{i\sqrt {6}}{6}S_{\alpha\beta\gamma ,A}Y^A
\nonumber\\
&&+\epsilon_{ABCDE} \left(\frac{\sqrt 2}{3}B_{[\alpha}^{AB} (B_\beta\cdot Y)^C
(B_{\gamma ]}\cdot Y)^D \frac{(T\cdot Y)^E}{Y\cdot T\cdot Y}\right.\nonumber\\
&&\left.+\frac{1}{2\sqrt 2}(\partial_{[\alpha} B_\beta^{AB} +
\frac{4}{3} (B_{[\alpha}\cdot B_\beta)^{AB})B_{\gamma ]}^{CD} Y^E\right)
\eea

Also, to ease the comparison between the ansatz necessary to
achieve a consistent truncation  (given in
(\ref{mnpr}-\ref{abcd})) and the geometrical ansatz of \cite{hmm}
we present the expression of $F_{\Lambda\Pi\Sigma\Omega}$ in form
language. It is gauge invariant. \bea
\frac{\sqrt{2}}{3}F_{(4)}&=&\epsilon_{ABCDE}
\left(-\frac{1}{3}DY^A DY^B DY^C DY^D\frac{(T\cdot Y)^E}{Y\cdot
T\cdot Y}\right.\nonumber\\&&\left. +\frac{4}{3} DY^A DY^B DY^C
D(\frac{(T\cdot Y)^D}{Y\cdot T\cdot Y})Y^E \right.\nonumber\\
&&\left.+2 F_{(2)}^{AB} DY^C DY^D \frac{(T\cdot Y)^E} {Y\cdot
T\cdot Y}+F_{(2)}^{AB} F_{(2)}^{CD} Y^E\right)+d({\cal A})
\label{f4} \eea where $F_{(2)}^{AB}= 2(dB^{AB} + 2(B\cdot
B)^{AB}))$ and $DY^A=dY^A+2(B \cdot Y)^A$.

To find the  ansatz for ${\cal A}$ and $\tilde{\cal B}$ , we
require that we  obtain  all the $S_{\alpha\beta\gamma}$ terms
in the d=7 gravitino susy transformation law from KK reduction.
Consider the  linearized gravitino
transformation law. This should already display the self-duality
mechanism, and also will fix  the free parameters $a$ and $b$ in the d=11 susy
transformation law of the auxiliary field ${\cal B}$.
The $F_{\Lambda\Pi\Sigma\Omega}$ term in
$\delta\Psi_{\Lambda}$ reads:
\bea
\delta\Psi_{\zeta}\;|_{lin.,F\;term,B=0 }&
=&\frac{\sqrt{2}}{288}(\Gamma^{\alpha
\beta\gamma\delta}\;_{\zeta}-8\delta_{\zeta}^{[\alpha}
\Gamma^{\beta\gamma\delta]})\varepsilon F_{\alpha\beta\gamma\delta}\nonumber\\
&&+\frac{\sqrt{2}}{288}(4\Gamma^{\mu\alpha\beta\gamma}\;_{\zeta}+3\cdot8
\delta_{\zeta}^{\alpha}\Gamma^{\mu\beta\gamma})\varepsilon F_{\mu
\alpha\beta\gamma}\\
\delta\Psi_m\;|_{lin.,F\;term,B=0 }&
=&\frac{\sqrt{2}}{288}\Gamma^{\alpha\beta\gamma\delta}\;_m
\varepsilon F_{\alpha\beta\gamma\delta}\nonumber\\&&
 +\frac{\sqrt{2}}{288} (4
\Gamma^{\mu\alpha\beta\gamma}\;_m-8e^{\mu}_m\Gamma^{\alpha\beta\gamma})
\varepsilon F_{\mu\alpha\beta\gamma}
\eea
The result for $\delta\psi_{\epsilon}$ can be worked out further by decomposing
the 11
dimensional gamma matrices and using the commutation relations of $\sqrt{
\gamma_5}$
\bea
\delta\psi_{\epsilon}|_{lin.,F\;term,B=0 }&
=&\frac{\sqrt{2}}{288}\frac{\gamma_5}{5}
\{3(3\tau^{\alpha\beta\gamma\delta}\;_{\epsilon}-8\delta_{\epsilon}^{[\alpha}
\tau^{\beta\gamma\delta]})F_{\alpha\beta\gamma\delta}\nonumber\\
&&-24i(\tau^{\alpha\beta\gamma}\;_{\epsilon}+\frac{9}{2}\delta_{\epsilon}
^{[\alpha}\tau^{\beta\gamma ]})\gamma^{\mu}F_{\mu\alpha\beta\gamma}\}\epsilon
\label{epslin}
\eea
We introduce a normalization constant c for $S_{\alpha\beta\gamma ,A}$
by embedding the 7d field $S$ into the 11d curvature $F$ as follows:
\be
F_{\mu\alpha\beta\gamma}|_{B=0}=cS_{\alpha\beta\gamma ,A}\partial_{\mu}Y^A
\;\; F_{\alpha\beta\gamma\delta}|_{B=0}=
4c\partial_{[\alpha}S_{\beta\gamma\delta],A}
Y^A\label{constant}
\ee
and a corresponding relation for the auxiliary field,
\be
\tilde{{\cal B}}_{\alpha\beta\gamma}= B_{\alpha\beta\gamma ,A} Y^A
\ee
Substituting the linearized ansatz for
$ B_{\alpha\beta\gamma ,A}$ in (\ref{auxili})
into the transformation law of $\psi_\epsilon$, we find
\bea
\delta \psi_{\epsilon}|_{lin.,{\cal B}\; term}&=&\frac{1}{24}
(b\Gamma_{\epsilon}\;
^{\alpha\beta
\gamma \delta}B_{\alpha\beta\gamma\delta}-a\Gamma^{\alpha\beta\gamma}
B_{\epsilon\alpha\beta\gamma})\epsilon\nonumber\\&&
+\frac{1}{5}\tau_{\epsilon}
\gamma_5\gamma^m\frac{b}{24}\Gamma^{\alpha\beta\gamma\delta}\;_m B_{\alpha
\beta\gamma\delta}\epsilon
\nonumber\\
&=&\frac{k}{24}\epsilon_{\alpha\beta\gamma\delta}\;
^{\epsilon\eta\zeta}B_{\epsilon\eta\zeta ,A}Y^A
(\frac{9}{5}b\tau_{\epsilon}\;^{\alpha\beta\gamma}-(a-\frac{16}{5}b)
\delta_{\epsilon}^{[\alpha}\tau^{\beta\gamma\delta]})
\gamma_5\epsilon\nonumber\\
&=&\frac{\gamma_5}{5} \{ -m\frac{9}{5}\frac{kb}{24}
\frac{4!3!}{2}\delta_{\epsilon}^{[\alpha}
\tau^{\beta\gamma]}S_{\alpha\beta\gamma, A}Y^A\nonumber\\&&
+\frac{9}{5}\frac{kb}{24}
 6\tau_{\epsilon}\;^{\alpha\beta\gamma\delta}
4D_{\alpha}S_{\beta\gamma\delta ,A}Y^A\nonumber\\&&
-m \frac{k}{24}(a-\frac{16}{5}b)
3!\tau_{\epsilon}\;^{\alpha\beta\gamma}S_{\alpha\beta\gamma, A}Y^A\nonumber\\
&&-\frac{k}{24}(a-\frac{16}{5}b) 6\tau^{\alpha
\beta\gamma}4D_{[\epsilon}S_{\alpha\beta\gamma ] ,A}Y^A\}\epsilon
\label{epsb}
\eea
In order to reproduce the 7 dimensional sugra result (and have a
consistent truncation of 11 dimensional sugra to the massless 7 dimensional
fields), we must reproduce the gauged sugra result for $\delta\psi_{\epsilon}
^I$, which does not have any $DS$ terms, so we must require $kb=-\frac{5
\sqrt{2}c}{72}, ka=-\frac{5\sqrt{2}c}{9}$. Then we obtain
\be
\delta\psi_{\epsilon}|_{lin., S\; terms}=-c\frac{\sqrt{2}}{60}[\tau_{\epsilon}
\;^{\alpha\beta\gamma}-\frac{9}{2}\delta_{\epsilon}^{[\alpha}\tau^{\beta\gamma
]}]S_{\alpha\beta\gamma , A}
\gamma_5(-i\gamma^{\mu}D_{\mu}+m)Y^A\epsilon
\label{epss}
\ee
Using (\ref{fierz}), we get by contracting with $(\gamma^A)_{JK}$ and using
$1/4\phi_5^{JK}(\gamma^A)_{JK}=Y^A$, according to section (3.1),
\be
\frac{1}{m}(m+i\gamma^{\mu}D_{\mu})Y^A\gamma_5\eta^I\epsilon _I
=-\epsilon_I (\gamma^A)^I
\;_K \eta^K
\ee
Hence
\be
\delta\psi_{\epsilon}|_{lin., S\; terms} = c\left[ \frac{\sqrt{2}}{60}
m\left( \tau_{\epsilon}\; ^{\alpha\beta\gamma}-\frac{9}{2}\delta _{\epsilon}^{[
\alpha}\tau^{\beta\gamma ]}\right) (\epsilon\gamma^A)_K\eta^K\right]
S_{\alpha\beta\gamma , A} \label{fixes}
\ee

Incidentally, we  note that here if we would use the p=q=+1/2 solution in
(\ref{epsi})  instead of $p=q=-1/2$ we would get $(i\gamma^{\mu}D_{\mu}+m)$
in (\ref{epss}) (because the commutation relation of $\gamma_5^{\pm 1/2}$
past $\gamma_{\mu}$ gives a $\pm$ sign). That means that at this moment
the sign is fixed by requiring a consistent truncation.

In order to match this result
with the $S$ term in $\delta\psi_{\epsilon}$ in (\ref{gravi}),
we fix $c=-i\sqrt{6}$. We can then check this value of $c$ by putting
(\ref{constant}) in the 11d action for $F_{\Lambda\Pi\Sigma\Omega}$. We
obtain the correct normalization for the 7d action of
$S_{\alpha\beta\gamma ,A}$.

At this moment, one can see why a first order formulation for gravity instead
of the antisymmetric tensor field does not work. Consider the spin connection
 $\Omega$ as an independent field, as done in \cite{cfgpp}; then the
11 dimensional gravitino transformation law is the same, with $D_{\Lambda}
(\hat{\Omega})\varepsilon$ replaced  by $D_{\Lambda}(\Omega)\varepsilon$, where
\be
\Omega_{\Lambda}\;^{MN}=\hat{\Omega}_{\Lambda}\;^{MN}(E,\Psi)+
{\cal B}_{\Lambda\Pi\Sigma}E^{M\Pi}E^{N\Sigma}+
\Delta\Omega '_{\Lambda}\;^{MN}
\ee
and $\Delta\Omega '_{[\Lambda}\;^{MN}E^M_{\Pi}E^N_{\Sigma]}=0$. Thus the
${\cal B}_{\alpha\beta\gamma}$ term we get in $\delta_{susy,11-dim}
\Psi_{\alpha}$ is
at the linearized level is $\frac{1}{4}{\cal B}_{\alpha\beta\gamma}
\tau^{\beta\gamma}\varepsilon$. But this has no $\gamma_5$ with respect
to (\ref{epsb}), and since the susy law is fixed by the definition of
${\cal B}$, we can't introduce by hand the missing $\gamma_5$.
Therfore  it will be
impossible to cancel the F terms in (\ref{epslin}),
as we did in (\ref{epss}). In conclusion, we need a 4-index tensor to mix
with the independent fluctuation $S_{\alpha\beta\gamma}$.

Now, let us analyze the nonlinear level. The nonlinear KK reduction gives for
the terms involving $S$
\bea
\delta \psi_{\epsilon}|_{S\;term}&=&\Delta^{-1/10}\gamma_5^{-1/2}
e'_{\epsilon}\;^a[E_a^{\Lambda}\delta\Psi_{\Lambda}|_{S\; term}
+\frac{1}{5}\tau_a\gamma_5\gamma^m E_m^{\Lambda}\delta\Psi_{\Lambda}|_{S\; term
}]\nonumber\\
&=&\frac{\sqrt{2}}{288}\Delta^{-1/10}\gamma_5^{-1/2}e'_{\epsilon}\;^a\{
(\Gamma^{\alpha\beta\gamma\delta}\;_a-8\delta_a^{[\alpha}\Gamma^
{\beta\gamma\delta]})\varepsilon F_{\alpha\beta\gamma\delta}|_{S\; term}
\nonumber\\
&&+\frac{12}{\sqrt{2}}(b\Gamma^{\alpha\beta\gamma\delta}\;_a-a\delta_a
^{[\alpha}\Gamma^{\beta\gamma\delta]})\varepsilon
\frac{{\cal B}_{\alpha\beta\gamma
\delta}}{\sqrt{E}}\nonumber\\
&&+4(\Gamma^{\mu\alpha\beta\gamma}\;_a+8\delta_a^{[\alpha}\Gamma^{\mu\beta
\gamma]})\varepsilon F_{\mu\alpha\beta\gamma}|_{S\; term}\nonumber\\
&&+\frac{1}{5}\tau_a\gamma_5\gamma^m[\Gamma^{\alpha\beta\gamma\delta}\;_m
\varepsilon (F_{\alpha\beta\gamma\delta}+\frac{12 b}
{\sqrt{2}}\frac{{\cal B}_{\alpha
\beta\gamma\delta}}{\sqrt{E}})\nonumber\\
&&+4(\Gamma^{\mu\alpha\beta\gamma}\;_m -2\delta_m^{\mu}\Gamma^
{\alpha\beta\gamma})\varepsilon F_{\mu\alpha\beta\gamma}]\}
\eea

In the same way as it happened at the linear level, the $\partial_{\alpha}
S_{\beta\gamma\delta}$ terms in ${\cal B}_{\alpha\beta\gamma\delta}$
will cancel the $F_{\alpha\beta\gamma\delta}$ terms, whereas the
$S_{\alpha\beta\gamma}$ terms in ${\cal B}_{\alpha\beta\gamma\delta}$
and $F_{\mu\alpha\beta\gamma}$ add, and moreover the term
$B_{\alpha}^{AB}S_{\beta\gamma\delta B}$ in
${\cal B}_{\alpha\beta\gamma}$ cancel the $B_{\delta}^{\mu}
F_{\mu\alpha\beta\gamma}$ term, provided we choose the ansatz (neglecting
``massive'' fields which are put to zero)
\bea
\frac{{\cal B}_{\alpha\beta\gamma\delta}}{\sqrt{E}}&=&\frac{6k}{5}
(F_{\alpha\beta\gamma\delta}+4B^{\mu}_{\delta}F_{\mu\alpha\beta\gamma})
-\frac{k}{5}\epsilon_{\alpha\beta\gamma\delta}\;^{\epsilon\eta\zeta}
\tilde{\tilde{\cal B}}_{\epsilon\eta\zeta}\nonumber\\
&=&\frac{24kc}{5}\bigtriangledown_{[\alpha }S_{\beta\gamma\delta ],A}Y^A
-\frac{k}{5}\epsilon_{\alpha\beta\gamma\delta}\;^{\epsilon\eta\zeta}
\tilde{\tilde{\cal B}}_{\epsilon\eta\zeta}\label{aux}
\eea
where $\tilde{\tilde{\cal B}}$ contains only $S$ terms (no
$\partial_{\alpha} S_{\beta\gamma\delta, A}$ or
$B_{\alpha}^{AB}S_{\beta\gamma\delta, B}$) and we have used (\ref{compl})
and (\ref{vmu}). In the following, the parameters $a, b, c$ take
the values which we determined previously. Then we obtain
\bea
\delta \psi_{\epsilon}|_{S\; term}&=&-\frac{\sqrt{2}}{60}(\tau
_{\epsilon}\;^{\alpha\beta\gamma}-\frac{9}{2}\delta_{\epsilon}^{[\alpha}
\tau^{\beta\gamma ]})\gamma_5\Delta^{3/5}\nonumber\\
&&[(\Delta^{-1/5}E^{\mu}_m)(-i\gamma^m)F_{\mu\alpha\beta\gamma}
+\tilde{\tilde{\cal B}}_{\alpha\beta\gamma}]\epsilon
\eea

We further work out the $F_{\mu\alpha\beta\gamma}$ term in $\delta\psi_
\epsilon$ by substituting our ans\"{a}tze for various fields, where in
$F_{\mu\alpha\beta\gamma}$ we consider only the term with S
\be
F_{\mu\alpha\beta\gamma}|_{B=0}=cS_{\alpha\beta\gamma,A}\partial_\mu Y^A
\ee

So, by substituting for $F_{\mu\alpha\beta\gamma}$, using the ansatz
for $E^{\mu}_m$ in (\ref{viel}) and introducing the identity as
$\bar{\eta}^N_{\alpha}
\eta_N^{\beta}=\delta_{\alpha}^{\beta}$, we get
\bea
&&\Delta^{-1/5} E^\mu_m \gamma_5 (-i\gamma^m)\Delta^{3/5}
F_{\mu\alpha\beta\gamma} \epsilon\nonumber\\
&=& -\frac{1}{4}\Delta^{3/5} cS_{\alpha\beta\gamma,C}
C_m^C Tr(\gamma^{ij}UV^n U^T
\Omega)\nonumber\\&&
{(\Pi^{-1})_i}^A {(\Pi^{-1})_j}^B V^{m\; AB}C^{n\; NI}
 \eta_N {U^{I'}}_I\epsilon_{I'}\label{intermediar}
\eea

Then, using (\ref{fierz}), we have
\be
V_n^{PQ}C_n^{NI}=4\phi_5^{P[I}\Omega^{N]Q}-(\Omega^{PQ}\phi_5^{NI}+
\Omega^{NI}\phi_5^{PQ})\label{VC}
\ee
and substituting this relation for the two products $V_nC_n$ in
(\ref{intermediar}), and doing a bit of algebra, we get
\bea
&&\Delta^{3/5} {(\Pi^{-1})_i}^A {(\Pi^{-1})_j}^B Y^A cS_{\alpha\beta\gamma,B}
\nonumber\\&&
\frac{1}{2}\left[
\phi_5^{PI}(U^T\gamma^{ij}U)_{PN}-\phi_5^{PN}(U^T\gamma^{ij}U)_{PI}\right]
\eta_N{U^{I'}}_I\epsilon_{I'}
\eea

But now we can use the condition we got on $U$ from matching the scalar
transformation law (\ref{urel}), and get
\be
c S_{\alpha\beta\gamma,A}{(\Pi^{-1})_i}^A{{\gamma^j}^{I'}}_{L'}
{U^{L'}}_N \eta_N \epsilon_{I'}-\Delta^{3/5} T^{AB} cS_{\alpha\beta\gamma,B}
Y_A \gamma_5\eta^I{U^{I'}}_I \epsilon_{I'}
\ee
We now see that the first term is exactly what we want to have in 7 dimensional
supergravity (the $S_{\alpha\beta\gamma ,A}$ term of $\delta\psi_{\epsilon}$
in \cite{ppn}), whereas the second term gets canceled by the
$\tilde{\tilde {\cal B}}_{\alpha\beta\gamma}$ term, if we choose the
ansatz for $\tilde{\tilde{\cal B}}_{\alpha\beta\gamma}$ to be:
\be
\tilde{\tilde {\cal B}}_{\alpha\beta\gamma}=c T^{AB} S_{\alpha\beta\gamma, B}Y_A
\ee

However, let us pause and comment on the ansatz for ${\cal B}_{MNPQ}$.
We know that ${\cal B}_{MNPQ}=0$ is a solution of the 11d equations of
motion. So we can say that the ansatz for the various components of
${\cal B}_{MNPQ}$ has to be such that the various components of
${\cal B}_{MNPQ}=0 $ correspond to various 7d equations of motion. Then the
first remark is that we have to add to (\ref{aux}) the necessary bilinears
in fermions and gauge field strength to complete the 7-dimensional
equation of motion for $S_{\alpha\beta\gamma ,A}$ (remember that we
always dropped in the susy transformation rules  3- and 4-fermi terms).
Using that $k=5(6\sqrt 2)$ (as it will be determined below (\ref{detk})),
the ansatz of the auxiliary field ${\cal B}_{MNPQ}$ is
\bea
\frac{{\cal B}_{\alpha\beta\gamma\delta}}{\sqrt E}&=&
-24\sqrt{3}i\nabla_{[\alpha}S_{\beta\gamma\delta ],A}Y^A+\sqrt{3}i
{\epsilon_{\alpha\beta\gamma\delta}}^{\epsilon\eta\zeta}T^{AB}
S_{\epsilon\eta\zeta ,B}Y_A\nonumber\\
&+&9\epsilon_{ABCDE}F_{[\alpha\beta}^{BC}
F_{\gamma\delta ]}^{DE}Y^A+2-fermi \;\;terms
\eea
The necessity of adding the
gauge field terms in the ansatz for the auxiliary field  ${\cal B}_{MNPQ}$
can be seen if we try to derive the 7-dimensional action from the
11-dimensional one. The ansatz for $F_{\alpha\beta\gamma\delta}$
is already bilinear in the gauge field strength, and when integrating out
$(F_{\alpha\beta\gamma\delta})^2$ on $S_4$ we get as a result terms with
$(F_{\alpha\beta}^{AB})^4$ which are not present in the $d=7$ action.
However, they are precisely canceled by the gauge field strengths which enter
in the ansatz of ${\cal B}_{MNPQ}$, after integrating out
$({\cal B}_{MNPQ})^2$. The second observation is that
no other bosonic equation of motion
can appear in ${\cal B}_{MNPQ}$, since all other bosonic
equations  have 2 derivatives, and so we would get terms with 4 derivatives
in the 7d action.
After finding the correct ansatz for ${\cal B}_{\alpha\beta\gamma\delta}$,
we can
easily find the nonlinear terms in $\delta\lambda_i$:
\bea
\delta \psi_m|_{S \;term}&=&\Delta^{-1/10}\gamma_5^{1/2}E_m^{\Lambda}\delta
\Psi_{\Lambda}\nonumber\\
&=&\frac{\sqrt{2}}{288}\Delta^{-1/10}\gamma_5^{1/2}[\Gamma^{\alpha\beta
\gamma\delta}\;_m F_{\alpha\beta\gamma\delta}+4(\Gamma^{\mu\alpha\beta
\gamma}\;_m-2E^{\mu}_m\Gamma^{\alpha\beta\gamma})F_{\mu\alpha
\beta\gamma}\nonumber\\
&&-\frac{5}{6k}\Gamma^{\alpha\beta\gamma\delta}\;_m\frac{{\cal B}_{\alpha\beta
\gamma\delta}}{\sqrt{E}}]\varepsilon
\eea
where we have substituted the value of $b$ found before. If we now use the
ansatz for $\frac{{\cal B}_{\alpha\beta\gamma}}{\sqrt{E}}$, the $F_{\alpha
\beta\gamma\delta}$ and $B_{\delta}^{\mu}F_{\mu\alpha\beta\gamma}$ terms
cancel again, as in $\delta\psi_{\epsilon}$, and we are left with
\bea
\delta\psi_m |_{S\; term}&=&\frac{i\sqrt{2}}{288}4\tau^{\alpha\beta\gamma
}\Delta^{3/5}\nonumber\\
&&[(\Delta^{-1/5}E^{\mu}_n)(\gamma_m\;^n-2\delta_m^n)(-iF_{\mu\alpha\beta
\gamma})-\gamma_m\tilde{\tilde{\cal B}}_{\alpha\beta\gamma}]\epsilon
\label{delpsim}
\eea
We next substitute for $F_{\mu\alpha\beta\gamma}$ and $\tilde{\tilde{B}}_
{\alpha\beta\gamma}$. Then we write $E^{\mu}_n$ in terms of scalars via
(\ref{vielinvsim}). Using (\ref{dydy}) for the contraction of $C_{\mu}$'s
which we get, and the identity
\be
{(\Pi^{-1})_i}^CY_CTr(U^{-1}\gamma^iU\gamma^D)C_n^D=0
\ee
which we easily get by using (\ref{urel}), we can rewrite (\ref{delpsim})
as follows
\bea
\delta\psi_m^{(11d)}&=&
\frac{1}{48\sqrt{3}}\tau^{\alpha\beta\gamma}S_{\alpha\beta\gamma ,B}
\left[-i{(\Pi^{-1})_i}^BC_n^DTr(U^{-1}\gamma_iU\gamma^D)\right.\nonumber\\
&&\left.({\gamma_m}^n-2\delta_m^n)-4\Delta^{3/5}T^{AB}Y_B\gamma_m\right]\eta^I
\epsilon_{I'}{U^{I'}}_I
\eea

Next we want to put in a similar form the expression which we get for
$\delta\psi_m$ from the 7 dimensional result for $\delta\lambda^i_{J'}$.
>From the ansatz in (\ref{psim}) and the normalization in (\ref{normaliz}),
and substituting the 7 dimensional transformation law of $\delta\lambda^i_{
J'}$, we get
\bea
\delta\psi_m^{(7d)}&=&-\frac{1}{240\sqrt{3}}\tau^{\alpha\beta\gamma}
S_{\alpha\beta\gamma ,A}{(\Pi^{-1})_j}^A{(\gamma^{ij}+4\delta^{ij})^{I'}}
_{J'}\nonumber\\
&&(\gamma_i)_{K'L'}\epsilon_{I'}{U^{J'}}_J{U^{K'}}_K{U^{L'}}_L\eta_m^{JKL}
\eea
Then we write the spherical harmonic $\eta_m^{JKL}$ as
\be
\eta_m^{JKL}=[i(\gamma_{mp}-2\delta_{mp})C^{pA}+\gamma_mY^A]\eta^J
(\gamma_A)^{KL}
\ee
Using the U matrix relation (\ref{urel}), we find the identity
\be
Tr(U\gamma_AU^{-1}\gamma_i)Y_A=4\Delta^{3/5}{(\Pi^{-1})_i}^AY_A\label{tracerel}
\ee
which will be heavily used in the following. Using this relation and the
expression derived for $\eta_m^{JKL}$, we find that
\bea
\delta\psi_m^{(7d)}&=&\frac{1}{48\sqrt{3}}
\tau^{\alpha\beta\gamma}S_{\alpha\beta\gamma ,B}
[iTr(U\gamma^AU^{-1}\gamma_i){(\Pi^{-1})_i}^B\nonumber\\
&&(\gamma_{mp}-2\delta_{mp})C^{pA}+4\Delta^{3/5}T^{AB}Y_A\gamma_m]
\epsilon\nonumber\\&&
-\frac{1}{60\sqrt{3}}\tau^{\alpha\beta\gamma}S_{\alpha\beta\gamma ,B}
{(\Pi^{-1})_j}^B[i(\gamma_{mp}-2\delta_{mp})C^{pA} \nonumber\\
&&+\gamma_m Y^A]\eta^J {(\gamma_jU\gamma_AU^{-1})^{I'}}_{J'}{U^{J'}}_J
\epsilon_{I'}\label{delpsim7}
\eea
Now we work on the last term (the last two lines). For that, we need a
few identities: By multiplying the Fierzing relation (\ref{fierz})
with $Y_A(\gamma_A)_{JK}$ we deduce that
\be
\gamma_5\eta^I=-Y_A{(\gamma^A)^I}_J\eta^J\label{5etai}
\ee
If we multiply the same relation (\ref{fierz}) with $\gamma_5(\gamma_A)_{JK}
$, we get
\be
-i\gamma_p\eta^IC_p^A+Y^A\eta^I=Y_B{(\gamma^A\gamma^B)^I}_J
\eta^J\label{gapcp}
\ee
from which we also derive, by multiplying with $C_A^n$, that
\be
\gamma_m\eta^I=iC_m^AY^B{(\gamma_A\gamma_B)^I}_J\eta^J\label{gametai}
\ee

>From (\ref{gapcp}) and (\ref{gametai}) we also deduce
\be
\gamma_m\gamma^p C_p^A\eta^J {(U\gamma_A)^I}_J=
4C_m^A\eta^J{(U\gamma_A)^I}_J\label{somejunk}
\ee

Using (\ref{gametai}) and (\ref{somejunk}) it follows that the last term in
(\ref{delpsim7}) is zero, so that we are left with the correct 11 dimensional
result.

\subsection{The gravitino and $S_{\alpha\beta\gamma, A}$
 transformation laws}

Let us recapitulate what we have done so far. We have checked that using
our ans\"{a}tze for the 11 dimensional fields we reproduce the 7 dimensional
transformation laws for the graviton $e_{\alpha}^a$, the gauge field $B_{\mu}
^{AB}$, the scalars ${\Pi_A}^i$,
and the terms involving $S_{\alpha\beta\gamma}$ in the transformation
laws of the gravitini $\psi_{\alpha}^I$ and fermions $\lambda^{iI}$.
We have also checked the gauge field dependence of $F_{\Lambda\Pi\Sigma\Omega}
$ by matching terms in $\delta_{susy}^{(11d)}F_{\Lambda\Pi\Sigma\Omega}$
with the variation of all the fields in the ansatz for $F_{\Lambda\Pi
\Sigma\Omega}$. The analysis of the independent fluctuation $S_{\alpha
\beta\gamma ,A}$ was implicitely done when we studied the $F_{\alpha\beta
\gamma\mu}$ sector. However, for completeness of our work we give a
separate check on $S_{\alpha\beta\gamma}$ susy transformation law.

We begin by considering the case when
$B_{\alpha}^{AB}=0$ in the $S_{\alpha\beta\gamma ,A}$
transformation law and we find
$\delta S_{\alpha\beta\gamma ,A}$ from
\be
\frac{i}{\sqrt{6}}\delta F_{\mu\alpha\beta\gamma}|_{B=0}=
\delta S_{\alpha\beta\gamma ,A}|_{B=0}\partial_{\mu}Y^A
\ee
because the $B_{\alpha}^{AB}$ terms in $F_{\mu\alpha\beta\gamma}$
appear at least quadratically, (so $\delta B_{\alpha}^{AB}$ will always be
multiplied by $B_{\beta}^{CD}$, which we put to zero). Using the
transformation law for $A_{\Lambda\Pi\Sigma}$  in (\ref{deltaa}) and the fact
that
\be
\Psi_{\alpha}|_{B=0}=\Delta^{-1/10}(\gamma_5^{1/2}\psi_{\alpha
}+\frac{1}{5}\tau_{\alpha}\gamma^m \gamma_5^{1/2}\psi_m)
\ee
we get
\bea
&&\delta S_{\alpha\beta\gamma ,A}|_{B=0}\partial_{\mu}Y^A=
-i\sqrt{3}\partial_{[\mu}\bar{\varepsilon}\Gamma_{\alpha
\beta}\Psi_{\gamma]}\nonumber\\
&&=-\frac{i\sqrt{3}}{4}\{\partial_{\mu}[\bar{\epsilon}\Delta^{-3/5 }
(\tau_{[\alpha\beta}\gamma_5 \psi_{\gamma]}+\frac{i}{5}\tau_{\alpha\beta
\gamma}\gamma^m\psi_m)]\nonumber\\
&&\!\!\!-\partial_{[\alpha}[E_{\mu}^n \Delta^{-2/5}
\bar{\epsilon}(-2i\gamma_n
\gamma_5\tau_{\beta}\psi_{\gamma]}+\frac{1}{5}\tau_{\beta\gamma]}(-2\gamma
_{nm}+3\delta_{nm})\psi^m)]\}\label{deltasy}
\eea
We split this contribution into 3 terms, corresponding to the terms
in the 7 dimensional transformation law, one with $\partial_{\alpha}$ acting
on $\psi_{\gamma I'}$ and $\lambda_{i I'}$, one with $\partial_{\alpha}$
acting on ${\Pi_A}^i$, and one with no $\partial_{\alpha}$. Indeed, in 7
dimensions we have
\bea
\delta S_{\alpha\beta\gamma ,A}|_{B=0}&=&\frac{i\sqrt{3}}{12}\delta_{AB}
{(\Pi^{-1})_i}^B (3\bar{\epsilon}\tau_{[\alpha\beta}\gamma^i\psi_{\gamma]}
-\bar{\epsilon}\tau_{\alpha\beta\gamma}\lambda^i)\nonumber\\
&&-\frac{i\sqrt{3}}{4}\delta_{ij}{\Pi_A}^j\bigtriangledown_{[\alpha}
(2\bar{\epsilon}\tau_{\beta}\gamma^i\psi_{\gamma]}+\bar{\epsilon}
\tau_{\beta\gamma]}\lambda^i)
\eea
where $\bigtriangledown_{\alpha}(...)_i$ acts as $\partial_{\alpha}(...)_i
+Q_{\alpha\;i}\;^j(...)_j=\partial_{\alpha}(...)_i+{(\Pi^{-1})_{[i}}^A
(\partial_{\alpha}{\Pi_A}^{j]})(...)_j$. (If $B_{\alpha}^{AB}$ is not zero,
 $Q_{\alpha ij}$ will contain also a B term).

The term with no $\partial_{\alpha}$ in (\ref{deltasy}) is given by
\be
-\frac{i\sqrt{3}}{4}\Delta^{-3/5}[\Delta^{3/5}(\partial_{\mu}
\Delta^{-3/5})+\partial_{\mu}](\bar{\epsilon}
\tau_{[\alpha\beta}\gamma_5\psi_{\gamma]}
+\frac{i}{5}\bar{\epsilon}\tau_{\alpha\beta\gamma}\gamma_m\psi^m)\label{nodal}
\ee
To evaluate the  contribution with the the gravitino $\psi_{\gamma}$,
we make use of the equations (\ref{del}) and (\ref{urel}). From $\bar{\epsilon}
\tau_{\alpha\beta}\gamma_5\psi_{\gamma}$ we obtain a factor $U\bar{\eta}
\gamma_5\eta U^T\sim U\phi_5 U^T$. This is the combination which appears
on the l.h.s. of (\ref{urel}) (because $Y_A\gamma^A\sim \phi_5$).
Substituting (\ref{urel}), the $\partial_{\mu}$ derivatives of $\Delta^{3/5}$
cancel and we are left with
\be
\frac{-i\sqrt{3}}{4} \bar{\epsilon}_{I'}\tau_{[\alpha\beta}
{\gamma^i}^{I'J'}\psi_{\gamma] J'}{(\Pi^{-1})_i}^A \partial_\mu Y_A
\ee
Similarly, the contribution of the spin $1/2$ fermions in (\ref{nodal}) reads
\be
\frac{i\sqrt{3}}{4\cdot 3}{(\Pi^{-1})_i}^A (\tilde{\Omega})^{I'J'}
\bar\epsilon_{I'}
\tau_{\alpha\beta\gamma}\lambda^{i}_{J'} \partial_\mu Y_A.
\ee
We compare the sum of these two contributions with the 7-dimensional result
\be
\frac{i\sqrt{3}}{12}{(\Pi^{-1})_i}^A\bar\epsilon(3\tau_{[\alpha\beta}
\gamma^i
\psi_{\gamma}-\tau_{\alpha\beta\gamma}\lambda^{i})\partial_{\mu}
Y^A
\ee
and we can conclude that we find agreement.

The term with $\partial_{\alpha}$ acting on the scalars is given by
\bea
&&\frac{i\sqrt{3}}{4}[\partial_{[\alpha}(E_\mu^n\Delta^{-2/5})
\left(-2i\bar\epsilon \gamma_n\gamma_5\tau_{\beta}\psi_{\gamma]}+
\frac{1}{5}\bar\epsilon \tau_{\beta\gamma]}
(-2\gamma_{nm}+3\delta_{nm})\psi^m\right)\nonumber\\
&&+\Delta^{-2/5}E_\mu^n
\left(-2 iC_n^{IJ}\bar\epsilon_{I'}\tau_{[\beta}\psi_{\gamma J'}
\partial_{\alpha ]}(U^{I'}\;_I U^{J'}\;_J )+\right.\nonumber\\
&&\left.\frac{1}{5}\bar\epsilon_{I'}\tau_{[\beta\gamma}\lambda_{J'K'L'}
\bar\eta^I(-2\gamma^{nm}+3\delta^{nm})\partial_{\alpha]}(U^{I'}\;_I U^{J'}\;_J
U^{K'}\;_K U^{L'}\;_L)\eta_m^{JKL}\right)]\label{dalsca}
\eea
The spherical harmonic of the spin 1/2 fields, contracted with
$\lambda_{J'K'L'}$, yields
\bea
&&\frac{1}{5}\bar\eta^I (-2\gamma_n\gamma_m +5 \delta_{nm})\eta_m^{JKL} \lambda
_{J'K'L'}U^{J'}\;_J U^{K'}\;_K U^{L'}\;_L \nonumber\\&&
= 3\Omega^{IJ} C_n^{KL}
\lambda_{J'K'L'}U^{J'}\;_J U^{K'}\;_K U^{L'}\;_L \label{spheric}
\eea

 Together with (\ref{spheric}) we find for (\ref{dalsca})
\be
-\frac{i\sqrt{3}}{4}(\partial_{[\alpha} {\Pi_A}^j)
(-\bar\epsilon_{I'}2\tau_{\beta}\gamma_j^{I'J'}\psi_{\gamma]J'}+
\bar\epsilon_{I'}\tau_{\beta\gamma]}\lambda
_{j J'} \Omega^{I'J'})\partial_{\mu}Y^A
\ee
This agrees with the result in 7 dimensional
gauged supergravity
\be
-\frac{i\sqrt{3}}{4}{\Pi_A}^i{(\Pi^{-1})_i}^B
(\partial_{[\alpha} {\Pi_B}^j)
\bar{\epsilon}(2\tau_{\beta}\gamma_j\psi_{\gamma]}+\tau_{\beta\gamma]}\lambda
_j)\partial_{\mu}Y^A
\ee

Finally, we consider the terms with $\partial_{\alpha}$ acting on
$\psi_{\alpha I'}$ and $\lambda_{iI'}$
\bea
&&\frac{i\sqrt{3}}{4}\Delta^{-3/5}(\Delta^{1/5}E_{\mu}^n)
\left[(-\partial_{[\alpha}\bar\epsilon 2i\gamma_n\gamma_5\tau_{\beta}
\psi_{\gamma]I'})U^{I'}\;
_I\eta^I +\right.\nonumber\\
&&\left.\frac{1}{5}(\partial_{[\alpha}\bar\epsilon(-2\gamma_{nm}+3\delta_{nm})
(\tau_{\beta\gamma
]}\lambda_{J'K'L'})U^{J'}\;_JU^{K'}\;_KU^{L'}\;_L\eta_m^{JKL}\right]
\label{dalfa}
\eea
In the last term we use (\ref{spheric}) and get
\bea
&&-i\frac{\sqrt{3}}{4}2i \Delta^{-2/5} E_\mu^n
( \partial_{[\alpha}\bar\epsilon_{I'}\tau_\beta\psi_{\gamma ]})
U^{I'}\;_I U^{J'}\;_J C_n^{IJ} \nonumber\\
&&\!\!\!\! +\frac{i\sqrt{3}}{4}3 \Delta^{-2/5} E_\mu^n  \Omega^{IJ} C_n^{KL}
U^{I'}\;_I  U^{J'}\;_J U^{ K'}\;_K U^{L'}\;_L (\partial_{[\alpha}
\bar\epsilon_{I'} \tau_{\beta\gamma ]}\lambda_{J' K' L'})
\eea
Using (\ref{EC}) (and $U\tilde{\Omega}U^T=\tilde{\Omega}$) we obtain
\be
\frac{-i\sqrt{3}}{2}{\Pi_A}^j \partial_{[\alpha} \bar\epsilon_{I'}
(\gamma_j)^{I'J'}\tau_\beta \psi_{\gamma ] J'} \partial_\mu Y^A -
\frac{i\sqrt{3}}{4}
{\Pi_A}^j \partial_{[\alpha}\bar\epsilon_{I'}\tau_{\beta\gamma ]}
\lambda^j_{J'}\Omega^{I'J'}\partial_\mu Y^A\nonumber
\ee
This agrees with the d=7 result
\be
-i\frac{\sqrt{3}}{4}{\Pi_A}^j\partial_{[\alpha}(2\bar{\epsilon}
\tau_{\beta}\gamma^i\psi_{\gamma ]}+\bar{\epsilon}\tau_{\beta\gamma]}
\lambda^i)\partial_{\mu}Y^A\delta_{ij}\label{ss12}
\ee
This was the last term to be checked in the transformation law of the
3 index tensor field $S_{\alpha\beta\gamma ,A}$ with
$B_{\alpha}^{AB}$ set to zero, which therefore is in complete agreement
with the 11 dimensional transformation laws.

Now, we let the gauge fields to take non-zero values. Since the dependence on
the gravitini and gauge fields was already checked (see the analysis performed
for the susy law in the $F_{\alpha\beta\gamma\mu}$ sector), we will only check
the $F_{\alpha\beta} \lambda $ terms in the $S_{\alpha\beta\gamma}$
7-dimensional susy law. Hence we keep from the 11-dimensional susy variation of
$S_{\alpha\beta\gamma}$ only the terms with potential contribution to the
spin 1/2 and $F_{\alpha\beta}^{AB}$ dependent terms in the d=7 susy law:
\bea
-i\sqrt{6}\delta S_{\alpha\beta\gamma , A} Y^A|_{F\lambda\;terms}&=&
\frac{6}{\sqrt{2}}\epsilon_{ABCDE}F_{[\alpha\beta}^{AB} \delta (
B_{\gamma ]}^{CF})|_{\lambda\;terms} Y_F C_\mu^D \frac{(T\cdot Y)^E}{Y\cdot T
\cdot Y}\nonumber\\
&&-\frac{3i\sqrt{2}}{10}\Delta^{-1/5}\bar\epsilon\gamma_{mn}\tau_{[\gamma}
\gamma^p\psi_m F_{\alpha\beta ]}^{AB} V^{\nu}_{AB} E_\nu^n E_\mu^m
\nonumber\\
&&+3i\sqrt{2}\Delta^{-1/5}\bar\epsilon\tau_{[\gamma}
\gamma_{[n}\psi_{m ]}
F_{\alpha\beta ]}^{AB} V^\nu_{AB} E_\nu^n E_\mu^m\label{ss9}
\eea
After we substitute the ans\"atze of various 11-dimensional fields,
we need to evaluate the spherical harmonic of
$$\bar\eta^I\left(\frac{1}{5}\gamma_{mn}\gamma^p\psi_p +2\gamma_{[m}\psi_{n]}
\right)$$ which is $$3(\bar\eta^I \gamma_{mn}\eta^J \phi_5^{KL}+
2\bar\eta^I\gamma_{[m}\eta^J C_{n ]}^{KL}).$$
The contribution of the last two terms in (\ref{ss9}) is
\bea
&&-\frac{9i\sqrt{2}}{2}\Delta^{-1/5}\bar\epsilon_{I'}\tau_{[\gamma}\lambda_{I'J'K'}
U^{I'}_{\;\;I} U^{J'}_{\;\;J} U^{K'}_{\;\;K} U^{L'}_{\;\;L}
F_{\alpha\beta ]}^{AB} V^\nu_{AB} E_\nu^{[n} E_\mu^{m ]}
\nonumber\\
&&(\bar\eta^I\gamma_{mn}
\eta^J \phi_5^{KL} +2\bar\eta^I\gamma_{[m}\eta^J C_n^{KL})\nonumber\\
&&=\frac{3\sqrt{2}}{2Y\cdot T\cdot Y}F_{[\alpha\beta}^{DE}\bar\epsilon
\tau_{\gamma ]}\gamma^{[ij}\lambda^{k]}(\Pi_A^{\;\;i}\Pi_B^{\;\;j}
(-\Pi^{-1})_k^{\;\;C}+2\Pi_A^{\;\;i}\Pi_B^{\;\;k}(\Pi^{-1})_j^{\;\;C})
V_\nu^{DE} C_\mu^A C_\nu^B Y^C\nonumber\\\label{ss10}
\eea
Use now the identity
\be
\gamma^{[ij}\lambda^{k]}=\frac{1}{3}(\gamma^l\gamma^{ijk}\lambda_l-
\gamma^{lijk}\lambda_l)
\ee
to rewrite (\ref{ss10})
\bea
&&-\sqrt{2}F_{[\alpha\beta}^{DB}
(\bar\epsilon\tau_{\gamma ]}\gamma^l\gamma^{ijk}
\lambda_l+\bar\epsilon\tau_{\gamma ]}\gamma^{ijkl}\lambda_l)
\Pi_A^{\;\;i}\Pi_B^{\;\;j}(\Pi^{-1})_k^{\;\;C}C_\mu^A Y_C Y_D\nonumber\\
&&=\frac{6}{\sqrt 2}\epsilon_{ABCDE} (F_{[\alpha\beta}\cdot Y)^A
\delta(B_{\gamma ]}^{BC}) C_\mu^D \frac{(T\cdot Y)^E}{Y\cdot T\cdot Y}
\label{ss11}
\eea
where to reach the final result in (\ref{ss11}) we replaced $\gamma^{ijk}$
by $-1/2 \epsilon^{ijkmn} \gamma_{mn}$ and we also used the property of
the scalar fields $\Pi_A^{\;\;m}$ of having determinant one.

The first term in (\ref{ss9}) can be written as the sum of two terms upon
using the Schouten identity to release the variation $\delta(B_{\gamma}^{CF})$
from its contraction with $Y_F$.

One of these terms cancels precisely
the contribution of (\ref{ss11}), and the other, $(F_{[\alpha\beta}
\cdot Y)^A \delta B_{\gamma ]}^{BC} C_\mu^D Y^E\epsilon_{ABCDE}$ will
give the 7-dimensional susy variation $\delta(d=7)\\S_{\alpha\beta\gamma , A}
|_{\lambda F\; terms}Y^A$. Since the susy variation of a gauge invariant object
must be gauge invariant, we will not show explicitely the cancelation of terms
produced by the 11-dimensional susy variation which are proportional with
bare $B_\alpha^{AB}$ gauge fields.
The same argument implies that the partial derivative $\partial_\alpha$
which we obtained in (\ref{ss12}) is in fact, a gauge covariant $D_\alpha$
derivative. (That this is the case, we already know from the analysis of
the sector $F_{\alpha\beta\gamma\mu}$ where we derived the gravitino
dependent terms in (\ref{ss12}), and we obtained in fact a gauge covariant
derivative).

Let us now turn back to $\delta\psi_{\alpha}$. In
7 dimensions, we have:
\bea
\delta\psi_{\alpha I'}|_{S=B=0}&=&D_{\alpha}\epsilon_{I'}
+\frac{1}{4}
{(\Pi^{-1})_i}^A(\partial_{\alpha}{\Pi_A}^j)(\gamma^i\;_j)_{I'}\;^{J'}
\epsilon_{J'}\nonumber\\
&&-\frac{1}{20}{(\Pi^{-1})_i}^A{(\Pi^{-1})_i}^A\tau_{\alpha}\epsilon_{I'}
\eea
But  in 11 dimensions we have got
\be
\delta\Psi_{\Lambda}|_{B=S=0}=\partial_{\Lambda}\varepsilon+\frac{1}{4}
\Omega_{\Lambda}\;^{MN}(E)\Gamma_{MN}|_{B=0}\varepsilon+
F_{\cdot\cdot\cdot\cdot} terms +3{\rm -fermi\;\; terms}
\ee
By direct substitutions one obtains
\bea
\Omega_{\alpha}\;^{MN}\Gamma_{MN}|_{B=0}&=&\omega_{\alpha}\;^{ab}\tau_{ab}
+\gamma_{mn}E^{m\mu}\partial_{\alpha}E_{\mu}^n\nonumber\\
&&-2\tau_{\alpha}\gamma_m\gamma_5(\Delta^{-1/5}E^{m\mu})\Delta^{1/5}(\partial_{
\mu}\Delta^{-1/5})\nonumber\\
&&-2\tau^{\beta}\;_{\alpha}(\Delta^{1/5}\partial_{\beta}\Delta^{-1/5})
\nonumber\\
\Omega_{\mu}\;^{MN}(E)\Gamma_{MN}|_{B=0}&=&\gamma_{mn}E^{m\nu}(2\partial_{
[\mu}E_{\nu]}^n-E^{n\rho}E_{\mu}^p\partial_{\nu}E_{p\rho})\nonumber\\
&&+\tau^{\alpha}\gamma_m\gamma_5[\Delta^{1/5}\partial_{\alpha}E_{\mu}^m
+(\Delta^{1/5}E_{\mu}^n)(E^{m\rho}\partial_{\alpha}E_{n\rho})]
\eea
For the terms involving $F_{\mu\nu\rho\sigma}$ and $F_{\alpha\mu\nu\rho}$,
we find
\bea
&&E_a^{\Lambda}\delta\Psi_{\Lambda}|_{F_{\mu\nu\rho\sigma} term, B=S=0}
=\frac{3}{288}4!\Delta^{-1}\tau_a\varepsilon\left[1+
\frac{1}{3}\left(\frac{T}{Y_A Y_B T^{AB}}-5\right)\right.\nonumber\\&&
-\frac{2}{3}\left. \left(\frac{Y_A (T^2)^{AB} Y_B}{(Y_A T^{AB} Y_B)^2}-1\right)
\right]
-\frac{4\cdot 3!}{288}\Delta^{-1}({\tau^{\alpha}}_a+2e^{\alpha
}_a)\gamma_q\gamma_5\varepsilon
(\Delta^{1/5}E_{\sigma}^q)C^\sigma_A\nonumber\\
&&\left(\frac{\partial_\alpha T^{AB} Y_B}{Y_A T^{AB} Y_B}-
\frac{T^{AB} Y_B}{(Y_A T^{AB} Y_B)^2} (Y_C \partial_\alpha T^{CD} Y_D)\right)\\
&&E_m^{\Lambda}\delta\Psi_{\Lambda}|_{F_{\mu\nu\rho\sigma} term, B=S=0}=
-\frac{3}{288}3!\cdot 8\Delta^{-1}\gamma_m\gamma_5\varepsilon
\left[1+
\frac{1}{3}\left(\frac{T}{Y_A Y_B T^{AB}}-5\right)\right.\nonumber\\
&&-\frac{2}{3}\left. \left(\frac{Y_A (T^2)^{AB} Y_B}{(Y_A T^{AB}
Y_B)^2}-1\right)\right]
-\frac{1}{4}\Delta^{1/5}(E_{\sigma m}-2E_{\sigma}^r\gamma_{mr})\tau^{\alpha}
C^\sigma_A\nonumber\\&&
\left(\frac{\partial_\alpha T^{AB} Y_B}{Y_A T^{AB} Y_B}-
\frac{T^{AB} Y_B}{(Y_A T^{AB} Y_B)^2} (Y_C \partial_\alpha T^{CD} Y_D)\right)
\eea
We also have that
\be
\delta\psi_{\alpha}|_{S=B=0}=\Delta^{-1/10}\gamma_5^{-1/2}e_{\alpha}^a
[E_a^{\Lambda}\delta\Psi_{\Lambda}+\frac{1}{5}\tau_a\gamma_5\gamma_m E_m^{
\Lambda}\delta\Psi_{\Lambda}]|_{B=S=0}
\ee
Putting everything together, we find, besides the usual term
$\partial_{\alpha}
\epsilon_{I'}+\omega_{\alpha}\;^{ab}\tau_{ab}\epsilon_{I'}$, terms
with $\tau^{\beta}\;_{\alpha}$, terms with $\tau_{\alpha}$, and terms
without any $\tau_{\alpha}$'s. After some algebra, the terms with $\tau^{\beta}
\;_{\alpha}$ cancel, as they should, whereas the terms without any $\tau
_{\alpha}$'s simplify to
\bea
&&\frac{1}{4}\gamma_{mn}E^{m\mu}(\partial_{\alpha}E_{\mu}^n)\epsilon
+\epsilon_{I'}(\partial_{\alpha}U^{I'}\;_I)\eta^I
+\frac{3i}{4}\gamma_m(\Delta^{1/5}E_{\sigma}^m)\epsilon \Delta^{-6/5}
C^\sigma_A\nonumber\\&&
\left(\frac{\partial_\alpha T^{AB} Y_B}{Y_A T^{AB} Y_B}-
\frac{T^{AB} Y_B}{(Y_A T^{AB} Y_B)^2} (Y_C \partial_\alpha T^{CD} Y_D)\right)
\label{term1}
\eea
to be compared to
\be
-\frac{1}{4}{(\Pi^{-1})_i}^A(\partial_{\alpha}{\Pi_A}^j)
(\gamma^i\;_j)^{J'}\;_{I'}\epsilon_{J'
}U^{I'}\;_I\eta^I\label{compare}
\ee

We will first of all work on the first term in (\ref{term1}). Using the
Fierzing relation (\ref{fierz}) twice for each $\gamma$ matrix acting on
$\eta^I$, we prove that
\be
C^{mB}C^{nD}\gamma_{mn}\eta^I=\left[{(\gamma_{DB})^I}_J+Y_B{(\gamma_D)^I}_J
\gamma_5-Y_D{(\gamma_B)^I}_J\gamma_5\right]\eta^J\label{ccgamn}
\ee
Using (\ref{vielsim}) and (\ref{vielinvsim}) to rewrite the vielbeins, the
previous relation to get rid of $\gamma_{mn}$ and (\ref{gg}) and
(\ref{urel}) to get rid of the resulting traces we obtain, after some
algebra,
\bea
\frac{1}{4}E^{m\mu}\partial_{\alpha}E_{\mu}^n\gamma_{mn}\epsilon&=&
-\partial_{\alpha}{U^{I'}}_I\eta^I\epsilon_{I'} -\frac{1}{4}{\Pi_A}^i
\partial_{\alpha}{(\Pi^{-1})_j}^A{(\gamma_{ij})^{I'}}_{J'}{U^{J'}}_I
\eta^I\epsilon_{I'}\nonumber\\
&&-\frac{1}{2}Y_A\partial_{\alpha}(\Delta^{3/5}{(\Pi^{-1})_i}^A)
{(\gamma_i)^{I'}}_{J'}{U^{J'}}_I\gamma_5\eta^I\epsilon_{I'}\nonumber\\
&&+\frac{1}{2}{\Pi_A}^{[i}\partial_{\alpha}{(\Pi^{-1})_{j]}}^A
{(\Pi^{-1})_j}^BY_B\nonumber\\&&
\Delta^{3/5}{(\gamma^j)^{I'}}_{J'}{U^{J'}}_{I}
\gamma_5\eta^I\epsilon_{I'}\label{someterm}
\eea
The first term cancels against the second term in (\ref{term1}), and the
second term gives the correct 7 dimensional result in (\ref{compare}). That
means that the last two terms, together with the last term in (\ref{term1})
should give zero.

Using that $\Delta^{-6/5}=Y\cdot T\cdot Y$ and the identity
\bea
\Delta^{6/5}\partial_{\alpha}(Y\cdot T\cdot Y)T^{AD}Y_D-\partial_{\alpha}
T^{AD}Y_D&=&\Delta^{6/5}\partial_{\alpha}(Y\cdot T\cdot Y)T^{BD}Y_D(\delta^{AB}
-Y_AY_B)\nonumber\\&&-\partial_{\alpha}T^{BD}Y_D(\delta^{AB}-Y^AY^B)
\eea
and rewrite the last two terms in (\ref{someterm}) as
\bea
&&
\Delta^{-3/5}(\delta^{AB}-Y^AY^B)\frac{1}{4Y\cdot T\cdot Y}\left[
\frac{\partial_{\alpha}(Y\cdot T\cdot Y)}{Y\cdot T\cdot Y}T^{BD}Y_D
\nonumber\right.\\&&
-\left.\partial_{\alpha}T^{BD}Y_D\right] {\Pi_A}^i{(\gamma_i)^{I'}}_{J'}
{U^{J'}}_{I}\gamma_5\eta^I\epsilon_{I'}\label{good}
\eea
Now taking the last term in (\ref{term1}), substituting for $E_{\sigma}^m $
from (\ref{vielsim}) , using the summation relation (\ref{gg}) and the
Fierzing relation
\be
C^{mB}\gamma_m\eta^I =-i[{(\gamma_B)^I}_J\gamma_5\eta^J+Y_B\eta^I]
\label{fierz1}
\ee
which we can prove by using (\ref{fierz}), we get the same result
as in (\ref{good}), but with the opposite sign, as we should. So all the
extra contributions cancel and we are left with the 7 dimensional result
(\ref{compare}). Note that at this moment if we chose the  a=0 solution
for (\ref{calS}), it will not work. So, as promised, the ansatz for the
scalar field dependence of $F_{\mu\nu\rho\sigma}$ is fixed by the consistency
of the gravitino transformation law.

The hardest term in $\delta\psi_{\alpha}$
is the $\tau_{\alpha}$ term, which becomes (after some algebra)
\bea
&&\tau_{\alpha}(\Delta^{-1/5}E^{m\mu}\gamma_m)[\frac{i}{50}(\Delta^{-1}
\partial_{\mu}\Delta)\epsilon +\frac{1}{10}\sirc{\gamma}_{\mu}\epsilon
-\frac{i}{5}\epsilon_{I'}(\partial_{\mu} U^{I'}\;_I)\eta^I]\nonumber\\
&&-\frac{3}
{20}\Delta^{-6/5}\tau_{\alpha}\left[1+
\frac{1}{3}\left(\frac{T}{Y_A Y_B T^{AB}}-5\right)-
\frac{2}{3} \left(\frac{Y_A (T^2)^{AB} Y_B}{(Y_A T^{AB} Y_B)^2}
-1\right)\right]\epsilon\nonumber\\
&&-\frac{i}{20}\tau_{\alpha}\left(\Delta^{-1/5}E^{m\mu}E^{n\rho}
(\sirc D_\mu
E_{\rho}^q)(\gamma_{mnq}-2\delta_{mq}\gamma_n)\right)\epsilon
\label{term2}
\eea
to be compared with
\be
\frac{1}{20}{(\Pi^{-1})_i}^A{(\Pi^{-1})_i}^A\tau_{\alpha}\epsilon\label{tterm}
\ee

This last problem turns out to be surprisingly complicated. We have only
partial results; they seem to involve the explicit expression of the matrix $U$
and we prefer to devote a separate paper to this issue. The same holds for
the scalar dependent terms in the susy law of the 7 dimensional spin 1/2
fields.

\section{Bosonic equations of motion and action}
In this section we will do some further checks on our ansatze by looking at
the bosonic action and equations of motion, and reproducing some of the
corresponding equations of motion and terms in the action in seven
dimensions.

First, we give the 11 dimensional bosonic equations of motion following
from the action in (\ref{11action}):
\bea
R_{\Lambda\Pi}&=&-\frac{1}{6}( F_{\Lambda\Lambda_1\Lambda_2\Lambda_3
}{ F_{\Pi}}^{\Lambda_1\Lambda_2\Lambda_3}-\frac{1}{12}G_{\Lambda\Pi}
 F^2)\label{einstein}\\
\partial_{\Lambda}(E F^{\Lambda\Lambda_1\Lambda_2\Lambda_3})
&=&\frac{k}{\sqrt{2} (24)^2}
\epsilon^{\Lambda_1...\Lambda_{11}} F_{\Lambda_4
...\Lambda_7} F_{\Lambda_8...\Lambda_{11}}\label{maxwell}\\
B_{\Lambda\Pi\Sigma\Omega}&=&0\label{auxiliary}
\eea

In seven dimensions we have the bosonic action
\bea
e^{-1}{\cal L}'_{7d}
&=&-\frac{1}{2}R+\frac{1}{4}m^2(T^2-2T_{ij}T^{ij})-\frac{1}{2}
P_{\alpha ij}P^{\alpha ij}-\frac{1}{4}({\Pi_A}^i{\Pi_B}^j
F_{\alpha\beta}^{AB})^2
\nonumber\\
&+&\frac{1}{2}({{\Pi^{-1}}_i}^AS_{\alpha\beta\gamma ,A})^2
+\frac{1}{48}me^{-1}\epsilon^{\alpha\beta\gamma\delta\epsilon\eta\zeta}
\delta^{AB}S_{\alpha\beta\gamma ,A}F_{\delta\epsilon\eta\zeta ,B}
+\frac{m^{-1}}{8}e^{-1}\Omega_5[B]
\nonumber\\
&-&\frac{m^{-1}}{16}e^{-1}\Omega_3[B]
+\frac{ie^{-1}}{16\sqrt{3}}\epsilon^{\alpha\beta\gamma\delta\epsilon\eta\zeta}
\epsilon_{ABCDE}\delta^{AG}S_{\alpha\beta\gamma, G}F_{\delta\epsilon}^{BC}
F_{\eta\zeta}^{DE}
\label{bosonic}
\eea
For the equations of motion, we will put the gauge field $B_{\alpha}^{AB}$
to zero, so in the action we need to keep only terms at most linear
in $B_{\alpha}^{AB}$ which we can rewrite as follows.
\bea
e^{-1}{\cal L}'_{7d}
&=&-\frac{1}{2}R+\frac{1}{4}m^2(T^2-2T_{ij}T^{ij})-\frac{1}{2}
P_{\alpha ij}P^{\alpha ij}
\nonumber\\
&+&\frac{1}{2}({{\Pi^{-1}}_i}^AS_{\alpha\beta\gamma ,A})^2
+\frac{1}{48}me^{-1}\epsilon^{\alpha\beta\gamma\delta\epsilon\eta\zeta}
\delta^{AB}S_{\alpha\beta\gamma ,A}F_{\delta\epsilon\eta\zeta ,B}
\nonumber\\&=&-\frac{1}{2}R+\frac{1}{4}m^2(T^2-2T_{AB}T^{AB})
+\frac{1}{8}Tr(\partial_{\alpha}T^{-1}\partial^{\alpha}T)\nonumber\\
&-&\frac{1}{2}B_{\alpha}^{AB}(T\partial_{\alpha}T^{-1})_{AB}+\frac{1}{2}
T^{AB}S_{\alpha\beta\gamma, A}{S^{\alpha\beta\gamma}}_{B}\nonumber\\
&+&\frac{1}{48}me^{-1}\epsilon^{\alpha\beta\gamma\delta\epsilon\eta\zeta}
\delta^{AB}S_{\alpha\beta\gamma ,A}F_{\delta\epsilon\eta\zeta ,B}
\label{bosonic7}
\eea
We notice that the action is now expressed in terms of $T_{AB}$ instead of
${\Pi_A}^i$, so we obtain the following bosonic equations of motion
for $T_{AB}$, $B_{\alpha}^{AB}$ and $S_{\alpha\beta\gamma ,A}$ and the
metric $g_{\alpha\beta}$, respectively
\bea
&&\frac{1}{2}
m^2(T\delta_{AB}-2 T_{AB})+\frac{1}{4}(T^{-1}\partial_{\alpha}(\partial
_{\alpha}T T^{-1}))_{AB}+\frac{1}{2}S_{\alpha\beta\gamma ,A}
{S^{\alpha\beta\gamma}}_B\nonumber\\&&
-\frac{1}{5}T^{-1}_{AB}[\frac{1}{2}
m^2(T^2-2TrT^2)+\frac{1}{4}\partial_{\alpha}Tr(\partial_{\alpha}T T^{-1})+
\frac{1}{2}S_{\alpha\beta\gamma,A}{S^{\alpha\beta\gamma}}_BT^{AB}]
=0\label{scalarul}\\
&&-\frac{1}{12}me^{-1}\epsilon^{\alpha\beta\gamma\delta\epsilon\eta\zeta}
S_{\beta\gamma\delta ,A}S_{\epsilon\eta\zeta ,B}-\frac{1}{2}(T\partial_
{\alpha}T^{-1})_{[AB]}=0\label{gaugeul}\\
&&T^{AB}S_{\alpha\beta\gamma ,B}+\frac{1}{24}me^{-1}
\epsilon^{\alpha\beta\gamma\delta\epsilon\eta\zeta}
F_{\delta\epsilon\eta\zeta ,A}=0\label{antisimetricul}\\&&
R^{(7)}_{\alpha\beta}-\frac{1}{10}g_{\alpha\beta}
[m^2(T^2-2T_{AB}^2)-4T^{AB}S_{\alpha '\beta '\gamma ' ,A}{S^{\alpha '\beta '
\gamma '}}_B]\nonumber\\
&&-\frac{1}{4}Tr(\partial_{\alpha}T^{-1}\partial_{\beta}T)-3T^{AB}S_{\alpha
\gamma\delta ,A}{S^{\beta\gamma\delta }}_B=0\label{einsteinul}
\eea
We note here that in (\ref{scalarul}) we have used Lagrange multipliers for
the condition $\det T_{AB}=1$.
From (\ref{scalarul}) we obtain also
\be
\partial_{\alpha}(\partial_{\alpha}T T^{-1})_{[AB]}=0
\label{a-sim}
\ee
and from (\ref{a-sim}), together with (\ref{gaugeul}) we also get that
\be
\epsilon^{\alpha\beta\gamma\delta\epsilon\eta\zeta}
\delta^{AB}S_{\alpha\beta\gamma ,[A}\partial_{\delta}S_{\epsilon\eta\zeta ,B]}
=0\label{zeroul}
\ee
Now we want to see that we reproduce these equations from the 11 dimensional
equations of motion. First, we notice that (\ref{auxiliary})
was already discussed. The only nontrivial component is the $\{\alpha
\beta\gamma\delta \}$, which is just the equation of motion of the
antisymmetric tensor, $S_{\alpha\beta\gamma, A}$. Let's look now at
(\ref{maxwell}), setting the gauge field to zero.

Substituting the ansatze for $F_{\Lambda\Pi\Sigma\Omega}$ and $g_{\Lambda\Pi}$
the $\{ \nu\rho\sigma \}$ component of (\ref{maxwell}) becomes
\bea
&&Y_{(A}\partial_{\mu}Y_{B)}T_{CB}\left[2(T\delta_{AC}-2T_{AC})
+(T^{-1}\partial^{\alpha}(\partial_{\alpha}T T^{-1}))_{AC}+2 S_{\alpha
\beta\gamma ,A}S^{\alpha\beta\gamma ,C}\right]\nonumber\\
&&-Y_{(A}\partial_{\mu}Y_{B)}S_{\alpha\beta\gamma ,A}(S_{\alpha
\beta\gamma ,C}T_{CB}+\frac{1}{6}m e^{-1}
\epsilon^{\alpha\beta\gamma\delta\epsilon\eta\zeta}
\partial_{\delta}S_{\epsilon\eta\zeta ,B})\nonumber\\
&&-\frac{1}{3}Y^{[A} \partial_{\mu}Y^{B]}
me^{-1}\epsilon^{\alpha\beta\gamma\delta\epsilon\eta\zeta}
S_{\alpha\beta\gamma ,A}\partial_{\delta}S_{\epsilon\eta\zeta ,B}
=0
\eea
We notice that the first line is the scalar equation of motion
(\ref{scalarul}), the second line is the antisymmetric tensor equation
of motion in (\ref{antisimetricul})and the third is (\ref{zeroul}) which is
a combination of the scalar and gauge field equations of motion.

Similarly, by substituting the ansatze for $F_{\Lambda\Pi\Sigma\Omega}$ and
$g_{\Lambda\Pi}$, the $\{ \nu\rho\alpha \}$ component of (\ref{maxwell})
becomes
\be
\partial_{[\mu}Y^A\partial_{\sigma ]}Y^B[(T^{-1}\partial_{\alpha}T)_{BA}
+\frac{1}{6}\epsilon^{\alpha_1...\alpha_6\alpha}S_{\alpha_1\alpha_2\alpha_3 ,A}
S_{\alpha_4\alpha_5\alpha_6 ,B}]=0
\ee
which is just the equation of motion for the gauge field, (\ref{gaugeul}).
The $\{ \beta\gamma\delta \}$ component of (\ref{maxwell}) is
\bea
&&Y^A [4\partial^{\delta}\partial_{[\delta}S_{\alpha\beta\gamma] ,A}+
\frac{1}{6}\epsilon^{\alpha\beta\gamma\alpha_1 ...\alpha_4}(\partial_{\alpha_1}
T^{AB})S_{\alpha_2\alpha_3\alpha_4 ,B}-S_{\alpha\beta\gamma ,A}T^{2 AB}Y_B]
\nonumber\\
&&+\frac{Y^A}{Y\cdot T\cdot Y}[-4\partial_{[\delta}S_{\alpha\beta\gamma] ,A}
Y\cdot \partial_{\delta}\cdot Y +\frac{1}{6}\epsilon^
{\alpha\beta\gamma\alpha_1 ...\alpha_4}S_{\alpha_2\alpha_3\alpha_4 ,B}T^{AB}
(Y\cdot \partial_{\alpha_1}T\cdot Y)]\nonumber\\
&&-\left(T-2\frac{Y\cdot T^2\cdot Y}{Y\cdot T\cdot Y}\right)Y^A
\left[\frac{1}{6}
\epsilon^{\alpha\beta\gamma\alpha_1 ...\alpha_4}\partial_{\alpha_1}S_{\alpha_2
\alpha_3\alpha_4 ,A}+ S_{\alpha\beta\gamma , B}T^{AB}\right]=0
\eea
where all three lines are now zero due to the antisymmetric tensor equation
of motion, (\ref{antisimetricul}). The $\{ \alpha\beta\mu \}$ component of
(\ref{maxwell}) gives
\bea
&&C_{\mu}^A\left\{ \left(T_{AC}-\frac{T^{AD}Y_D T^{CE}Y_E}{Y\cdot T\cdot Y}
\right)\left(\partial_{\gamma}S_{\alpha\beta\gamma ,B}\delta^{BC}+
S_{\alpha\beta\gamma ,B}(\partial_{\gamma}T_{BD})T^{-1}_{DC}\right)\right.
\nonumber\\&&\left.
-(S_{\alpha\beta\gamma ,C}T^{BC}+\frac{1}{6}
\partial_{\alpha_2}S_{\alpha_3\alpha_4
\alpha_5 ,B}\epsilon^{\alpha\beta\alpha_1 ...\alpha_5})
\right.\nonumber\\&&\left.
\frac{1}{Y\cdot T\cdot Y
}\left(\partial_{\gamma}T^{AD}Y_D Y_B-T^{AD}Y_DY_B\frac{Y\cdot \partial_{
\gamma}T\cdot Y}{Y\cdot T\cdot Y}\right)\right\}
\eea
In the first line, the second bracket is zero upon using the equation of motion
for the 7d antisymmetric tensor, (\ref{antisimetricul}), to convert both
$S_{\alpha\beta\gamma ,B}$'s into $\partial_{\alpha_1}S_{\alpha_2\alpha_3
\alpha_4 ,C}T^{-1}_{BC}$, whereas the second line gives directly the
antisymmetric tensor equation.

The Einstein's equations (\ref{einstein}) are considerably more involved.
We have computed all the terms for zero gauge field, but the task of
reconstructing the seven dimensional field equations is quite laborious, and
we have not completed it, but we can see that we get nontrivial combinations
of these seven dimensional field equations. The fact that already  from
the antisymmetric field equation (\ref{maxwell}) we get all the seven
dimensional field equations (except Einstein's equation) is already quite
nontrivial. Moreover, the eleven dimensional Einstein's equation for
$\Lambda\Pi=\alpha\beta$ contains the seven dimensional Einstein's equation,
(\ref{einsteinul}), together with many more terms. So we can say that the
seven dimensional bosonic equations of motion solve the eleven dimensional
equations of motion in all the sectors we have checked. We leave the complete
check of the Einstein's equations to the diligent reader.

Let's now turn to the bosonic action in (\ref{bosonic}).
 Part of the action was already calculated.
 We have also reproduced the scalar field potential, and this
was how we fixed the dependence of $F_{\Lambda\Pi\Sigma\Omega}$
on the scalars ${\Pi_A}^i$. The condition that the Einstein action in 11
dimensions gives us the Einstein action in 7 dimensions  gave us the ansatz for
the 11 dimensional vielbein  component $E_{\alpha}^a$. Incidentally,
we note also that we
found the kinetic terms for the gravitini $\psi_{\alpha}^{I'}$ and spin 1/2
fermions $\lambda^i_{I'}$ (without the Q-connection piece).
 By requiring that we reproduce them we have
fixed the ansatz for the components of the 11 dimensional gravitino.
So we need to reproduce the scalar field and
gauge field kinetic terms, $-\frac{1}{2}
P_{\alpha ij}P^{\alpha ij}-\frac{1}{4}({\Pi_A}^i{\Pi_B}^j
F_{\alpha\beta}^{AB})^2$, and also the $S_{\alpha\beta\gamma ,A}$ terms. The
$S_{\alpha\beta\gamma ,A}$ terms give us the mechanism for 'self-duality
in odd dimensions' at the level of the action (we have deduced this
mechanism from the
susy laws).

Let us first look at the $P_{\alpha ij}P^{\alpha ij}$ term. It is equal to
\be
+\frac{1}{8}Tr(\partial_{\alpha}T^{-1}\partial^{\alpha}T)
-\frac{1}{2}B_{\alpha}^{AB}(T^{-1}\partial_{\alpha}T)_{AB}
-\frac{1}{4}(T_{AC}T^{-1}_{BD}B_{\alpha}^{AB}B_{\alpha}^{CD}
-(B_{\alpha}^{AB})^2)\label{palfa}
\ee
We will look only at the first two terms, the last two being inferred as
being the gauge invariant completion (already the term linear in B should
be dictated by gauge invariance, so this will be a nontrivial check on the
algebra).
The contributions to this term come from the 11 dimensional Einstein
action and the Maxwell term, $-1/48 {\cal F}_{\Lambda\Pi\Sigma\Omega}^2$.

The contributions to $P_{\alpha ij}P^{\alpha ij}$ from
$-1/48 {\cal F}_{\Lambda\Pi\Sigma\Omega}^2$ come from the components
\bea
-\frac{1}{48}\times 4E&&\left[ F_{\alpha\mu\nu\rho} F_{\alpha '\mu '\nu '
\rho '}G^{\alpha\alpha '}G^{\mu\mu '}G^{\nu\nu '}G^{\rho\rho '}\right.
\nonumber\\&&\left.
+2 F_{\alpha\mu\nu\rho}F_{\sigma ' \mu '\nu '\rho '}G^{\alpha\sigma '}
G^{\mu\mu '}G^{\nu\nu '}G^{\rho\rho '}\right. \nonumber\\&&\left.
+\frac{1}{4}F_{\mu\nu\rho\sigma}F_{\mu '\nu '\rho '\sigma '}
G^{\mu\nu '}G^{\nu \nu '}G^{\rho\rho '}G^{\sigma\sigma '}\right]
\label{wow}
\eea
We note that the metric $G^{\mu\nu}$ gives also a gauge field dependence:
\bea
G^{\mu\nu}&=&g^{\mu\nu}+\Delta^{2/5}B_{\alpha}^{\mu}B_{\alpha}^{\nu}
\nonumber\\
&=&
\Delta^{2/5} \left( C^\mu_A C^\nu_B T^{AB} Y_C Y_D T^{CD}
-C^\mu_A Y_B T^{AB} C^\nu_C Y_D T^{CD}+B_{\alpha}^{\mu}B_{\alpha}^{\nu}
\right)\label{metricul}
\eea
but since it already has two gauge fields it can contribute only in the
last term in (\ref{wow}).
The first term contributes (after substituting the ansatze for the various
fields)
\be
\frac{V_4}{4}\left[ \frac{1}{7}Tr(\partial_{\alpha}T^{-1}\partial^{
\alpha}T)+\frac{1}{7\cdot 5}(Tr(T^{-1}\partial_{\alpha}T))^2
+\frac{12}{5}B_{\alpha}^{AB}(T^{-1}\partial_{\alpha }T)_{AB}\right]
\ee
whereas the second term contributes
\be
+ \frac{4}{5\cdot 7}B_{\alpha}^{AB}(T^{-1}\partial_{\alpha}T
)_{AB}
\ee
Finally, the third term in (\ref{wow}) contributes only to the $B^2$ term in
$P_{\alpha ij}^2$, so we will neglect it. We 'see' that the two contributions
add up to the combination required by gauge invariance,
\be
\frac{V_4}{4\cdot 7}\left[ Tr(\partial_{\alpha}T^{-1}\partial^{
\alpha}T)-8B_{\alpha}^{AB}(T^{-1}\partial_{\alpha }T)_{AB}\right]
+\frac{1}{7\cdot 5\cdot 4}(Tr(T^{-1}\partial_{\alpha}T))^2
\ee
where the last term is gauge invariant by itself (we can freely replace
$\partial_{\alpha}$ with $\bigtriangledown_{\alpha}$).

One would think that we need to add also the term $
F_{\alpha\mu\nu\rho} F_{\alpha '\mu '\nu '
\rho '}G^{\alpha\mu '}G^{\mu\alpha '}G^{\nu\nu '}G^{\rho\rho '}$ which has also
two gauge fields and scalars, however it has both two gauge fields and two
derivatives on scalars, so it should cancel with similar terms coming from the
11 dimensional Einstein action. For the same reason,
 the gauge field contribution of $G^{\mu\nu} $ to the first term in (\ref{wow})
has not been taken into account.

The Einstein action $\int -\frac{1}{2}R$, gives for the term with
two derivatives on scalars and no gauge fields
\be
-\frac{1}{5\cdot 7\cdot 4}Tr(T^{-1}\partial_{\alpha}T)^2+\frac{5}{
7\cdot 8}Tr (\partial_{\alpha}T\partial^{\alpha}T^{-1})
\ee
But we know that the metric is gauge invariant, and so we expect that $\int R$
is a gauge invariant object too, that means that we can complete the previous
terms in a gauge invariant way. Then the sum of the $\int F^2$ and $\int R$
contributions reproduces (\ref{palfa}).

Next we try to reproduce the kinetic term for the gauge field, $
-\frac{1}{4}({\Pi_A}^i{\Pi_B}^j F_{\alpha\beta}^{AB})^2$. It has also
contributions from the Einstein action and from the antisymmetric tensor
kinetic term, more precisely the component
\be
-\frac{1}{48}EF_{\mu\nu\alpha\beta}F_{\mu '\nu '\alpha '\beta '}G^{\mu\mu '
}G^{\nu\nu '}G^{\alpha\alpha '}G^{\beta\beta '}
\ee
which gives
\be
-\frac{3V_4}{20}T^{-1}_{AC}T^{-1}_{BD}F_{\alpha\beta}
^{AB}F_{\alpha\beta}^{CD}
\ee
We note that if we didn't have any scalar field dependence in
$F_{\mu\nu\alpha\beta}$, we would get a term of the type
\be
\left[ 4T^{-1}_{AC}T^{-1}_{BD}F_{\alpha\beta}
^{AB}F_{\alpha\beta}^{CD}+ F_{\alpha\beta}^{AC}F_{\alpha\beta}^{CB}
(T^{-1}_{AB}-TrT^{-1} T^{-1}_{AB})\right]
\ee

From the Einstein action, if we look only at $(\partial  B)^2$ terms, we get
\be
-\frac{V_4}{10}4\int T^{-1}_{AC}T^{-1}_{BD}\partial_{\alpha} B_{\beta}^{AB}
\partial_{[\alpha}B_{\beta ]}^{CD}
\ee
In the same manner as before, we assume that we can complete this term
in a gauge invariant manner to an $F^2$ term. When we add the two
contributions, we get the correct term,
\be
-\frac{V_4}{4}T^{-1}_{AC}T^{-1}_{BD}F_{\alpha\beta}
^{AB}F_{\alpha\beta}^{CD}
\ee

Finally, let's come to the S  terms in the bosonic action. They come
from the $F^2, \epsilon FFA$ and ${\cal B}^2$ terms in the 11 dimensional
action. We will treat first the terms with no $F_{\alpha\beta}^{AB}$,
and afterwards
the term with  $F_{\alpha\beta}^{AB}$, namely $\epsilon^{\alpha\beta\gamma
\delta\epsilon\eta\zeta}\epsilon^{ABCDE} S_{\alpha\beta\gamma ,A}F_{\delta
\epsilon}^{BC}F_{\eta\zeta}^{DE}$.

The $F^2$ terms contributing to the part with no $F_{\alpha\beta}^{AB}$ are
\bea
&&-\frac{1}{48 }E [ 4F_{\mu\alpha\beta\gamma}F_{\mu '\alpha '\beta '\gamma '}
(G^{\mu\mu '}G^{\alpha\alpha '}G^{\beta\beta '}G^{\gamma\gamma '}
-3G^{\mu\alpha '}G^{\mu '\alpha} G^{\beta\beta '}G^{\gamma\gamma '})
\nonumber\\&&+F_{\alpha\beta\gamma\delta}F_{\alpha '\beta '\gamma '\delta '}
G^{\alpha\alpha '}G^{\beta\beta '}G^{\gamma\gamma '} G^{\delta \delta '}
+F_{\mu\alpha\beta\gamma}F_{\delta '\alpha '\beta '\gamma '}G^{\mu\delta '}
G^{\alpha\alpha '}G^{\beta\beta '}G^{\gamma\gamma '}]
\eea
When we substitute the ansatz for the $S_{\alpha\beta\gamma ,A}$ dependence
of F and the ansatz for the metric ($G^{\mu\nu}$ from (\ref{wow})), we
get
\be
\frac{2}{5}[S_{\alpha\beta\gamma ,A}{S^{\alpha\beta\gamma}}_BT_{AB}
+T^{-1}_{AB}(\bigtriangledown_{[\alpha}S_{\beta\gamma\delta ] ,A}
\bigtriangledown^{[\alpha}{S^{\beta\gamma\delta ] }}_B]
\ee
The ${\cal B}^2$ terms give
\bea
&&\frac{1}{5}(\frac{6k}{5})^2[S_{\alpha\beta\gamma ,A}
{S^{\alpha\beta\gamma}}_BT_{AB}
-4T^{-1}_{AB}(\bigtriangledown_{[\alpha}S_{\beta\gamma\delta ] ,A}
\bigtriangledown^{[\alpha}{S^{\beta\gamma\delta ] }}_B
\nonumber\\&&
+\frac{1}{3}\epsilon^{\alpha\beta\gamma\delta\epsilon\eta\zeta}
S_{\epsilon\eta\zeta ,A}\bigtriangledown_{\alpha}S^{\beta\gamma\delta ,A}
]
\eea
Finally, the $\epsilon FFA$ term contributes
\be
\frac{1}{20}\epsilon^{\alpha\beta\gamma\delta\epsilon\eta\zeta}
S_{\epsilon\eta\zeta ,A}\bigtriangledown_{\alpha}S^{\beta\gamma\delta ,A}
\label{detk}
\ee
The condition of cancelation of the terms with 2 derivatives gives
$k=5/(6\sqrt{2})$, and we are left with
\be
\frac{1}{2}
S_{\alpha\beta\gamma ,A}
S^{\alpha\beta\gamma}_BT_{AB}
+\frac{1}{12}\epsilon^{\alpha\beta\gamma\delta\epsilon\eta\zeta}
S_{\epsilon\eta\zeta ,A}\bigtriangledown_{\alpha}S^{\beta\gamma\delta ,A}
\ee
exactly as we have in the seven dimensional action.
So we have still to recover only the $\epsilon SFF$ term. It comes from
two contributions,
\be
-\frac{3\sqrt{2}}{6(24)^2}\epsilon^{\mu\nu\rho\sigma}F_{\mu\nu\rho\sigma}
\epsilon^{\alpha\beta\gamma\delta\epsilon\eta\zeta}
{\cal A}_{\alpha\beta\gamma}F_{\delta\epsilon\eta\zeta}+\frac{{\cal B}^2}{48}
\ee
and they  sum up to
\be
\frac{i}{16\sqrt{3}}\epsilon^{\alpha\beta\gamma\delta\epsilon\eta\zeta}
S_{\alpha\beta\gamma ,A}\epsilon_{ABCDE}F_{\delta\epsilon}^{BC}F_{\eta\zeta}
^{DE}
\ee
as it should. This finishes the analysis of the bosonic action.

In the last part of this section we will make some  comments on other
ansatze for bosonic fields which appeared in the literature.
Most notably, we will try to see the relation with the truncation considered
in \cite{lupo} to the bosonic sector of
N=1 gauged supergravity with gauge group SU(2). A series
of papers \cite{cetal,cllp,cglp} also looked at other bosonic truncations
of 11d sugra to 7 dimensions with abelian gauge groups. Consistent truncations
to other dimensions were considered in \cite{cjlp,lpt,volk,chavo}.

The SU(2) truncation has the metric, one scalar X, an SU(2) gauge field
$A_{(1)}^i$ and a 3-form field $A_{(3)}$. The reduction ansatz in \cite{lupo},
written in form language, is
\bea
ds_{11}^2&=&\tilde{\Delta}^{1/3}ds_7^2+2g^{-2}X^3\tilde{\Delta}
^{1/3}d\xi^2+\frac{1}{2}g^{-2}\tilde{\Delta}
^{-2/3}X^{-1}cos ^2\xi \sum_i(\sigma^i-gA_{(1)}^i)^2\label{wow22}\\
\hat{F}_{(4)}&=&-\frac{1}{2\sqrt{2}}g^{-3}(X^{-8}sin ^2\xi-2X^2 cos ^2\xi
+3X^{-3} cos ^2\xi -4 X^{-3})\tilde{\Delta}
 ^{-2} cos^3\xi d\xi\wedge \epsilon_{(3)}
\nonumber\\&&
-\frac{5}{2\sqrt{2}}g^{-3}\tilde{\Delta}^{-2}X^{-4} sin \xi \cos ^4\xi dX\wedge
\epsilon_{(3)} +sin \xi F_{(4)}\nonumber\\&&
+\sqrt{2}g^{-1} cos \xi X^4 *F_{(4)}\wedge d\xi -\frac{1}{\sqrt{2}}g^{-2}
cos \xi F_{(2)}^i\wedge d\xi \wedge h^i\nonumber\\&&
-\frac{1}{4\sqrt{2}}g^{-2}X^{-4}\tilde{\Delta}^{-1}sin \xi cos ^2\xi F_{(2)}^i
\wedge h^j\wedge h^k\epsilon_{ijk}\label{truncation}
\eea
where $h^i=\sigma^i-gA_{(1)}^i,\;\; \epsilon_{(3)}=h^1\wedge h^2\wedge h^3$,
$\sigma^i$ are the three left-invariant forms on $S_3$, and
\be
\tilde{\Delta}=X^{-4}sin^2\xi +X cos^2\xi
\ee
and where the selfduality of $F_{(4)}$ is imposed by hand
\be
X^4 *F_{(4)}=-\frac{1}{\sqrt{2}}gA_{(3)}+\frac{1}{2}\omega_{(3)},
\;\;\; \omega_{(3)}=A_{(1)}^i\wedge F_{(2)}^i-\frac{1}{6}g\epsilon_{ijk}
A_{(1)}^i\wedge A_{(1)}^j\wedge A_{(1)}^k
\ee
The corresponding seven dimensional action is
\bea
{\cal L}&=&R*\id -\frac{1}{2} *d\phi\wedge d\phi -g^2(\frac{1}{4}e^{\frac{8}{
\sqrt{10}}\phi} -2e^{\frac{3}{\sqrt{10}}\phi}-2e^{-\frac{2}{\sqrt{10}}\phi}
) *\id \nonumber\\&&
-\frac{1}{2}e^{-\frac{4}{\sqrt{10}}\phi}*F_{(4)}\wedge F_{(4)}
-\frac{1}{2}e^{\frac{2}{\sqrt{10}}\phi}*F_{(2)}^i\wedge F_{(2)}^i
\nonumber\\&&
+\frac{1}{2}F_{(2)}^i\wedge F_{(2)}^i\wedge A_{(3)}-\frac{1}{2\sqrt{2}}
gF_{(4)}\wedge A_{(3)}
\label{sometrunc}
\eea
where $X=e^{-\frac{\phi}{\sqrt{10}}}$. On the other hand our bosonic action
is
\bea
e^{-1}{\cal L}'_{7d}
&=&-\frac{1}{2}R+\frac{1}{4}m^2(T^2-2T_{AB}T^{AB})
+\frac{1}{8}Tr(\partial_{\alpha}T^{-1}\partial^{\alpha}T)\nonumber\\
&-&\frac{1}{2}B_{\alpha}^{AB}(T^{-1}\partial_{\alpha}T)_{AB}+\frac{1}{2}
T^{AB}S_{\alpha\beta\gamma, A}S^{\alpha\beta\gamma}_{B}\nonumber\\
&+&\frac{1}{48}me^{-1}\epsilon^{\alpha\beta\gamma\delta\epsilon\eta\zeta}
\delta^{AB}S_{\alpha\beta\gamma ,A}F_{\delta\epsilon\eta\zeta ,B}
\nonumber\\
&-&\frac{1}{4}(T_{AC}T^{-1}_{BD}B_{\alpha}^{AB}B_{\alpha}^{CD}
-(B_{\alpha}^{AB})^2)+ \frac{m^{-1}}{8}e^{-1}\Omega_5[B]-\frac{m^{-1}}{16}
e^{-1}\Omega_3[B]
\nonumber\\&+&
\frac{i}{16\sqrt{3}}\epsilon^{\alpha\beta\gamma\delta\epsilon\eta\zeta}
S_{\alpha\beta\gamma ,A}\epsilon_{ABCDE}F_{\delta\epsilon}^{BC}F_{\eta\zeta}
^{DE}
\label{othertrunc}
\eea
and the metric and field strength  in form language are given in
(\ref{nicemetric}) and (\ref{f4}), respectively. A comparison of the
seven dimensional part of the metric tells us that $\tilde{\Delta}
=\Delta^{-6/5}$ and suggests the ansatz for embedding X into $T_{AB}$:
$T_{AB}=diag \{ X,X,X,X, X^{-4} \}$, and breaking the SO(5) invariance by
writing the 4-sphere in terms of  3-spheres as $Y^A=\{cos\xi
\tilde{Y}^{\hat{\mu}},
sin\xi\}$, with $\tilde{Y}^{\hat{\mu}}\;^2=1$.
Indeed, then we reproduce the form
of $\tilde{\Delta}$, and if we also choose $B^{\hat{\mu} 5}=0$, we get
\be
ds_{11}^2=\tilde{\Delta}^{1/3}ds_7^2+\tilde{\Delta}^{1/3}X^3d\xi^2 +
\tilde{\Delta}^{-2/3}X^{-1} cos^2\xi (d\tilde{Y}^{\hat{\mu}}+2 B^{\hat{\mu}
\hat{\nu}}\tilde{Y}^{\hat{\nu}})^2
\label{wow2}
\ee
and we also reproduce the ansatz for the
11 dimensional antisymmetric tensor at zero gauge field. The gauge field
dependence looks somewhat different. We are thankful to C.N. Pope, H.L\"u
and A. Sadrzadeh
for pointing to us that the ansatz in (\ref{wow22}) is in fact
contained in our general ansatz and is the same as the
one given in (\ref{wow2}).
We know that $SO(4)\simeq SU(2)\times SU(2)$.
One can restrict then the set of six gauge fields $B^{\hat\mu\hat\nu}$ to
one of the two sets of $SU(2)$ gauge fields by imposing a (anti)self-duality
condition
\be
B^{\hat\mu\hat\nu}=-\frac{1}{2}\epsilon^{\hat\mu\hat\nu\hat\rho\hat\sigma}
B_{\hat\rho\hat\sigma}\label{wow23}
\ee
As a consequence, with this constraint imposed on our initial set of gauge
fields, we are left with only three independent gauge fields, namely
$B^{\hat i\hat 4}=-1/2 A_{(1)}^i$, with $i=1,2,3.$ Also, (\ref{wow23}) implies
that
\be
B^{\hat\mu\hat\nu} B^{\hat\rho\hat\nu}=\frac{1}{4} \delta^{\hat\mu\hat\rho}B^2
=\frac{1}{4}\delta^{\hat\mu\hat\rho} B^{\hat\sigma\hat\tau} B^{\hat\sigma
\hat\tau}
\ee
Thus, when squaring $B^{\hat\mu\hat\nu}Y^{\hat\nu}$ in (\ref{wow2})
we get
\be
4B^{\hat\mu\hat\nu}Y^{\hat\nu}B^{\hat\mu\hat\rho}Y^{\hat\rho}=
\delta_{\hat\mu\hat\rho}Y^{\hat\mu}Y^{\hat\rho} B^2
=\sum_{i=1}^{3}(A_{(1)}^i)^2
\ee
Using the connection between Euclidean coordinates on $S_3$ and Euler angles
\bea
Y_{\hat 1}&=& \cos{\frac{\varphi+\psi}{2}}\cos{\frac{\theta}{2}}
\;\;\; Y_{\hat 2}=\sin{\frac{\varphi+\psi}{2}}\cos{\frac{\theta}{2}}
\nonumber\\
Y_{\hat 3}&=& \cos{\frac{\psi-\varphi}{2}}\sin{\frac{\theta}{2}}
\;\;\;Y_{\hat 4}=\sin{\frac{\psi-\varphi}{2}}\sin{\frac{\theta}{2}}
\label{relations}
\eea
we can now easily check that in (\ref{wow2}) another term in the
4-dimensional line element, namely $\sum_{\hat\mu=1}^{4} (d Y_{\hat\mu})^2$
corresponds to the term $1/4(d \theta^2+d\varphi^2 +d\psi^2+2d\varphi
d\psi \cos{\theta})=1/4\sum_{i=1}^3 d\sigma_i^2$ in (\ref{wow22}).
What remains to be shown to is that the cross terms, linear in the gauge field
also coincide in both relations.
This is indeed the case, as we can deduce by using the set of relations
(\ref{relations})
\bea
4B^{\hat\mu\hat\nu}dY^{\hat\mu} Y^{\hat\nu}&=&
4B^{\hat 1\hat 4}(dY^{\hat 1}Y^{\hat 4}-dY^{\hat 4}Y^{\hat 1}+
dY^{\hat 3} Y^{\hat 2} -dY^{\hat 2} Y^{\hat 3})+B^{\hat 2\hat 4}(\dots)
+B^{\hat 3\hat 4}(\dots)
\nonumber\\&=&-\sum_{i=1}^{3}A_{(1)}^i\sigma_i
\eea
In fact, we can understand the previous result by organizing the
four $Y_{\hat\mu}$ in a $2\times 2$ $SU(2)$ matrix
called ${\cal G}$. Then we obtain the $SU(2)$ invariant one-forms $\sigma_i$
by considering the product ${\cal G}^{-1} d{\cal G}$.

Our ansatz for the field strength of the 11-dimensional 3-index antisymmetric
tensor up to one gauge field and no $S_{(3)}$ reduces in the truncated case to
\bea
&&-\frac{1}{\sqrt{2}}g^{-3}(X^{-8}sin ^2\xi-2X^2 cos ^2\xi
+3X^{-3} cos ^2\xi -4 X^{-3})\tilde{\Delta}^{-2}\nonumber\\
&&\epsilon_{A_1...A_5}
(dY^{A_1}...dY^{A_4} Y^{A_5}+dY^{A_1}dY^{A_2}dY^{A_3}B^{A_4A_5})
\nonumber\\&&
-\frac{5}{\sqrt{2}}\tilde{\Delta}^{-2}X^{-4}\sin\xi\cos\xi dX\epsilon_{
A_1...A_5}dY^{A_1}dY^{A_2}dY^{A_3}M^{A_4}Y^{A_5}
\eea
where $M^{A_4}\equiv(\cos \xi,\sin\xi \tilde{Y}^{\hat{\mu}})$ and
 we have used here a certain rewriting of $F_{\mu\nu\rho\alpha}$, namely:
\bea
&&\frac{\sqrt{2}}{3}F_{\mu\nu\rho\alpha}=\frac{1}{3}\epsilon_{ABCDE}C_{\mu}^A
C_{\nu}^BC_{\rho}^C\left[
Y^D\partial_{\alpha}\left(\frac{T^{EF}Y_F}{Y\cdot T\cdot Y}
\right)-B_{\alpha}^{DE}\right]\nonumber\\&&
+\frac{2}{3}\epsilon_{\mu\nu\rho\sigma}B_{\alpha}^{AB}\partial^{\sigma}
\left(\frac{Y^AT^{BC}Y_C }{Y\cdot T\cdot Y}\right)
\eea
Working along the same lines as before, it can be shown that this ansatz
coincides with the one proposed by \cite{lupo} (for a complete discussion 
see \cite{nv}). The field equations obtained from our action 
(\ref{othertrunc}) also coincide with the field equations and the 
constraints of (\ref{sometrunc}). 
Hence, as observed by C.N.Pope, H.L\"u and A. Sadrzadeh, our general KK
ansatz \cite{nvv} for the maximal sugra in d=7  contains as a special 
case the N=2 model in \cite{lupo}. In conclusion, there is one 
consistent embedding which contains all (known) subcases in $d=7$ \cite{nv}.

\section{Discussion and conclusions}
In this paper we discuss the consistency of the
KK reduction
of the original formulation of 11d sugra \cite{cjs} on
$AdS_7\times S_4$ \cite{pnt} using the nonlinear ansatz presented
in a previous
letter \cite{nvv} for the embedding of d=7 fields in d=11.
Our ansatz for the metric factorizes into a rescaled 7 dimensional metric
and a gauge invariant two form which depends on the  composite
tensor $T^{AB}={{\Pi^{-1}}_i}^A{{\Pi^{-1}}_i}^B$, where ${\Pi_A}^i$ are
the coset elements for the scalar fields.
\bea
ds_{11}^2&=&\Delta^{-2/5}\left[ g_{\alpha\beta} dy^{\alpha}dy^{\beta}
+(DY)^A \frac{T^{-1}_{AB}}{Y\cdot T\cdot Y}(DY)^B\right]
\\
DY^A&=&dY^A+2B^{AB} Y_B\nonumber\\
\eea
The scale factor satisfies $\Delta^{-6/5}=Y\cdot T\cdot Y$.
We note that the full  internal metric $g_{\mu\nu}= \Delta^{4/5}\partial
_{\mu}Y^A T^{-1}_{AB}\partial_{\nu}Y^B$
has the geometrical meaning of a metric on an ellipsoid multiplied by a
conformal factor, namely $g_{\mu\nu}=\Delta^{4/5}\partial_{\mu}Z^A
\partial_{\nu}Z_A$ where $Z^A$ is constrained to lie on an ellipsoid.
Therefore the overall effect of all scalar fluctuations on the internal
metric is to deform the background sphere into a conformally
rescaled ellipsoid.

The 4-form field strength is given by
\bea
\frac{\sqrt{2}}{3}F_{(4)}&=&\epsilon_{A_1...A_5}\left[-\frac{1}{3}
(DY)^{A_1} ...(DY)^{A_4}\frac{(T\cdot Y)^{A_5}}{Y\cdot T\cdot Y}
\right.\nonumber\\&&\left.
+\frac{4}{3} (DY)^{A_1}(DY)^{A_2}(DY)^{A_3}D\left(\frac{(T\cdot Y)^{A_4}}{
Y\cdot T\cdot Y}\right)Y^{A_5}\right.\nonumber\\
&&\left.+2F_{(2)}^{A_1A_2}(DY)^{A_3}(DY)^{A_4}\frac{(T\cdot Y)^{A_5}}{Y\cdot
T\cdot Y}+ F_{(2)}^{A_1A_2}F_{(2)}^{A_3A_4}Y^{A_5}\right]
+d(S_{(3)B}Y^B)\nonumber\\
\eea
where $S_{(3)B\alpha\beta\gamma}=-\frac{8i}{\sqrt{3}}S_{\alpha\beta\gamma ,B}
$ is real. It is again gauge invariant but differs from the geometric
proposal of Freed, Harvey, Minasian and Moore \cite{hmm}
by the dependence on T.
Setting T=I we recover their result but our expression follows from
consistency
of the KK program  and still satisfies $DF_{(4)}=0$. To prove $DF_{(4)}=0$,
one may use the Schouten identity $\epsilon _{[A_1...A_5}Y_{B]}=0$. This
fixes all relative coefficients in $F_{(4)}$. Because in this process we
can at most convert one factor $T\cdot Y/(Y\cdot T \cdot Y)$ into a $Y$,
there do not seem possible deformations with two factors
 $T\cdot Y/(Y\cdot T \cdot Y)$.

A confirmation of our ansatz for $F^{(4)}$
is obtained by evaluating the term $\epsilon FFA$ in the 11 dimensional
action. We begin with $\epsilon FFF $ in d=12. The terms without bare B's
contain $F_{(2)}$, Y, $\partial_{\mu}Y$ and T's. Using (\ref{ey}) in reverse
order and the orthogonality relations in (\ref{4Y/YTY^2}), integration over
$S_4$ produces $2Tr F^4-(Tr F^2)^2$, which is indeed the exterior derivative
in the d=7 Chern Simons term. The B terms should not affect this result
because both $F^{(4)}$ and the final result are gauge invariant.

The 4-index antisymmetric auxiliary field ${\cal B}$ has only a
7-dimensional part,
\be
\frac{{\cal B}_{\alpha\beta\gamma\delta}}{\sqrt{E}}=
\frac{i}{2\sqrt{3}}\epsilon_{\alpha\beta\gamma\delta
\epsilon\eta\zeta}\frac{\delta S^{(7)}}{\delta S_{\epsilon\eta\zeta , A}}
Y^A\label{b}
\ee
On d=7 shell, ${\cal B}_{\alpha\beta\gamma\delta}$ should vanish and since it
should contain at most one derivative to exclude higher derivative terms
in the d=7 action, it can only be proportional to the field equation of
$S_{\alpha\beta\gamma ,A}$. Since it should mix with ${\cal A}_{\alpha\beta
\gamma}$ to produce selfduality  in odd dimensions, it should have the same
spherical harmonic as ${\cal A}_{\alpha\beta\gamma}$, and this explains
the factor $Y^A$ in (\ref{b}) and rules out an alternative
$(T\cdot Y)^A/(Y\cdot T\cdot Y)$.

Our ansatz for the fermions (\ref{psia}-\ref{epsi}) is the standard one
except for the following aspects:

(i) There are U matrices connecting the $SO(5)_c$ spinor indices of the
d=7 fermions to the $SO(5)_g$ indices of the Killing spinors.

(ii) Factors of $\sqrt{\gamma_5}$. They are fixed by requiring that $\psi_{
\alpha}(y,x)$ varies only into the Killing spinor ($\delta (d=11)\psi_{\alpha
}(y,x)=\delta (d=7)\psi_{\alpha I'}(y) {U^{I'}}_I(y,x) \eta^I (x)$).

The consistency of the
truncation is proven by obtaining the correct d=7 susy transformation laws
from d=11 ones.
We checked all transformation rules  except: (i) the variation with more than
one fermion field and (ii) some terms in the variation of the fermions which
depend only on scalars. No authors have ever determined the higher order
fermionic terms, but we would like to come back to them in the future.
Presumably, they fix the remaining freedom in $U$. As to the scalar terms in
$\delta\psi_\alpha$ and $\delta\lambda$, many consistency checks are
satisfied, but this is a complicated problem which deserves a separate study.
 Also here we expect that the remaining freedom in $U$ will get fixed.

We have followed de Wit and Nicolai \cite{dwn84,dwnw} by introducing a
matrix $U(y,x)$
which connects the d=7 fermions to the
Killing spinors in the ansatz for the gravitino\footnote{
Actually, in their work, the matrix $U$ interpolates between spinor indices,
whereas in our approach $U$ connects the $SO(5)_g$ indices $I$ of the
Killing spinors with the $SO(5)_c$ indices $I'$ of the $d=7$ fermions.}.
This matrix must satisfy equation (\ref{urel}),
$UY\llap/ =v\llap/ U$ with $v_i={{\Pi^{-1}}_i}^A Y_A \Delta^{3/5}$,
in order that the
 susy transformation laws  $\delta B_{\alpha}^{AB}$ for the gauge fields and
$\delta {\Pi_A}^i$ for the scalars come out correctly. We have found general
solutions to this equation, but for most of our results, the explicit form of
$U$ is not needed.

The ansatz for the antisymmetric tensor
field strength $F_{\Lambda\Pi\Sigma\Omega}$ was determined in 3 steps:
first, the dependence on the gauge fields was uniquely fixed by requiring
that the 11 dimensional susy variation of $F_{\Lambda\Pi\Sigma\Omega}$
matches the 7 dimensional susy variation of our ansatz for $F_{\Lambda\Pi
\Sigma\Omega}$.
Next, the ansatz for the embedding of the independent $d=7$
fluctuation $S_{\alpha\beta\gamma, A}$ was derived
together with the ansatz for the $d=11$ auxiliary field
${\cal B}_{MNPQ}$.
Finally, the dependence of $F_{\Lambda\Pi\Sigma\Omega}$ on the $d=7$ scalar
fields was fixed by requiring that one obtains the correct scalar
field potential in d=7 and the correct dependence of the d=7 sugra
transformation rules on the tensor $T_{ij}=(\Pi^{-1})_i\;^A (\Pi^{-1})_j\;^B
\delta_{AB}$ where $\Pi_A\;^i$ is the group vielbein for the scalars.

For completeness of our results, we have derived the 7 dimensional bosonic
action and bosonic equations of motion at zero gauge field from the
corresponding objects in 11 dimensions. Since in the bosonic sector all
criteria for consistency are satisfied (the consistency of the transformation
rules, equations of motion and even the action agrees) we infer that our
results also prove the consistency of the bosonic truncation.
We have shown the relation to other
ans\"{a}tze found in the literature for consistent truncation to subsets of
bosonic fields.

We have also explained the origin of self-duality in odd dimensions. One has
to use a first order action for $F_{\Lambda\Pi\Sigma\Omega}$ in d=11 to obtain
a self-dual  action for $S_{\alpha\beta\gamma ,A}$ in d=7 which is linear in
derivatives. The $d=7$ fluctuation $S_{\alpha\beta\gamma ,A}$
not only appears in the
$d=11$ curvatures
$F_{\mu\alpha\beta\gamma}$ and $F_{\alpha\beta\gamma\delta}$,
but also in the $d=11$ auxiliary field ${\cal B}$ in (\ref{b}), namely
in the form ${\cal B}\sim S+\epsilon DS$, and
substituting the ansatz for ${\cal B}$ and $F(S)$ into $d=11$ action, we
recovered the selfdual action in $d=7$.
We believe that in all cases, self-dual actions can be obtained
from KK reduction of a first order formalism for the corresponding field.

It may be useful to point out why in our opinion the original
$d=11$ sugra theory is to be preferred for purposes of
compactification over the $SU(8)$ version of de Wit and Nicolai.\\
(i)Their theory has a local $SO(3,1)\times SU(8)$ group in
tangent space. The $SO(3,1)$ becomes the Lorentz group in
$d=(3,1)$ dimensions, and thus it becomes immediately clear that
for compactifications to other dimensions we cannot use this
theory. (For compactification to $d=(6,1)$ we would need an 11d theory
with a $SO(6,1)\times SO(5)$ tangent group. For compactifications to other 
dimensions one would need other versions of 11d sugra than the SU(8) 
version, namely verisons with
$SO(2,1)\times SO(16), SO(4,1)\times USp(8)$ and $SO(5,1)\times SO(5)\times 
SO(5)$ tangent group. 
These were found in \cite{nicolai}.) On the other hand, the 
standard version of d=11 sugra can be used for compactifications to 
any dimension. 
\\ (ii)If we gauge fix
in the $SU(8)$ theory the group $SU(8)$ to $SO(7)$ and compare to
the standard $d=11$ sugra with $SO(10,1)$ gauged fixed to
$SO(3,1)\times SO(7)$ then in the $SU(8)$ theory the equations of
motion correspond to both equations of motion and Bianchi
identities in the usual $d=11$ sugra. This means that the two
gauged fixed theories are only equivalent on-shell. Of course, also in Cremmer
and Julia's theory for N=8 sugra in d=4, one needs to go temporarily on-shell
to dualize certain fields. Note that in our paper we never need to dualize and 
always stay off-shell.
\\ (iii)In fact, in the work of
de Wit and Nicolai there is no action for their SU(8) theory, only
field equations. To quote ref. \cite{dwn86}, page 389:
``...  the d=11 lagrangean depends explicitly on the antisymmetric 
tensor field $A_{MNP}$ for which no expression exists in terms of the 
SU(8) covariant expressions used in this paper.''
In the field equations only the field strength appears, but in the action
bare $A$'s appear.
\\(iv) To lift solutions of $d=4$ gauged
sugra to solutions of the usual 
$d=11$ sugra such as in \cite{cetal}, one would 
need the analog of our results for d=4.  The
ansatz of de Wit and Nicolai could be used to lift such solutions to 
 their $SU(8)$ theory. However, all
recent work on the consistency of truncations of the KK reductions
to $d=(3,1)$ has used  the standard d=11 theory
 \cite{clp}. In particular, all work on M theory  and membranes is based on 
the standard sugra theory.\\
 {\bf Acknowledgements.} We would like to thank I. Park for
collaboration at the early stages of this work, B. de Wit and H. Nicolai for 
useful discussions on their work and C.N. Pope, H. L\"u and A.
Sadrzadeh for pointing out that our bosonic ans\"atze reduce to
theirs \cite{lupo} when we further truncate ${\cal N}=4\; \;d=7$
to ${\cal N}=2\;\; d=7$ gauged sugra.

\appendix1
\subsection{Conventions and gamma matrix algebra}
We denote  in 7 dimensional Minkowski space the coordinates by $y$,
flat vector indices by a,b,c...
and curved vector indices by $\alpha, \beta, \gamma $...
Similarly, we denote in 4 dimensional Euclidean space the coordinates by $x$,
flat indices by $m,n,p,$ and curved vector indices by $\mu,\nu$...
Eleven dimensional vector indices are denoted by
$M,N,P$ for flat indices and $\Lambda, \Pi, \Sigma$... for curved indices.
The $SO(5)\simeq Usp(4)$ gauge group is denoted by $SO(5)_g$ and
$A,B... =1,5$ are $SO(5)_g$ vector indices and $I,J...=1,4$
are $Usp(4)$ indices (or $SO(5)_g$ spinor indices). The scalars form a coset
$Sl(5,R)/SO(5)$ and to avoid confusion we denote the composite subgroup by
$SO(5)_c$, with vector indices
$i,j...$ and spinor indices $I', J',...$=1,4.

The d=11
metric $g_{\Lambda\Pi}$ has signature mostly plus, $(-+...+)$.
The Clifford algebra in d=11 reads
\be
\Gamma_M \Gamma_N +\Gamma_N\Gamma_M=2\eta_{MN}\;\;;\;\; M,N=0,...,10
\ee
We introduce Dirac matrices in d=7 and d=4.
\bea
\tau_a\tau_b +\tau_b\tau_a&=&2\eta_{ab}\;\;;\;\; a,b=0,...,6\\
\gamma_m\gamma_n+\gamma_n\gamma_m&=&2\delta_{mn}\;\;;\;\; m,n=1,...,4
\eea
They are used to construct the d=11 Dirac matrices
\be
\Gamma_a = \tau_a\otimes\gamma_5 {\;\;\rm for\;\;} a=0,6 {\;\;\rm and\;\;}
\Gamma_{6+m}=1\otimes \gamma_m {\;\;\rm for\;\;} m=1,4
\ee
where $\gamma_5=\gamma_1...\gamma_4$ hence $\gamma_5^2=1$. We choose
$\tau_0=\tau_1...\tau_6$, hence
 $\Gamma_0=\Gamma_1...\Gamma_{10}$.
We normalize the $\epsilon$ symbols as
$\epsilon^{0...10}=\epsilon^{0...7}=\epsilon^{1...4}=+1$, so that
in d=7
\be
\tau^{[\alpha_1}...\tau^{\alpha_7]}\equiv
\tau^{\alpha_1...\alpha_7}=\epsilon^{\alpha_1...\alpha_7}\rightarrow
\tau^{\alpha_1...\alpha_k}=\epsilon^{\alpha_1...\alpha_7}\tau_{\alpha_{k+1}
...\alpha_7}\frac{(-1)^{[k/2+1]}}{(7-k)!}
\ee
where all antisymmetrizations are with strength one.
A similar duality relation between $\Gamma$ matrices holds in d=11:
\be
\Gamma^{\Lambda_1...\Lambda_k}=\frac{(-1)^{[k/2+1]}}{(11-k)!}\epsilon^
{\Lambda_1\Lambda_2...\Lambda_{11}}\Gamma_{\Lambda_{k+1}...\Lambda_{11}}
\ee

Also, for  general d,
\be
\epsilon^{\alpha_1...\alpha_k\alpha_{k+1}...\alpha_d}\epsilon_{\beta_1...
\beta_k\alpha_{k+1}...\alpha_d}=-k!(d-k)!\delta_{[\beta_1}^{[\alpha_1}...
\delta_{\beta_k]}^{\alpha_k]}
\ee

A definition of $\sqrt{\gamma_5}$ is obtained from $e^{i\alpha\gamma_5}
=\cos\alpha +i\sin\alpha \;\,\gamma_5$. Choosing $\alpha=\pi/2$ we get
 $\gamma_5=-ie^{i\frac{\pi}{2}\gamma_5}$, and thus
\be
 \gamma_5^{1/2}=
-\frac{1-i}{2}(1+i\gamma_5),\;\;\;  \gamma_5^{-1/2}=
-\frac{1+i}{2}(1-i\gamma_5)
\ee
 Then we also have
\bea
C\gamma_5^{1/2}C^{-1}&=&(\gamma_5^{1/2})^T\nonumber\\
\gamma_{\mu}\gamma_5^{1/2}&=&-i\gamma_5^{-1/2}\gamma_{\mu} \label{gamma5}
\eea
where $\gamma_5^{-1/2}=\gamma_5\gamma_5^{1/2}$ and $C\gamma_{\mu}C^{-1}=
-\gamma_{\mu}^T$, see next appendix.
Some useful formulas for gamma matrices in d dimensions are
\bea
\Gamma_a\Gamma^{b_1...b_n}\Gamma^a&=&(-)^n (d-2n)\Gamma^{b_1...b_n}\nonumber\\
\Gamma_a\Gamma^{b_1...b_n}&=&{\Gamma_a}^{b_1...b_n}+n\delta_a^{[b_1}
\Gamma^{b_2...b_n]}\nonumber\\
\Gamma_{a_1 a_2}\Gamma^{b_1...b_n}&=&{\Gamma_{a_1 a_2}}^{b_1...b_n}+
n\delta_{[a_2}^{[b_1}{\Gamma_{a_1]}}^{b_2...b_n]}\nonumber\\&&
+n(n-1)\delta_{a_2}^{[b_1}\delta_{a_1}^{b_2}\Gamma^{b_3...b_n]}\nonumber\\
\Gamma_{a_1...a_3}\Gamma^{b_1...b_n}&=&{\Gamma_{a_1a_2a_3}}^{b_1...b_n}+
2n\delta_{[a_1}^{[b_1}{\Gamma_{a_2a_3}}^{b_2...b_n]}\nonumber\\&&
+2n(n-1)\delta_{[a_3}^{[b_1}\delta_{a_2}^{b_2}{\Gamma_{a_1]}}^{b_3...b_n]}
\nonumber\\&&
+n(n-1)(n-2)\delta_{a_3}^{[b_1}\delta_{a_2}^{b_2}\delta_{a_1}^{b_3}\Gamma
^{b_4...b_n]}
\eea
 As representation for the
SO(5) Clifford algebra we take
\be
\gamma^A=\{i\gamma^{\mu}\gamma_5, \gamma_5\}\label{5gammas}
\ee
In 5 dimensions,
\bea
(C\gamma^A)_{IJ}(C\gamma^A)_{KL}&=&
+2(\Omega_{IK}\Omega_{JL}-\Omega_{JK}\Omega_{IL})-\Omega_{IJ}\Omega_{KL}
\label{gg}
\\
(C\gamma^{AB})_{IJ}(C\gamma^{AB})_{KL}&=&4(\Omega_{IK}\Omega_{JL}
+\Omega_{IL}\Omega_{JK})\label{gggg}
\eea
because these tensors are USp(4) invariant tensors with definite symmetry
properties. Hence they should be constructed from the symplectic metric
$C_{IJ}$.We choose a gamma matrix representation such that
$C_{IJ}=\Omega_{IJ}$.
(This matrix $C_{IJ}$ is thus the charge conjugation matrix
$C^{(5)}=C^{(5)}_+$ in 5 dimensions, see below. It equals $C^{(4)}_-$).

 We lower USp(4) indices  with $\Omega_{IJ}$ as follows
\be
\lambda_I=\lambda^J\Omega_{JI}\label{lower}
\ee
and raise them  with $\tilde{\Omega}^{IJ}$ as
\be
\lambda^I=\tilde{\Omega}^{IJ}\lambda_J
\ee
where $\tilde{\Omega}^{IJ}$ is defined by $\tilde{\Omega}
^{IJ}\Omega_{IK}=\delta^J_K$. Of course, $\tilde{\Omega}$ is $-\Omega^{-1}$
and can be obtained from $\Omega_{IJ}$ by raising indices with $\tilde
\Omega^{IJ}$. The result is $\tilde{\Omega}^{IJ}
=\Omega^{IJ}$. We have found it convenient to use a different
symbol for $\Omega^{IJ}$, namely $\tilde{\Omega}^{IJ}$.

\subsection{Charge conjugation matrices and modified Majorana spinors}
In even dimensions there are two charge conjugation matrices, $C_{(+)}$ and
$C_{(-)}$, satisfying $C_{(\pm)}\gamma^{\mu}C^{-1}_{(\pm})=\pm\gamma^{\mu ,T}
$. In odd dimensions there is only one charge conjugation matrix, either
$C_{(+)}$ or $C_{(-)}$. They are either symmetric or antisymmetric, and all
their properties are independent of the representation chosen. For a  general
discussion of Majorana and modified Majorana spinors in Minkowski or Euclidian
space and charge conjugation matrices, see \cite{houch}.

In 11 Minkowski dimensions, there is only one C matrix.
It satisfies
\be
C^{(11)}\Gamma_M C^{(11)-1}=-\Gamma_M^T
\ee
hence $C^{(11)}=C^{(11)}_{(-)}$. Furthermore, $C^{(11),T}=-C^{(11)}$.

In 7 Minkowski dimensions,  $C^{(7)}=C^{(7)}_{(-)}$, but is symmetric
\be
C^{(7)}\tau_a C^{(7)-1}=-\tau_a^T,\;\;\; C^{(7),T}=C^{(7)}
\ee
In d=4 both a $C^{(4)}_{(+)}$ and a $C^{(4)}_{(-)}$ exist. Both are
antisymmetric, but $C^{(11)}=C^{(7)}\otimes C^{(4)}$, where $C^{(4)}=C^{(4)}
_{(-)}$.

Then $C\gamma_{\mu}$ is symmetric and $C\gamma_5$ antisymmetric (since $C^T=
-C$). It also follows that for the  5-dimensional gamma matrices, $C\gamma^A$
in (\ref{gg}) are antisymmetric and the matrices $C\gamma^{[AB]}$ in
(\ref{gggg}) are symmetric.
This means that $(C\gamma^A)_{[IJ]}$ and $(C\gamma^{[AB]})_{(IJ)}$ give the
isomorphisms between the {\bf
5}, respectively the {\bf 10} representations of SO(5) and Usp(4) (SO(5)$\simeq
$Usp(4)).

The Majorana condition in 11 dimensions is
\be
\Psi^T C^{(11)}=i\Psi^{\dag}\Gamma^0
\ee
In 7 Minkowski dimensions we can only define a modified Majorana condition,
\be
(\lambda_I)^TC^{(7)}\tilde{\Omega}
^{JI}=i(\lambda_J)^{\dag}\tau^0\equiv \bar{\lambda}^J\label{moma7}
\ee

In the text we always define $\bar{\lambda}_I\equiv \lambda_I^T C_{(7)}$.
Using (\ref{moma7}), the spacetime spinors satisfy
\be
\bar{\lambda}_I =(\lambda_J)^\dag i\tau^0\Omega_{JI}
\ee
To determine which Majorana condition we need in d=4 Euclidean space,
 we decompose the anticommuting 11
dimensional Majorana spinors $\Psi$ into 4 anticommuting 7-dimensional modified
Majorana spinors times corresponding commuting 4-dimensional
hspinors $\eta^I$. The spinors $\eta$ must satisfy the following modified
Majorana condition
\be
(\eta^K)^TC_{(-)}^{(4)}\Omega_{KJ}=-(\eta^J)^{\dag}\gamma_5
\label{momaj}
\ee
Since in 4 Euclidian dimensions there are 2 choices for $C^{(4)}$, we can
use $C^{(4)}_{(+)}=C_{(-)}^{(4)}\gamma_5$, satisfying $C_{(+)}^{(4)}\gamma_m
C_{(+)}^{(4)-1}=\gamma_m^T$, in terms of which (\ref{momaj}) takes the same
form as (\ref{moma7}),
\be
(\eta^K)^T C_{(+)}^{(4)}\Omega_{KJ}=(\eta^K)^{\dag}
\label{momaj2}
\ee
Note that both (\ref{momaj2}) as well as the same condition with
$C_{(-)}^{(4)}$
instead of $C_{(+)}^{(4)}$ satisfy the consistency condition obtained by taking
 its complex conjugate and applying the relation twice \cite{houch}.

Also, in the text we always
define $\bar\eta^I\equiv (\eta^I )^T C_{(-)}^{(4)}$. Using (\ref{momaj}),
the Killing spinors satisfy
\be
 \bar{\eta}^I=-(\eta^J)^{\dag}
\gamma_5\tilde{\Omega}^{IJ}
\ee

In the fermionic ans\"atze we decompose the d=11 spinors into $\lambda
_{I'}(y) {U^{I'}}_I\eta^I(x)$. Using the Majorana conditions in the (10,1),
(6,1) and (4,0) dimensional spaces one obtains the condition
\be
({U^{I'}}_I)^*=\Omega_{I'J'}{U^{J'}}_J \tilde{\Omega}^{JI}
\ee

Combined with the relation $U\tilde{\Omega}U^T =\tilde{\Omega}$ derived
in the text, we deduce that U is unitary. Hence the matrices U are matrices
of the group USp(4), namely the intersection of U(4) and Sp(4,C).

\end{document}